\documentclass[twocolumn]{aastex631} 

\newcommand\omicron{o}

\usepackage{CJK}
\usepackage{float}
\usepackage[absolute]{textpos}
\usepackage{amsmath}
\usepackage{mathrsfs}
\usepackage{txfonts}
\usepackage{bm}  

\shorttitle{TMRT  Ka-Band Survey of the Orion KL} 
\begin{document}
\begin{CJK*}{UTF8}{gbsn}
\title{The first Ka-band (26.1-35 GHz) blind line survey towards Orion KL} 
\author[0000-0001-8315-4248]{Xunchuan Liu (刘 训川)}
\thanks{liuxunchuan@shao.ac.cn}
\affiliation{Shanghai Astronomical Observatory, Chinese Academy of Sciences, Shanghai 200030, PR China}
\author[0000-0002-5286-2564]{Tie Liu}
\thanks{liutie@shao.ac.cn}
\affiliation{Shanghai Astronomical Observatory, Chinese Academy of Sciences, Shanghai 200030, PR China}
\author[0000-0003-3540-8746]{Zhiqiang Shen}
\thanks{zshen@shao.ac.cn}
\affiliation{Shanghai Astronomical Observatory, Chinese Academy of Sciences, Shanghai 200030, PR China}
\author{Sheng-Li Qin}
\affiliation{School of Physics and Astronomy, Yunnan University, Kunming, 650091, PR China}
\author[0000-0003-4506-3171]{Qiuyi Luo}
\affiliation{Shanghai Astronomical Observatory, Chinese Academy of Sciences, Shanghai 200030, PR China}
\author{Yan Gong}
\affiliation{Max-Planck-Institut f\"ur Radioastronomie, Auf dem H\"ugel 69, 53125 Bonn, Germany}
\author[0000-0002-8691-4588]{Yu Cheng}
\affiliation{National Astronomical Observatory of Japan, 2-21-1 Osawa, Mitaka, Tokyo, 181-8588, Japan}
\author{Christian Henkel}
\affiliation{Max-Planck-Institut f\"ur Radioastronomie, Auf dem H\"ugel 69, 53125 Bonn, Germany}
\author{Qilao Gu}
\affiliation{Shanghai Astronomical Observatory, Chinese Academy of Sciences, Shanghai 200030, PR China}
\author{Fengyao Zhu}
\affiliation{Center for Intelligent Computing Platforms, Zhejiang Laboratory, Hangzhou, 311100, PR China}
\author[0000-0002-1466-3484]{Tianwei Zhang}
\affiliation{I. Physikalisches Institut, Universit{\"a}t zu K{\"o}ln, Z{\"u}lpicher Stra{\ss}e 77, 50937 K{\"o}ln, Germany}
\author{Rongbing Zhao}
\affiliation{Shanghai Astronomical Observatory, Chinese Academy of Sciences, Shanghai 200030, PR China}
\author{Yajun Wu}
\affiliation{Shanghai Astronomical Observatory, Chinese Academy of Sciences, Shanghai 200030, PR China}
\author{Bin Li}
\affiliation{Shanghai Astronomical Observatory, Chinese Academy of Sciences, Shanghai 200030, PR China}
\author{Juan Li} 
\affiliation{Shanghai Astronomical Observatory, Chinese Academy of Sciences, Shanghai 200030, PR China}
\author{Zhang Zhao}
\affiliation{Shanghai Astronomical Observatory, Chinese Academy of Sciences, Shanghai 200030, PR China}
\author{Jinqing Wang}
\affiliation{Shanghai Astronomical Observatory, Chinese Academy of Sciences, Shanghai 200030, PR China}
\author{Weiye Zhong}
\affiliation{Shanghai Astronomical Observatory, Chinese Academy of Sciences, Shanghai 200030, PR China}
\author{Qinghui Liu}
\affiliation{Shanghai Astronomical Observatory, Chinese Academy of Sciences, Shanghai 200030, PR China}
\author{Bo Xia}
\affiliation{Shanghai Astronomical Observatory, Chinese Academy of Sciences, Shanghai 200030, PR China}
\author{Li Fu}
\affiliation{Shanghai Astronomical Observatory, Chinese Academy of Sciences, Shanghai 200030, PR China}
\author{Zhen Yan}
\affiliation{Shanghai Astronomical Observatory, Chinese Academy of Sciences, Shanghai 200030, PR China}
\author{Chao Zhang}
\affiliation{Shanghai Astronomical Observatory, Chinese Academy of Sciences, Shanghai 200030, PR China}
\author{Lingling Wang}
\affiliation{Shanghai Astronomical Observatory, Chinese Academy of Sciences, Shanghai 200030, PR China}
\author{Qian Ye}
\affiliation{Shanghai Astronomical Observatory, Chinese Academy of Sciences, Shanghai 200030, PR China}
\author{Aiyuan Yang}
\affiliation{National Astronomical Observatories, Chinese Academy of Sciences, Beijing 100101, PR China}
\author[0000-0001-5950-1932]{Fengwei Xu}
\affiliation{Kavli Institute for Astronomy and Astrophysics, Peking University, 5 Yiheyuan Road, Haidian District, Beijing 100871, PR China}
\author[0000-0002-5682-2906]{Chao Zhang}
\affiliation{Institute of Astronomy and Astrophysics, School of Mathematics and Physics, Anqing Normal University, Anqing, PR China}
\author[0000-0002-2338-4583]{Somnath Dutta}
\affiliation{Institute of Astronomy and Astrophysics, Academia Sinica, Roosevelt Road, Taipei 10617, Taiwan (R.O.C)}
\author{Shanghuo Li}
\affiliation{Max Planck Institute for Astronomy, K\"onigstuhl 17, D-69117 Heidelberg, Germany}
\author{Meizhu Liu} 
\affiliation{School of Physics and Astronomy, Yunnan University, Kunming, 650091, PR China}
\author{Dongting Yang} 
\affiliation{School of Physics and Astronomy, Yunnan University, Kunming, 650091, PR China}
\author{Chuanshou Li} 
\affiliation{School of Physics and Astronomy, Yunnan University, Kunming, 650091, PR China}
\author{Li Chen} 
\affiliation{School of Physics and Astronomy, Yunnan University, Kunming, 650091, PR China}

\begin{abstract}
We conducted a Ka-band (26.1--35 GHz) line survey  
towards Orion KL using the TianMa 65-m Radio Telescope (TMRT). 
It is the first blind line survey in the Ka band, and achieves a sensitivity of
mK level (1--3 mK at a spectral resolution of $\sim$1 km s$^{-1}$).
In total, 592 Gaussian features are extracted.
Among them, 257 radio recombination lines (RRLs) are identified.
The maximum $\Delta n$ of RRLs of H, He and C are 20, 15, and 5, respectively.
Through stacking, we have detected the $\beta$  lines of ion RRLs (RRLs of C$^+$ 
with possible contribution of other ions like O$^+$) for the first time,
and tentative signal of the $\gamma$ lines of ion RRLs can  also be seen on the stacked spectrum.
Besides, 318 other line features were  assigned to  37 molecular species,
and ten of these species were not detected in the Q-band survey of TMRT.
The vibrationally excited states of nine species were also detected. 
Emission of most species can be modeled under LTE.
A number of transitions of E-CH3OH ($J_2-J_1$)  display
maser effects, which are confirmed by our modeling,  and
besides the bumping peak at $J\sim 6$  there is another peak at  $J\sim 13$.
Methylcyanoacetylene (CH$_3$C$_3$N) is detected in Orion KL for the first time.
This work emphasizes that the
Ka band, which was long-ignored for spectral line surveys, 
is very useful for surveying RRLs and molecular lines simultaneously.
\end{abstract}


\section{Introduction}\label{sec_intro}

\begin{table*}
\centering
\caption{The line surveys of Orion KL in low frequency range ($\nu<50$ GHz).
\label{tablelowfreqsurvey}}
\begin{tabular}{ccccc}
\hline\hline
Frequency coverage$^{(1)}$ & Telescope  & Spectral resolution  &  Sensitivity$^{(2)}$ & Reference$^{(4)}$ \\ 
\hline
17.9--26.2 GHz & Effelsberg 100 m & 61 kHz ($\sim$0.7 km s$^{-1}$) & 5--22 mK & \citet{Gong2015}  \\
26.1--35 GHz   & TMRT 65 m        & 92 kHz ($\sim$0.9 km s$^{-1}$)  & 1--3 mK  & This work \\
34.8--50 GHz           & TMRT 65 m      & 92 KHz ($\sim$0.65 km s$^{-1}$) &
1.5--5 mK$^{(3)}$ &  \citet{2022ApJS..263...13L} \\
41.5--50 GHz    & DSS-54 34 m     &    180 kHz ($\sim$1.2 km s$^{-1}$)   & 8--12 mK & \citet{Rizzo2017} \\
\hline
\end{tabular}\\
{\raggedright 
$^{(1)}$ Only wide-band blind line surveys are listed here.\\
$^{(2)}$ The line sensitivity in $T_{\rm MB}$ scale at the corresponding spectral resolution.\\
$^{(3)}$ The line sensitivity at 49--50 GHz is slightly worse with a value
of 5--8 mK.\\
$^{(4)}$ Little can be known about the line survey of \citet{Ohishi1986} 
from its relevant
published work. From the spectra of all of its covered frequency bands
(34.25--50, 83.5--84.5, and 86--91.5 GHz), 
\citet{Ohishi1986} detected 19 molecular species, much smaller than 
the values of 39 and 53 detected in Q band alone by \citet{Rizzo2017} and  \citet{2022ApJS..263...13L}. \\
}
\end{table*}

The low-frequency ($<$50 GHz) line surveys have  advantages
in searching for complex organic molecules (COMs) and radio recombination lines
(RRLs),
because low-frequency lines tend to be non-seriously blended and more easily excited compared with those in higher frequency ranges.
In  contrast to the ample surveys at 
the high-frequency  millimeter and submillimeter bands  
\citep[$\nu>70$ GHz; e.g.,][]{1984A&A...130..227J,
1989ApJS...70..539T,1997ApJS..108..301S,Schilke_2001,2003A&A...407..589W,
2010A&A...517A..96T},  there are only a scarcity of  line surveys  at low-frequency bands
towards Orion KL \citep[e.g.][]{2015MNRAS.452.3969C,Gong2015,Rizzo2017}, which 
is the nearest \citep[$\sim$414 pc;][]{2007A&A...474..515M} high-mass star forming region and
one of the most representative objects of line surveys.
The emission of Orion KL 
is complex \citep[e.g.,][]{2010A&A...517A..96T}, with contributions from several physical components including the
foreground {H}{\sc ii} region  M42 \citep{1997A&A...327.1177W}, the PDR
between M42 and the molecular cloud \citep[e.g.,][]{1994ApJ...428..209N}, the 
externally heated ``compact ridge'' \cite[e.g.,][]{1993ApJS...89..123M,2011A&A...527A..95W,2016ApJ...832...12T},
the hot cores \citep[e.g.,][]{1993ApJS...89..123M,2021A&A...647A..42J},
the extended ridge and plateau \citep[e.g.,][]{1989ARA&A..27...41G,2021ApJ...906...55B} 
as well as some other millimeter continuum sources
\citep[e.g.,][]{2014ApJ...791..123W}. It makes Orion KL rich in spectral lines at different frequency bands
\citep[e.g.,][]{1987ApJ...315..621B,1997ApJS..108..301S,2010A&A...517A..96T,2014ApJ...787..112C,Rizzo2017}.
This motivated us to conduct a multi-band low-frequency line survey using the Tianma 65-m Radio Telescope (TMRT) towards 
Orion KL  (referred as the TMRT line survey below), aiming to obtain an unprecedented  
sensitive template for low-frequency spectra  (with a line sensitivity of one or several mK), covering the whole
working frequency range (1--50 GHz) of the TMRT.

The ongoing TMRT line survey began with the
Q-band (35-50 GHz) pioneer survey \citep{2022ApJS..263...13L},
which is so far the most sensitive one compared with previous wide-band Q-band surveys of  Orion KL
\citep{Ohishi1986,Rizzo2017}. 
It reached a sensitivity at a level of 1.5-5 mK (Table \ref{tablelowfreqsurvey})
and 597 Gaussian features were extracted, proving 
the capability for the TMRT  of conducting deep line surveys.
In the Q band, the number of detected molecular lines, 395, is more than twice the number of RRLs, 153. 
Most of the Q-band molecular lines are emission of complex organic molecules (COMs). 
In contrast, the K-band (18--26 GHz) spectrum of Orion KL was found to be dominated by RRLs, 
as revealed by the survey of \citet{Gong2015} 
that have identified 164 RRLs among the 261 detected lines.
It is reasonable to expect that the Ka-band (26--35 GHz) spectrum of Orion KL
would have comparable numbers of RRLs and molecular lines, making the Ka band 
very suitable for observing COMs and RRLs simultaneously.

Despite its promising potential uses, the Ka band  was long-ignored for spectral line surveys, 
and acted as a frequency gap especially for single-dish observations.
This frequency range is usually preferred for satellite communication
\citep[e.g.,][]{VANELLICORALLI2015303},
and was historically avoided for radio astronomy.
It makes the Ka band a missing piece for previous line surveys (Table \ref{tablelowfreqsurvey}).
To fill in this blank  gap, we conducted a Ka-band (26--35 GHz)  survey of Orion KL 
as a part of the TMRT line survey, following the
Q-band one \citep{2022ApJS..263...13L}.  
Through long-time integrations, a line sensitivity of 1--3 mK was achieved
with a frequency resolution of 91.553 kHz.
This is the first blind line survey in the Ka band reaching a line sensitivity of
mK level.

In \citet{2023AA...671L...1L},
we have reported the successful detection of the 
ion RRLs (X$^+$ RRLs with X$^+$ referred to C and/or O) in the interstellar medium for the first time,
through combining the data of the TMRT surveys in three bands (Ka, Ku, and Q). 
This Ka-band survey is the most sensitive one, compared with the less sensitive Q-band survey and
the partly executed Ku-band survey.  
It expanded the scope of our understanding of ion RRLs, because prior to that work,
only two blended lines (105$\alpha$ and 121$\alpha$) of
\ion{He}{2} in planetary nebulae had been reported \citep{1976ApJ...210..108C,1980ASSL...80...75T,1990A&A...230..457V}.

In this work, we present the overall  results of the Ka-band (26.1--35 GHz) line survey  of Orion KL. 
The paper is structured as follows: In Section \ref{secobs} we
introduce the  observations.
In Section \ref{sec_datareduct_lineid} we describe the procedures of data reduction, line identification and emission modeling
of RRLs and molecular lines.
In Sections \ref{secrrl} and \ref{secmols}, we present more detailed results about the RRLs and molecular lines,
respectively.
We provide a summary in Section \ref{secsummary}.

%

\begin{figure*}[!htp]
\centering
\includegraphics[width=0.98\linewidth]{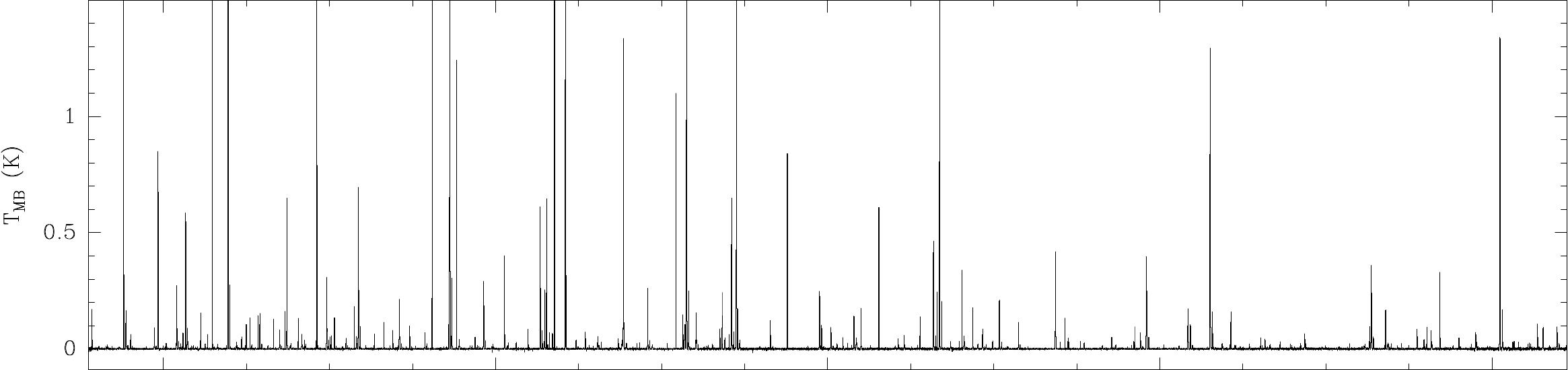}
\vskip 0.2cm
\includegraphics[width=0.98\linewidth]{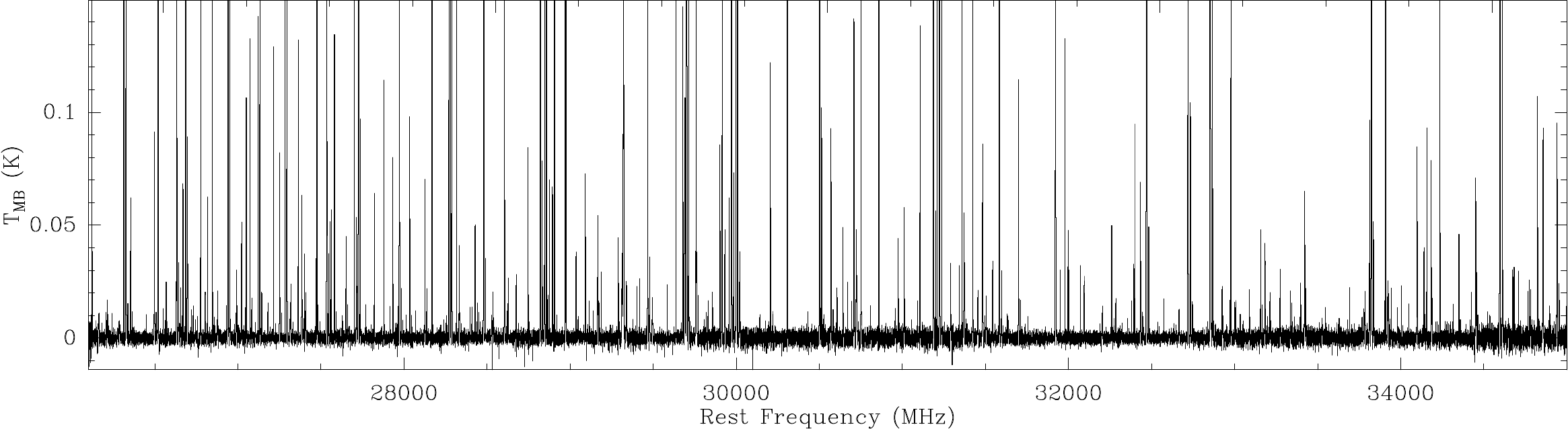}
\caption{The overview of the   Ka-band (26.1--35 GHz) spectrum of Orion KL
observed by the TMRT. Lower panel: A zoomed-in view of the same spectrum. \label{fig_overall_spe} 
(The data used to create this figure are available online.)}
\end{figure*}

\begin{figure}[!htb]
\centering
\includegraphics[width=0.99\linewidth]{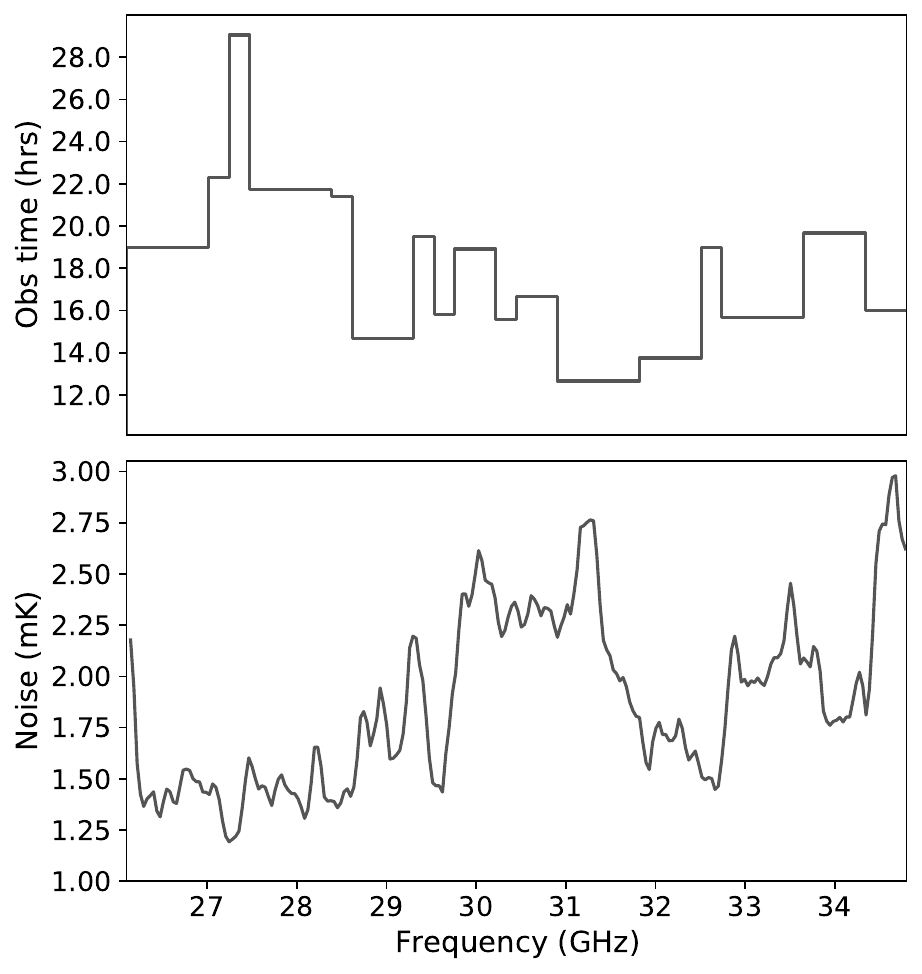}
\caption{The observation time (upper) and rms noise (lower) of the Ka-band
spectrum. Both the on-source and off-source integration time
have been included to calculate the observation time. 
\label{noise_freq} }
\end{figure}

\begin{figure*}[!thp]
\centering
\includegraphics[width=0.99\linewidth]{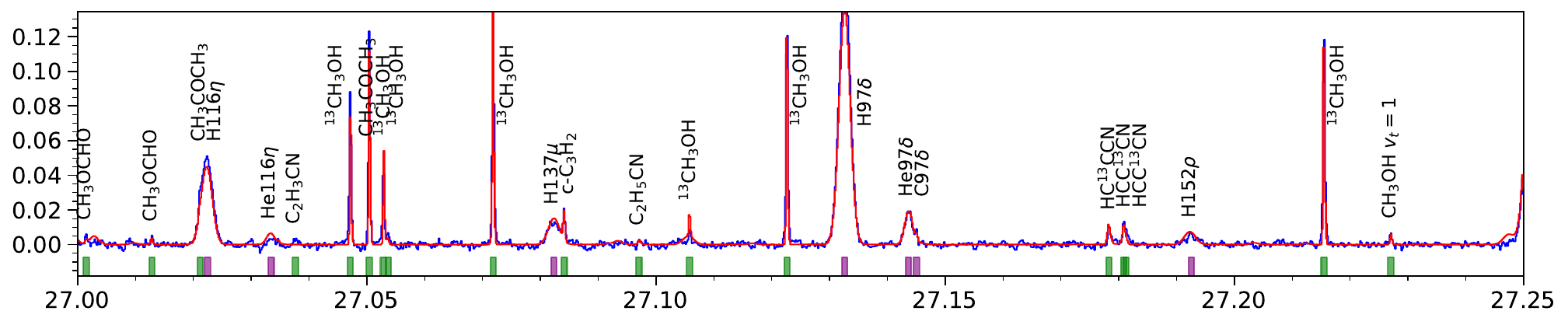}\\
\includegraphics[width=0.99\linewidth]{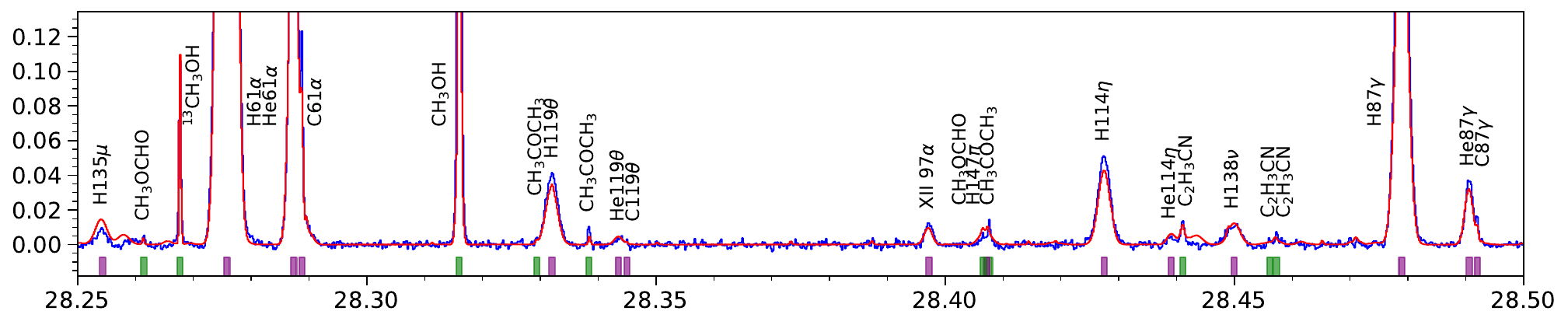}\\
\includegraphics[width=0.99\linewidth]{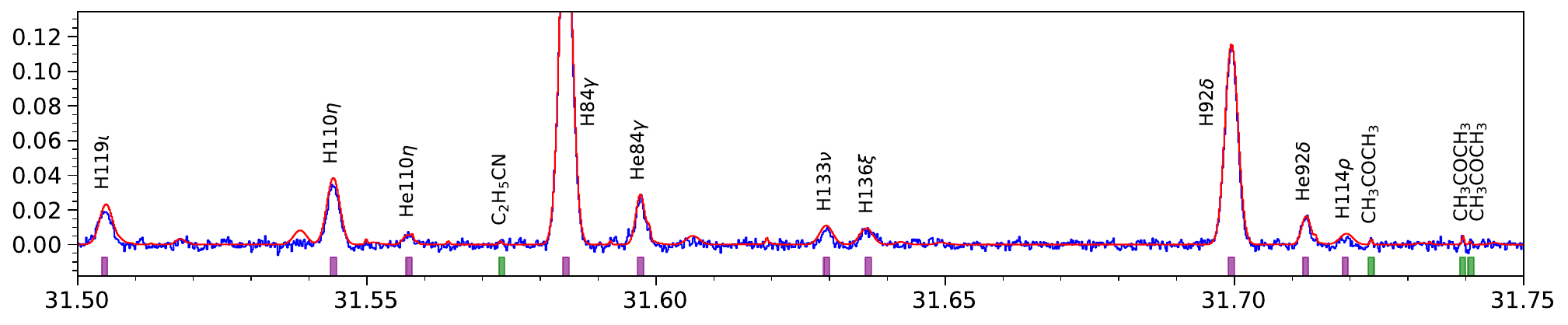}\\
\includegraphics[width=0.99\linewidth]{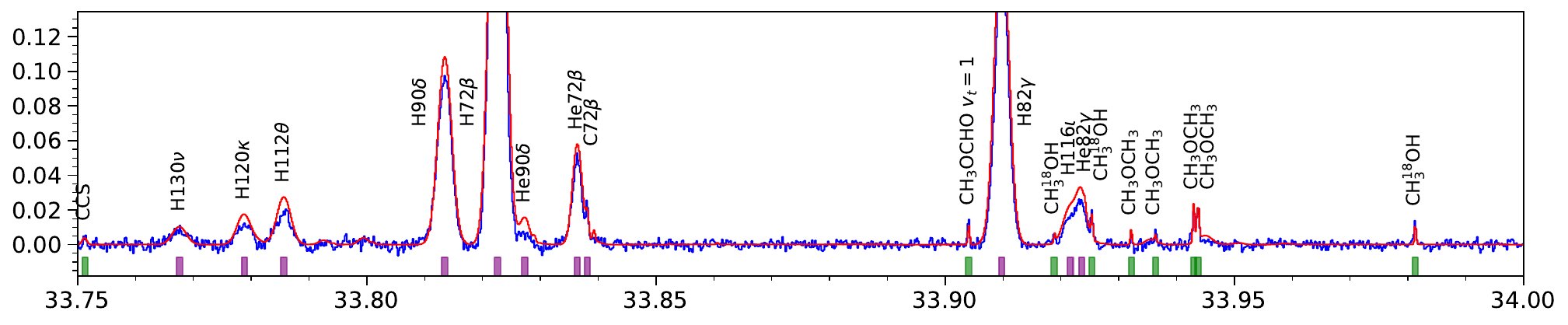}\\
\caption{The images of Ka-band spectrum of Orion KL. 
The x-axis is rest frequency in units of gigahertz 
(with a Doppler correction applied adopting a $V_{\rm LSR}$ of 6 km s$^{-1}$). 
The $y$-axis is $T_{\rm MB}$ in units of kelvin.
The blue line is the Orion KL spectrum 
observed by the TMRT 65 m,
which has been smoothed to have a frequency resolution of 183 kHz ($\sim$1.8 km s$^{-1}$ at 30 GHz). 
The red line represents the modeling fitting. The purple strips denote the detected RRLs.  
The green strips denote the 
molecular lines.
The gray strips mark the U lines.
See Figure \ref{allspectra} for all the spectral images.
\label{allspectra_first}}
\end{figure*}

\section{Observations} \label{secobs}
\subsection{Ka-band receiving system of TMRT}
The observations were carried out using the TMRT, built and operated by the Shanghai Astronomical Observatory\footnote{\url{http://english.shao.cas.cn/sbysys/}}. 
The TMRT is located in  the western suburb of Shanghai, China. It has an aperture diameter of 65 m, corresponding to a
full-width at half-maximum (FWHM) of the 
primary beam of $\sim$46\arcsec--34\arcsec~at 26--35 GHz. 
Its Ka receiver is a two-beam dual-polarization (LCP and RCP) cryogenic receiver,
covering 26--40 GHz, with the 35--40 GHz overlapping with the working frequency range of the Q-band receiver.
The beam 2 was employed for our single-pointing observations. The receiver noise temperatures were 10-30 K,
and the system temperatures ($T_{\rm sys}$) ranged from 60 K to 150 K, depending on the
frequency, elevation and especially the weather conditions \citep{2017AcASn..58...37W}. 

During the preparation for  this survey, the Ka-band pointing model of the TMRT 
was updated through tracking
strong continuum sources, including DR21 and 3C84, for one whole day, following the
method of \citet{2017SSPMA..47l9504W}. 
The pointing accuracy was better than 5 arcsec.
The aperture efficiency is nearly constant (0.5$\pm$0.1) thanks to the 
active surface control utilizing actuators which can compensate for the gravity deformation
of the main reflector during observations \citep{2018ITAP...66.2044D}.
For calibration, the signal of noise diodes was injected lasting for one second within each two-second period.
In summer, the high water vapour content and deformation by sunlight
make TMRT difficult to be used for Ka/Q-band observations. 
Under typical weather conditions at the TMRT in the winter semester,
the zenith atmospheric opacities were 0.1 in the Ka band
\citep{2017AcASn..58...37W}.
Calibration uncertainties were estimated to be
within 20\%.

\begin{figure*}[!thp]
\centering
\includegraphics[width=0.995\linewidth]{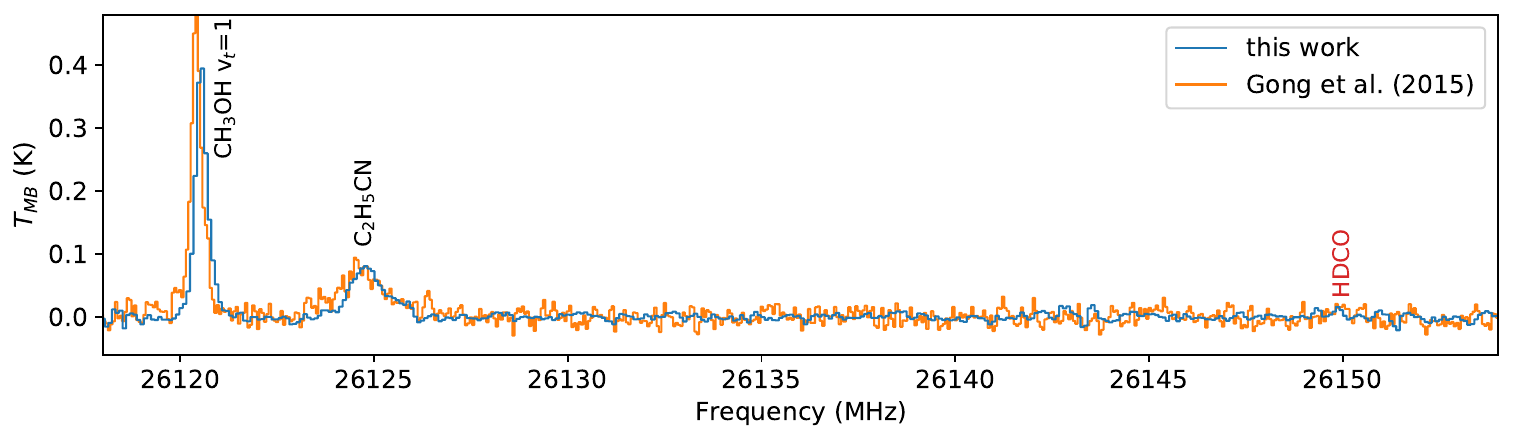}
\includegraphics[width=0.995\linewidth]{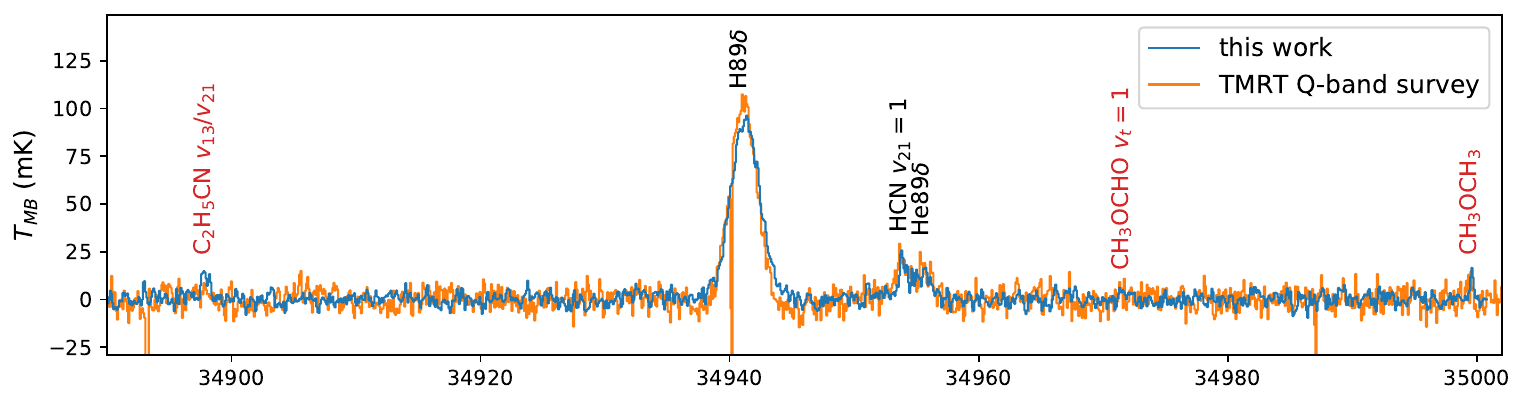}
\caption{Upper: Comparison between the spectra of this Ka-band (26.1--35 GHz) survey and of the K-band (17.9--26.2 GHz) 
survey of Effelsberg 100 m 
by \citet{Gong2015}.
See Section \ref{sec_datareduct} for possible reasons of the slight velocity discrepancy (by about two channels) between the two surveys.
Lower:  Comparison between the spectra of this Ka-band survey and the TMRT Q-band (34.8--50 GHz) survey \citep{2022ApJS..263...13L}. 
The overlapping frequency ranges of those surveys are located at the edges of the
frequency coverages, and consequently have relatively poor sensitiveties. 
Note that the lower panel has has wider frequency range than that of the upper panel.
\label{figcompareprevious}}
\end{figure*}

\subsection{The Ka-band survey} \label{sect_kasurveyobs}
This Ka-band line survey of Orion KL  covered 26.1--35 GHz.
Although the working frequency range of the K-band receiver is 26--40 GHz,
the 35--40 GHz was not covered by this survey, because it
has been done by our previous Q-band line survey \citep{2022ApJS..263...13L}.
The 26--26.1 GHz was not covered because of the limitation of the bandpass filter.
The digital backend
system (DIBAS) of TMRT is an FPGA-based spectrometer
based on the design of the Versatile GBT Astronomical
Spectrometer \citep{2012AAS...21944610B}, and it
provides 29 observing modes.
The  setup of the backend for the Ka-band survey is the same as that of the Q-band survey
\citep{2022ApJS..263...13L}. Two banks (the third one is not available at present) have a bandwidth of 1.5 GHz each,
and they work independently with no limitation on the separation between their central frequencies. 
Under the mode 2 of DIBAS, each polarization of each bank has 16384 channels, and the channel width is 91.553 kHz, 
corresponding to a velocity resolution ($\delta V$) of $\sim$0.92 km s$^{-1}$ at 30 GHz. 

The observations were conducted  from September 28th to 
October 30th, 2022. 
The targeting position is RA(J2000)=05:35:14.55, DEC(J2000)=$-$05:22:31.0.
Different physical components of Orion KL (see Section \ref{sec_intro})
can be covered by a single beam of the TMRT, as depicted in our previous Q-band survey \citep{2022ApJS..263...13L}.
Position-switching observation mode was adopted, with the off points 15\arcmin~away (in azimuth direction) 
from the target. The offset value is much larger than 4\arcmin,
the angular size of the brightest region of M42 \citep{1990ApJ...361L..19Y,1997A&A...327.1177W}. 
We periodically integrated two minutes in each position (on/off).
During observations of this survey, pointing    was checked every two hours.

For each frequency bank, its  frequency coverage  in the sky frequency scale was fixed.
The frequency of the local oscillator (LO) did not change during the observation for a single frequency setup.  
The spectrum of each on-off repeat was corrected from the topocentric frame to the frame of the
local standard of rest (LSR)
during data processing.  
The frequencies of the two banks were shifted in different setups  to cover 26.1--35 GHz.
For most frequency ranges of Ka band, there is no noticeable radio frequency interference (RFI) on the spectrum.
At some narrow frequency ranges (e.g., $\sim 26.4-26.8$ GHz), RFI can be sometimes very strong 
but it does not last for a long time.
Each frequency point was covered by at least two different frequency setups observed on different days.
The spectra with obvious  RFI were manually dropped.
Each of the observed frequency points is covered by a telescope time of 12--25 hours, 
depending on the weather conditions.

\section{Observational results}\label{sec_datareduct_lineid}
\subsection{Data reduction} \label{sec_datareduct}
The spectra of each polarization of each frequency were first averaged weighted by
$1/T_{\rm sys}^2$. 
The averaged spectrum was then chopped into segments of 100 MHz.  
For each segment,
we manually fit the spectral baseline and get the noise of each segment using the procedure
provided by Gildas/CLASS \citep{2000ASPC..217..299G}.
To conduct this procedure, the
frequency intervals that contain visible  line features or spectral emission 
predicted from the Q-band data \citep{2022ApJS..263...13L} have been masked out.
We further converted the frequency of the spectrum 
from the LSR frame to the rest frame of Orion KL 
assuming a systematic velocity of Orion KL ($V_{\rm LSR}$)  of 6 km s$^{-1}$. 
Finally, we spliced together all the segments weighted by their noise levels to obtain
the final Ka-band spectrum, covering  26.1--35 GHz with a frequency resolution of 91 kHz.  

The final spectrum (in $T_{\rm MB}$ scale) 
is shown in Figure \ref{fig_overall_spe}.
To estimate the rms noise of the spectrum, we first masked out all the extracted emission features and 
bad channels (Sect. \ref{sect_linefeature}), and then for each channel we calculated the
standard deviation of the intensities of its unmasked neighbour channels within 25 MHz.
As expected, the rms noise is roughly inversely proportional to the integration time.
The 26.1--26.2 GHz is close to the edge of the working frequency of the
TMRT in the Ka band, and can not be placed in the inner region of a frequency bank which has
better performance. Thus, the noise level of 26.1--26.2 GHz is slightly higher (Figure \ref{noise_freq}).  
The rms noise of the final spectrum  
ranges from 1 mK to 3 mK, with a mean value of 1.87 mK and a
median value of 1.80 mK.

\begin{table*}[!thb]
\caption{The number of RRLs and molecular lines detected by different surveys$^{(1)}$.\label{tablineover}}
\centering
\begin{tabular}{lccccccccc}
\hline
\hline
Ref. & $N_l$ & $\rho_l$ & $N_{\rm  RRL}^{(2)}$& $N_{\rm  mol}^{(3)}$& $N_{RRL}^{\rm mol}$ & $N_{\rm U}$ & $N_{\rm  RRL}$/$N_{\rm  mol}^{(4)}$\\
\hline
This work & 592 & 64  & 257(12) & 318(20) & 7 &10 &0.9\\
\citet{2022ApJS..263...13L} & 596    & 36  & 153(13) & 395(50) & 9 &39 & 0.6\\
\citet{Rizzo2017} & 227  & 26 & 66 & 143 & -- & 18 & 0.5\\
\citet{Gong2015} & 261& 32 & 164 & 97 & -- & --& 1.7\\
\hline
\end{tabular}\\
{\raggedright 
$^{(1)}$ $N_l$ is the total number of detected line features. $\rho_l$ is the averaged number of detected line features per GHz.
$N_{\rm RRL}$ is the number of RRLs, including those blended with other RRLs.
$N_{\rm mol}$ is the number of molecular lines, including those blended with other molecular lines.
$N_{\rm RRL}^{\rm mol}$ is the number of line features blended by both RRLs and molecular emission.
$N_{\rm U}$ is the number of U lines. \\
$^{(2)}$ The values in the brackets represent the number of line features blended by more than one RRL but no molecular transition.\\
$^{(3)}$ The values in the brackets represent the number of repeated molecular line features because of multiple velocity components.\\
$^{(4)}$ The ratio between the number of detected transitions of RRLs and molecular transitions.\\
}
\end{table*}

\begin{figure}
\centering
\includegraphics[width=0.98\linewidth]{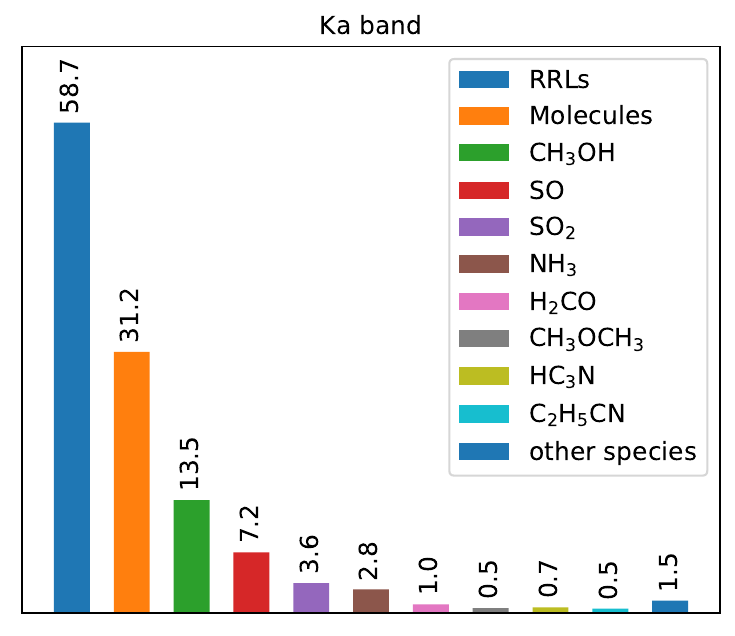}
\includegraphics[width=0.98\linewidth]{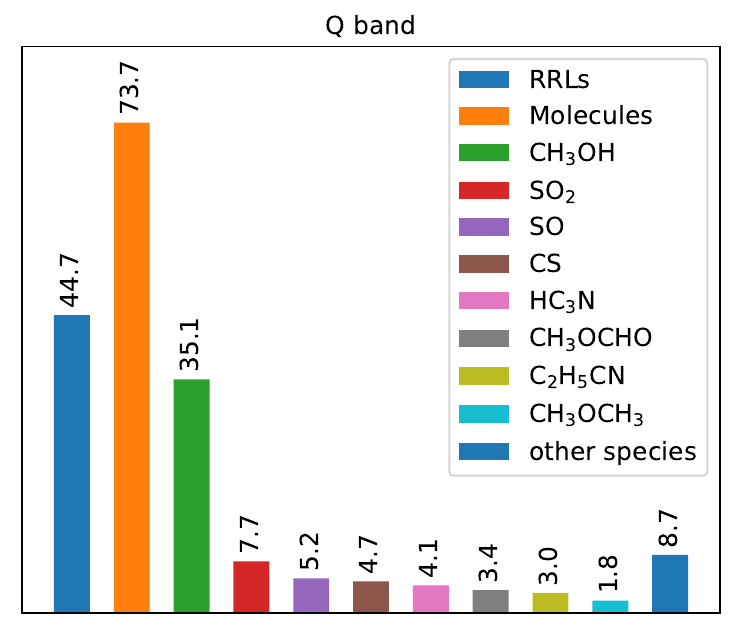}
\caption{The total Ka-band (upper) and Q-band (lower) emission flux (in units of K MHz) of detected species. \label{fig_flux_kaq}}
\end{figure}

Figure \ref{allspectra_first} shows examples of the zoom-in  
Ka-band spectrum. 
The lower end of the Ka-band spectrum (26.1--26.16 GHz) was also covered by the K-band survey of the Effelsberg 100 m 
telescope \citep{Gong2015}.
The upper end of the  Ka-band spectrum (34.88--50 GHz) was also covered by the Q-band survey of the TMRT \citep{2022ApJS..263...13L}.
Figure \ref{figcompareprevious} compares the  overlapped spectra of the three surveys.
Note that the transition of CH$_3$OH at $\sim$26120 MHz on the K-band spectrum is slightly
redder (by two channels) than that on the Ka-band spectrum.
This may be caused by the slightly different targeting centers of the two surveys.
The target of the K-band survey is located at 16\arcsec~south-west of the target of this survey,
which is smaller than the beam sizes of both the TMRT and the Effelsberg 100 m.
The intersection between the fields of view of the two surveys covers more than half the primary beam
of the TMRT (see Figure 1 of \citet{2022ApJS..263...13L}). 
We modeled the K-band spectral emission based on the Ka band spectra, and
compared the observed and modeled spectra of RRLs and CH$_3$OH in the K band.
The K-band spectra around 26 GHz are all redshifted by about two channels compared with the model.
However, this is not a systematical trend in the whole K band since in some frequency ranges the velocities match each other. 
The velocities of the Q-band spectra of RRLs and CH$_3$OH by  by \citet{2022ApJS..263...13L} and \citet{Rizzo2017} were
consistent with each other, as highlighted by the Figure 5 of \citet{2022ApJS..263...13L}.
The velocities of the Ka-band RRLs are self-consistent and are consistent with the modeled spectrum derived
based on Q-band emission parameters \citep{2022ApJS..263...13L}.  
Therefore there is no systematical frequency bias in the Ka-band survey.
The small velocity discrepancy between this survey and the K-band survey \citep{Gong2015}  
has  no influence on the comparison between the two surveys.  
 
Among the three surveys, the Ka-band survey has the best sensitivity (Table \ref{tablelowfreqsurvey}), 
and can detect emission lines that have not been detected by the previous Q-band and K-band surveys,
although the overlapped frequency ranges are very narrow. The main-beam temperatures of spectral lines from extended 
regions (e.g., RRLs) measured by different surveys should be consistent with each other. 
The aperture of Effelsberg  is larger than that of the TMRT.
Thus, for spectral lines originate from more compact regions, e.g., the transition of CH$_3$OH $v_t=1$ at 26120.6 MHz, 
their main-beam temperatures 
measured by the  TMRT are slightly lower than  that by the Effelsberg as expected due to potentially smaller filling factors
(see the upper panel of Figure \ref{figcompareprevious}).


\begin{figure}
\centering
\includegraphics[width=0.98\linewidth]{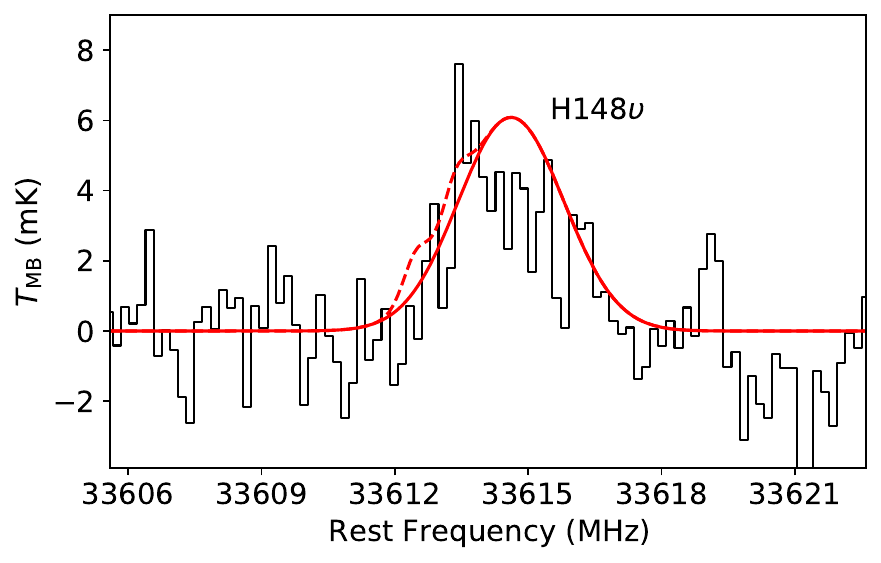}
\includegraphics[width=0.98\linewidth]{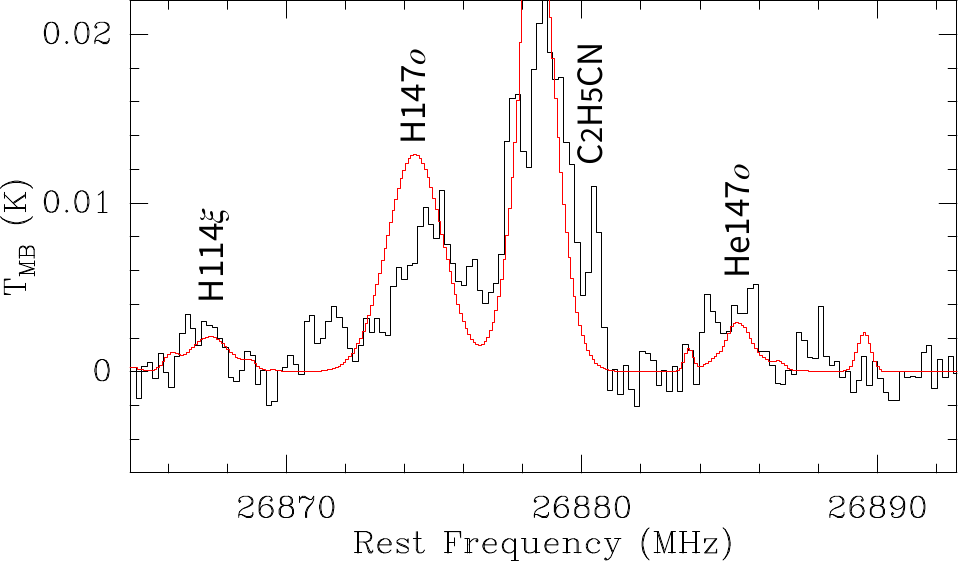}
\includegraphics[width=0.98\linewidth]{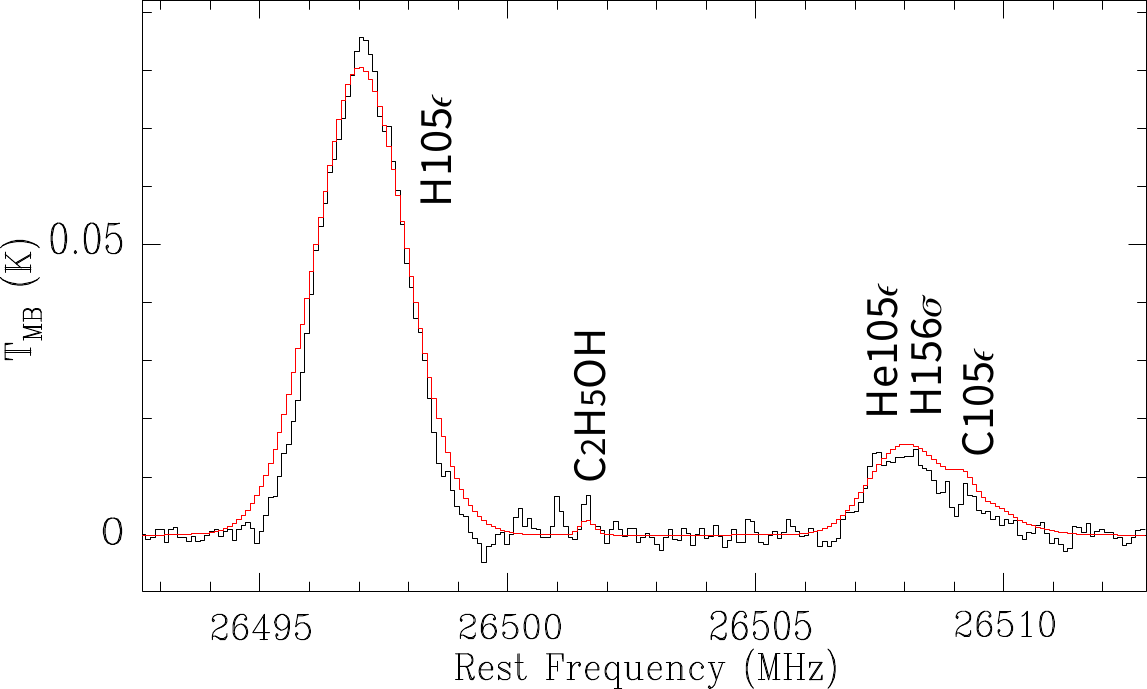}
\caption{Example of the detected RRLs with large $\Delta n$. The
red lines represent the result of model fitting of RRLs and molecular lines.
The dashed line in the upper panel has
taken into account the emission of C$_2$H$_5$CN $\varv_{13}$/$\varv_{21}$. \label{fig_eelexample}}
\end{figure}

\subsection{Line identification} \label{seclineid}\label{sect_linefeature}

We visually searched for line features on the Ka-band spectrum. Then, 
Gaussian fittings were manually conducted to those possible line features one by one using the fitting procedure provided by the GILDAS/CLASS. 
For blended lines or  strong lines with obvious non-Gaussian shapes,
multiple Gaussian fittings  were applied.
In total, 592 Gaussian components (referred to line features) were extracted (Tables \ref{tablineover}, \ref{linelist}, and
\ref{linelist_rrls}).

Then, we  modeled  the Ka-band spectrum based on the physical parameters
fitted from the Q-band observations following \citet{2022ApJS..263...13L}.
Most of the Ka-band line features can be reproduced by the physical parameters of
species detected in the Q band. 
We first coarsely crossmatched those line features with RRLs of H, He and C, and the transitions of 
molecules detected in the Q-band survey. 
Further, we tried to assign the unmatched lines to species undetected in the Q-band survey.
A species was marked as detected when  more than one of its transitions were detected above 3$\sigma$
and can be reproduced by the emission model with parameters compatible with values in the literatures
\citep[e.g.,][]{2011A&A...528A..26T,Gong2015,Rizzo2017,2023AA...671L...1L}.
The species with only one line feature in the Ka band is  also marked as detected if it is detected in the Q band.  
For vibrationally excited species,
it would be marked as detected even if only one of its
rotational lines is detected, as long as that transition is consistent with the model prediction
based on the physical parameters of its corresponding ground-state species (e.g., E-CH$_3$OH $v_t=1$ in Section
\ref{sec_ch3oh}).
See Sections \ref{secrrl} and \ref{secmols} for the details about the model fitting process of RRLs and molecular lines, respectively. 
We iterated the above procedures trying to properly assign as many as possible detected line features. 
Finally, the still unmatched line features are marked as  unidentified lines (marked as `U'). 

Table \ref{linelist} lists the molecular lines and  unidentified lines,
including those blended with RRLs.
Table \ref{linelist_rrls} lists the
RRLs which are not blended with any molecular emission.

\begin{table*}[!thb]
\centering
\caption{The maximum $\Delta n$ of RRLs detected in Orion KL. \label{tabmaxn}}
\begin{tabular}{lcccccc}
\hline
\hline
Ref. & band & frequency & H & He & C & X$^+$\\
\hline
This work& Ka & 26.1--35 GHz &  20 & 15 & 5 & 2,3$^{(1)}$ \\
\citet{2022ApJS..263...13L} & Q & 35--50 GHz &  16 & 7 & 3 & --\\
\citet{2023AA...671L...1L} & Ku/Ka/Q & 12--18 GHz, 26.1--50 GHz &-- & -- & -- & 1 \\
\citet{Rizzo2017} & Q & 41.5--50 GHz & 11 & 4 & 2  &--\\
\citet{Gong2015} & K & 17.9--26.2 GHz & 11 & 4 & 1 & --\\
\citet{2011ApSS.333..377B} & C& 5.98--6.05 GHz & 21,25$^{(2)}$ & 6 & -- & --\\
\hline
\end{tabular}\\
{\raggedright 
$^{(1)}$ Through stacking (Section \ref{sec_irrls_ldn}). 
The signal of the stacked $\gamma$ ($\Delta n=3$) is marginal, and the stacked $\beta$ ($\Delta n=2$) line is more reliable
(Section \ref{sec_irrls_ldn}).\\
$^{(2)}$ An H RRL with $\Delta n$ of
25 was reported but only has an intensity of 2$\sigma$.
Note that the H RRL with $\Delta n=21$ (H274$\phi$ at 6013.5 MHz) detected by \citet{2011ApSS.333..377B} is close to 
He230$\mu$ (6013.8 MHz). Although, under LTE, the intensity of He230$\mu$ is only one fourth the value of H274$\phi$.\\
}
\end{table*}

\begin{figure}
\centering
\includegraphics[width=0.98\linewidth]{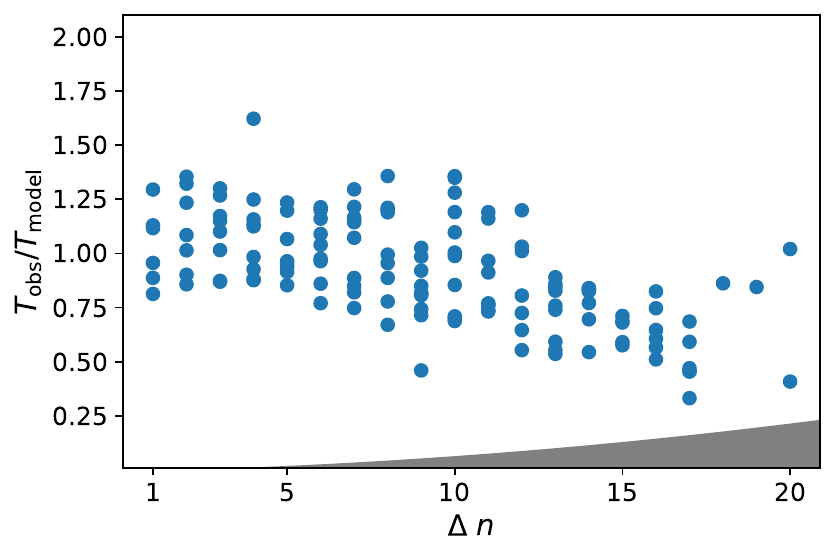}
\includegraphics[width=0.98\linewidth]{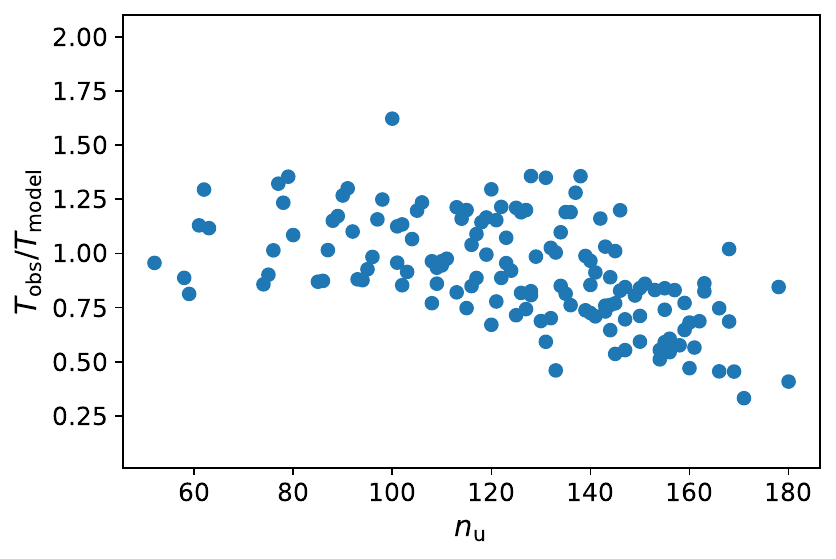}
\caption{Upper: The correlation between  $\Delta n$ 
and the intensity ratio between observation and model ($T_{\rm obs}/T_{\rm model}$).
The gray area represents the 1-$\sigma$ uncertainty  of $T_{\rm obs}/T_{\rm model}$.
Only the uncertainty of spectral noise is considered here, and the uncertainty contributed by calibration is not taken into account.
Lower: The correlation between  upper-level principle quantum  number ($n_{\rm u}$) and $T_{\rm obs}/T_{\rm model}$.
The two panels are only for H RRLs.
\label{fig_Tratio_rrls}}
\end{figure}

\begin{figure*}[!thb]
\centering
\includegraphics[width=0.99\linewidth]{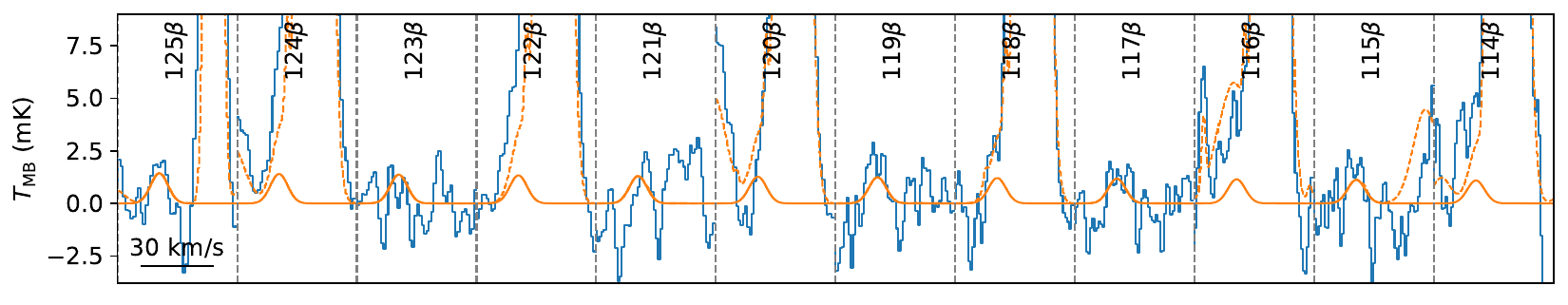}
\includegraphics[width=0.99\linewidth]{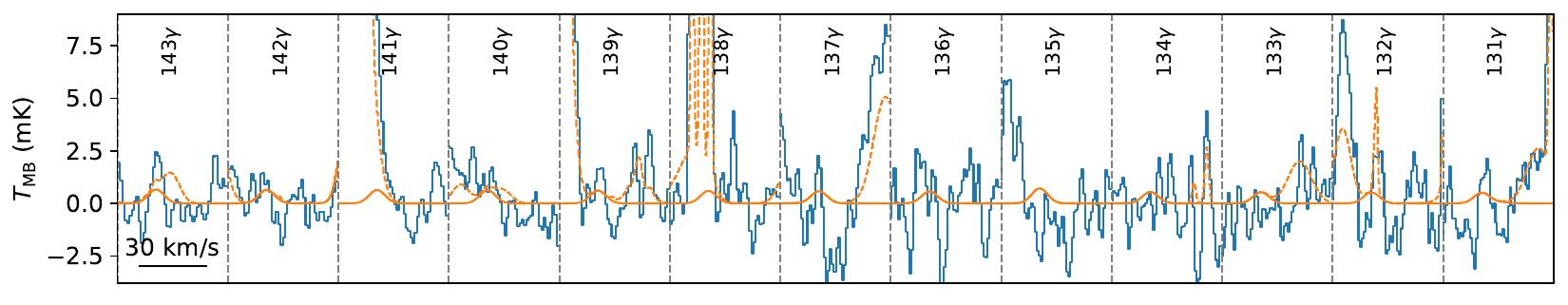}
\caption{The Ka-band X$^+n\beta$ (upper panel) and X$^+n\gamma$ (lower panel) lines of ion RRLs (blue lines).
The orange lines are the modeled spectrum. The  orange  dashed line takes emission of all species 
(molecular transitions and RRLs) into account,
and the orange solid line only considers the ion RRLs. 
We note that, these  X$^+n\beta$ and X$^+n\gamma$ 
RRLs  were not  detected individually, and some of them were at most marginally detected. See 
Figure \ref{fig_highorder} for the stacked spectra.
\label{fig_betagammalong}}
\end{figure*}

\subsection{Overview of the spectral lines}
Although the bandwidth of this Ka-band line survey ($\sim$9 GHz) is narrower than that of
the Q-band survey of the TMRT  (15 GHz), the number of line features detected by the two surveys are comparable,
partly because the Ka-band survey has a higher sensitivity and the weak line features can be detected (Table \ref{tablelowfreqsurvey}).  
The line density ($\rho_l$), defined as the averaged number of line features detected per GHz,
of this survey is nearly twice the $\rho_l$ of previous surveys in the K and Q bands (Table \ref{tablineover}). 
The Q-band spectrum of Orion KL is dominated by molecular transitions \citep{Rizzo2017,2022ApJS..263...13L}, 
and the ratio between the numbers of RRLs and molecular transitions, 0.5,
is weakly dependent on the sensitivity (Table \ref{tablineover}).
In contrast, the K-band spectrum is dominated by RRLs \citep{Gong2015}.
Among the 592 line features of this survey, the number of transitions of RRLs, 263, and molecular transitions, 305, are comparable
with each other.
The total emission flux contributed by RRLs, 58.7 K MHz, is also comparable with that by molecular transitions, 35.2 K MHz
(Figure \ref{fig_flux_kaq}).
Thus, the Ka band, a long-ignored window for spectral line surveys  located between the K band and Q band,
is very useful for studying RRLs and molecular lines simultaneously.

The U lines are all weak ($T_{\rm peak}<10$ mK) and some of them are 
narrow spikes (Table \ref{linelist} and Figure \ref{allspectra_first}). 
The Ka-band survey is deeper than the Q-band survey, and we have dropped the data of bad quality during the observations since
there were usually several different frequency setups for each frequency range (Section \ref{sect_kasurveyobs}).
This is why the Ka-band survey has much less spikes and bad channels than in the Q band (Table \ref{tablineover}).

\begin{figure*}[!thb]
\centering
\includegraphics[width=0.4\linewidth]{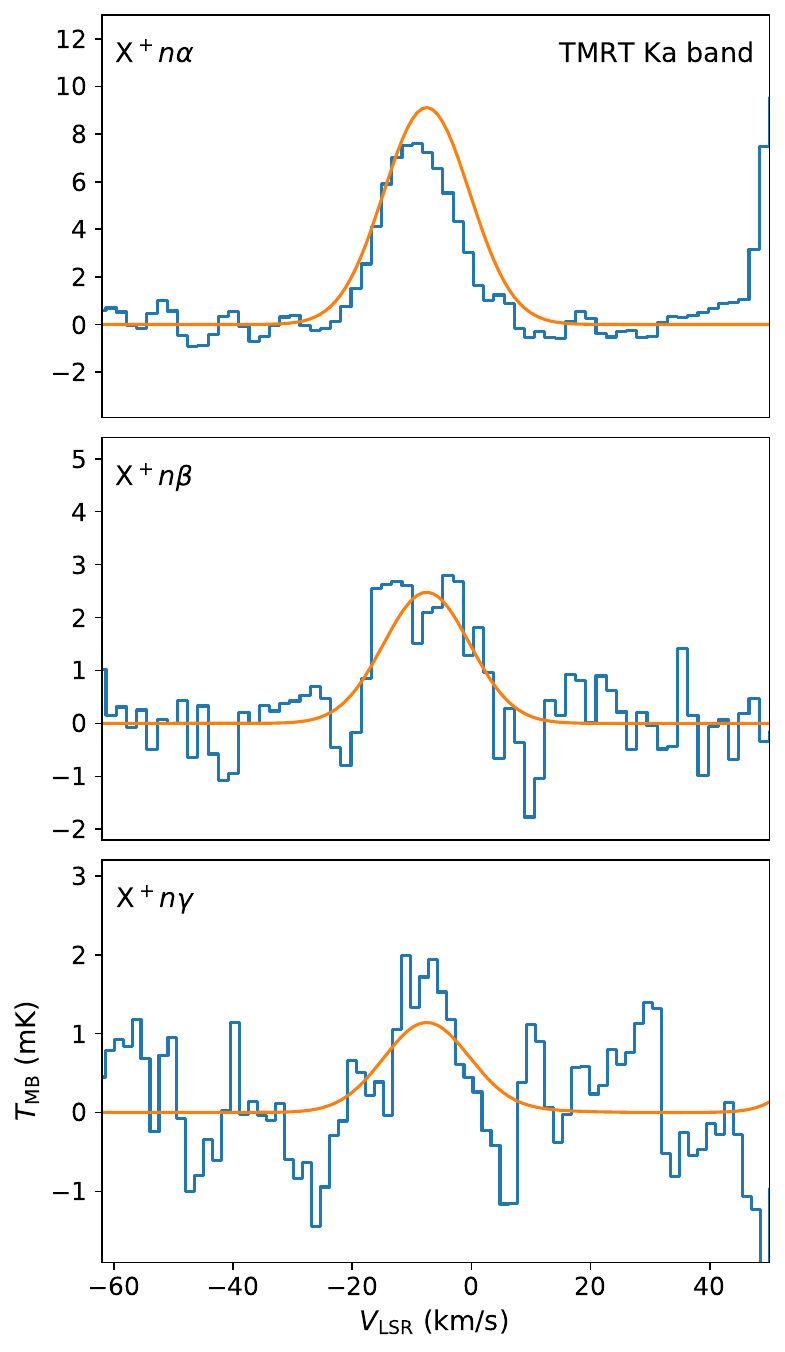}
\includegraphics[width=0.4\linewidth]{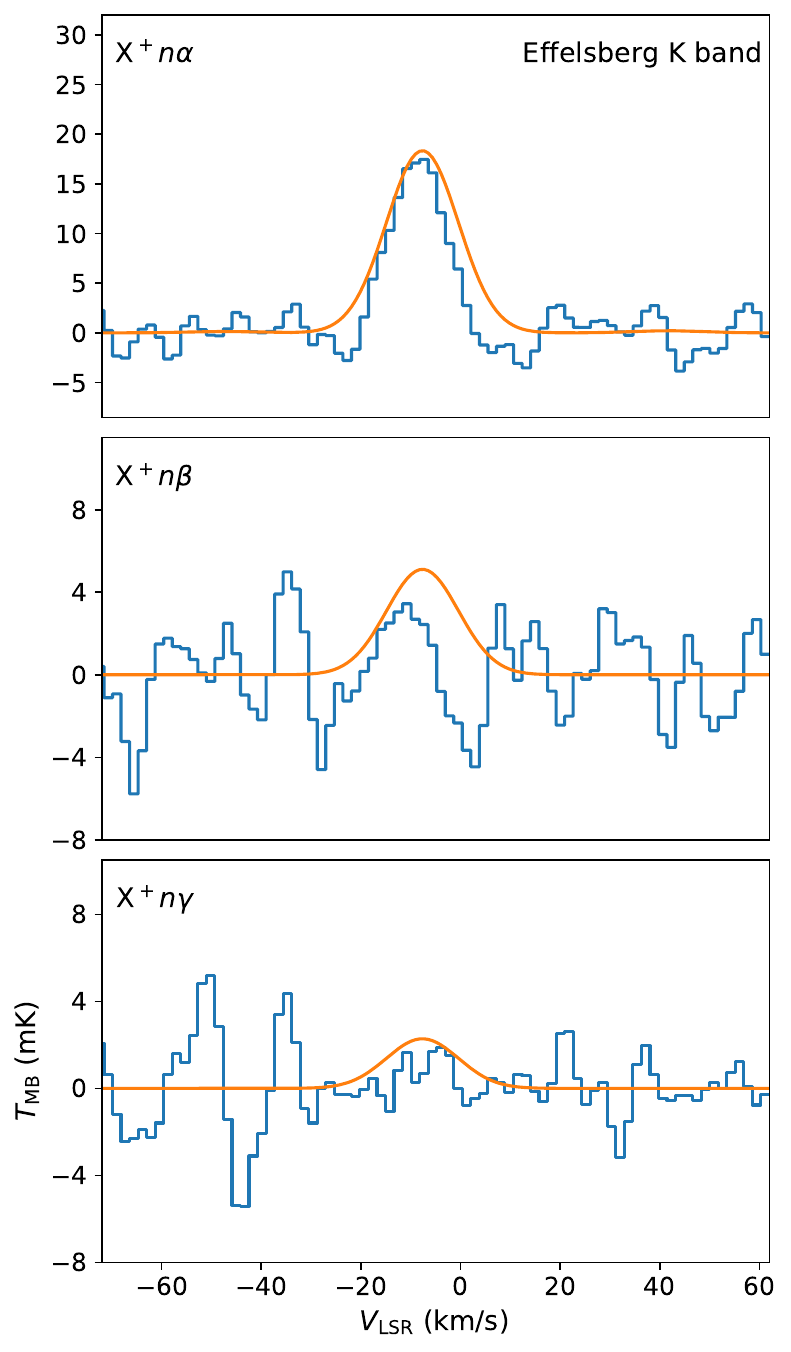}
\caption{The blue lines are the $\alpha$ (top), $\beta$ (middle) and $\gamma$ (bottom) lines of 
ion RRLs stacked  from the spectra of Orion KL measured in Ka band using the TMRT 65 m (left) and  in K band 
using the Effelsberg 100 m by \citet[][]{Gong2015} (right). 
The orange lines are the modeled spectra of ion RRLs.
The orange lines in the right panels have been  multiplied
by a factor of 0.8. 
Only the unblended transitions are stacked. Specifically,
in Ka band, we stacked ${\rm X}^+n\alpha$  ($n=$ 99, 97, 95, 92, 91), 
${\rm X}^+n\beta$ ($n=$ 123, 121, 119, 117, 115) and ${\rm X}^+n\gamma$ 
($n=$ 142, 137, 136,  134, 131). 
In K band, we stacked ${\rm X}^+n\alpha$  ($n=$ 112, 111, 110, 108, 107, 106, 105, 104, 103, 101, 100), 
${\rm X}^+n\beta$ ($n=$ 141, 137, 135, 131, 127) and ${\rm X}^+n\gamma$ 
($n=$ 162, 159, 158, 156, 153, 148, 147, 146, 145, 144).
\label{fig_highorder}
 }
\end{figure*}

\section{RRLs} \label{secrrl}
The rest frequencies of RRLs can be calculated from the Rydberg formula
\citep[e.g.,][]{1996ApJS..107..747T,2002ASSL..282.....G}
\begin{equation} \label{eq_rrlfreq}
\nu^{\rm RRL}_{\rm rest}(n+\Delta n,n) = \frac{m_e'}{m_e}R_\infty  cZ^2
\left( \frac{1}{n^2}-\frac{1}{(n+\Delta n)^2} \right)
\end{equation} 
with 
\begin{equation}
m_e'=\frac{M-Zm_e}{M-(Z-1)m_e}.
\end{equation}
Here, $M$ and $Z-1$ are the total mass and atomic charge of a hydrogenic emitter, respectively, $m_e$ is the electron mass, $m_e'$ is the
reduced mass of the electron, and
$R_\infty=109737.31568$ cm$^{-1}$ \citep{2021RvMP...93b5010T}.
The value of $n$ is large for RRLs and it is thus valid to treat the  emitters 
as hydrogenic.  
For RRLs of neutral atoms recombined from singly ionized ions ($X^+$ or \ion{X}{2}), $Z$ should be adopted as one.  
For RRLs of singly
ionized ions  recombined from doubly ionized ions ($X^{2+}$ or \ion{X}{3}), Z should be adopted as two.
In this work, the RRLs with $Z=1$ and $Z>1$ are refereed as neutral RRLs and ion RRLs, respectively.

In total, 257 emission features in this survey are assigned to  RRLs, including 245 unblended ones
(Tables  \ref{tablineover}, \ref{linelist}, and \ref{linelist_rrls}).
The RRLs contribute to 60 percent of the total emission flux of the spectral lines in Ka band (Figure \ref{fig_flux_kaq}). 
In \citet{2023AA...671L...1L}, we have reported the first detection of the $\alpha$ lines of RRLs of carbon and/or oxygen ions
in the interstellar medium.
In below sections, 
we would display all the  RRLs detected in the Ka band towards  Orion KL (Section \ref{sec_rrlfit}), 
including the high-order neutral RRLs with the maximum value of $\Delta n$
reaching 20 (Section \ref{sec_ldn}), reported the first detection of ion RRLs  with $\Delta n>1$  (the $\beta$ and $\gamma$ lines) 
through stacking all the unblended transitions  (Section \ref{sec_irrls_ldn}). 


\begin{table}[!t]
    \centering
    \caption{Model parameters of molecules $^{(1)}$ \label{model_pars_table}}
\begin{tabular}{llllll}
\hline
\hline
Species & size$^{(2)}$ & $T_{\rm ex}$ & $N_{\rm tot}^{(3)}$ & $\Delta V$ & $V_{\rm lsr}$\\
                  & (\arcsec)    & (K)          & (cm$^{-2}$)         & (km s$^{-1}$) &(km s$^{-1}$) \\
\hline
H$_2$CS                             & 20       & 100      &      1.8e+15 & 4.0      & 8.0     \\
CCS                                 & 30       & 100      &      1.2e+13 & 8.0      & 7.5     \\
SO                                  & 15       & 100      &      1.2e+17 & 15.0     & 6.0     \\
                                    & 15       & 100      &      2.2e+17 & 25.0     & 9.0     \\
$^{34}$SO                           & 15       & 100      &      4.8e+15 & 15.0     & 6.0     \\
                                    & 15       & 100      &      9.6e+15 & 25.0     & 9.0     \\
SO$_2$                              & 40       & 50       &      5.0e+15 & 4.0      & 7.0     \\
                                    & 40       & 200      &      6.0e+15 & 10.0     & 7.0     \\
                                    & 40       & 100      &      1.9e+16 & 25.0     & 7.5     \\
$^{34}$SO$_2$                       & 40       & 50       &      1.0e+14 & 4.0      & 7.0     \\
                                    & 40       & 200      &      6.0e+14 & 10.0     & 7.0     \\
                                    & 40       & 100      &      6.0e+14 & 25.0     & 7.5     \\
HCN $\varv_2=1$                     & 15       & 200      &      1.8e+17 & 4.0      & 5.0     \\
HC$_3$N                             & 15       & 100      &      3.0e+14 & 3.0      & 9.0     \\
                                    & 15       & 100      &      4.0e+14 & 7.0      & 5.5     \\
                                    & 15       & 100      &      6.0e+14 & 15.0     & 5.5     \\
                                    & 30       & 100      &      1.2e+14 & 25.0     & 6.0     \\
H$^{13}$CCCN                        & 20       & 100      &      1.4e+13 & 4.0      & 9.0     \\
                                    & 15       & 100      &      2.4e+13 & 7.0      & 5.5     \\
HC$^{13}$CCN                        & 20       & 100      &      1.4e+13 & 4.0      & 9.0     \\
                                    & 15       & 100      &      2.4e+13 & 7.0      & 5.5     \\
HCC$^{13}$CN                        & 20       & 100      &      1.4e+13 & 4.0      & 9.0     \\
                                    & 15       & 100      &      2.4e+13 & 7.0      & 5.5     \\
HC$_3$N $\varv_6=1$                 & 15       & 150      &      3.0e+15 & 7.0      & 5.5     \\
HC$_3$N $\varv_7=1$                 & 10       & 150      &      5.0e+15 & 7.0      & 4.5     \\
                                    & 10       & 100      &      4.0e+15 & 25.0     & 6.0     \\
HC$_3$N $\varv_7=2$                 & 10       & 150      &      3.0e+15 & 7.0      & 4.5     \\
HC$_5$N                             & 15       & 100      &      4.0e+13 & 4.0      & 8.5     \\
\hline
\end{tabular}\\
{\raggedright 
$^{(1)}$ Take caution if the modeled parameters (e.g. $N_{\rm tot}$) 
are compared with other work because of the uncertainties of size and $T_{\rm ex}$ and source size.\\
$^{(2)}$ The size of emission region can not be constrained by single-dish observation.
The size of 40 $\arcsec$ is corresponding to the extended case.
If another emission source size is adopted, the column densities could be 
recalculated through multiplying the values listed here by a factor of $\rm (size/size^{new})^2$.\\
$^{(3)}$ For species in vibrational state, the column density ($N_{\rm tot}$) has taken into account the
ground-state molecule. \\
$^{(\dag)}$  dnotes the species that have not been detected in the TMRT Q-band  line survey.
\par}
\end{table}

\begin{table}
    \centering
{\raggedright \centering  \textbf{Table  \thetable} {\it (continued)} \par}
\begin{tabular}{llllll}
\hline
\hline
{Species} & size & $T_{\rm ex}$ & $N_{\rm tot}$ & $\Delta V$ & $V_{\rm lsr}$\\
                  & (\arcsec)    & (K)          & (cm$^{-2}$)         & (km s$^{-1}$) &(km s$^{-1}$) \\
\hline
C$_2$H$_3$CN                        & 8        & 320      &      3.6e+14 & 6.0      & 5.0     \\
                                    & 15       & 100      &      1.2e+14 & 6.0      & 5.0     \\
                                    & 8        & 200      &      1.1e+14 & 20.0     & 3.0     \\
                                    & 15       & 90       &      1.6e+14 & 20.0     & 3.0     \\
C$_2$H$_5$CN                        & 8        & 275      &      1.9e+16 & 5.0      & 5.5     \\
                                    & 15       & 110      &      1.4e+15 & 13.0     & 4.0     \\
                                    & 30       & 65       &      3.0e+14 & 20.0     & 4.0     \\
CH$_3$NH$_2$                        & 20       & 100      &      2.4e+14 & 4.0      & 5.0     \\
CH$_3$OH                            & 40       & 100      &      1.2e+16 & 2.0      & 7.0     \\
                                    & 40       & 150      &      2.0e+16 & 4.0      & 6.0     \\
                                    & 40       & 150      &      1.0e+16 & 10.0     & 7.0     \\
$^{13}$CH$_3$OH                     & 40       & 100      &      8.0e+14 & 2.0      & 7.0     \\
                                    & 40       & 150      &      8.0e+14 & 4.0      & 6.0     \\
A-CH$_3$OH $\varv_t=1$              & 40       & 150      &      2.0e+16 & 4.0      & 6.0     \\
E-CH$_3$OH $\varv_t=1$              & 40       & 150      &      2.0e+16 & 4.0      & 6.0     \\
C$_2$H$_5$OH                        & 20       & 60       &      7.2e+14 & 4.0      & 7.0     \\
H$_2$CO                             & 40       & 50       &      6.0e+14 & 25.0     & 5.0     \\
                                    & 20       & 50       &      6.0e+15 & 3.0      & 7.0     \\
                                    & 15       & 150      &      1.0e+16 & 10.0     & 4.5     \\
H$_2^{13}$CO                        & 20       & 50       &      2.0e+14 & 3.5      & 7.0     \\
H$_2$CCO                            & 15       & 100      &      1.8e+15 & 3.0      & 7.0     \\
CH$_3$CHO                           & 15       & 50       &      2.4e+14 & 3.0      & 8.0     \\
                                    & 30       & 150      &      2.4e+14 & 25.0     & 9.0     \\
CH$_3$OCHO                          & 40       & 60       &      4.8e+14 & 4.0      & 8.0     \\
                                    & 40       & 150      &      1.9e+15 & 25.0     & 9.0     \\
                                    & 20       & 110      &      1.7e+16 & 4.0      & 7.5     \\
                                    & 15       & 300      &      2.4e+16 & 4.0      & 7.5     \\
                                    & 15       & 250      &      7.7e+15 & 10.0     & 5.5     \\
CH$_3$OCHO $\varv_t=1$              & 20       & 100      &      1.1e+16 & 3.0      & 7.0     \\
c-CH$_2$OCH$_2$                     & 20       & 50       &      8.0e+13 & 3.0      & 6.5     \\
                                    & 20       & 50       &      2.0e+13 & 1.5      & 6.5     \\
CH$_3$OCH$_3$                       & 20       & 100      &      2.0e+16 & 3.0      & 7.5     \\
CH$_3$COCH$_3$                      & 10       & 100      &      1.8e+15 & 4.0      & 5.5     \\
NH$_3$                              & 10       & 300      &      1.2e+16 & 20.0     & 8.0     \\
                                    & 20       & 250      &      2.4e+16 & 8.0      & 6.0     \\
NH$_2$D                             & 20       & 300      &      4.0e+14 & 5.0      & 6.0     \\
$^{15}$NH$_3^{(\dag)}$              & 20       & 300      &      1.0e+14 & 8.0      & 6.0     \\
$^{13}$CH$_3$OH $\varv_t=1^{(\dag)}$ & 40       & 150      &      1.0e+15 & 4.0      & 6.0     \\
CH$_3$C$_3$N$^{(\dag)}$             & 40       & 50       &      2.0e+12 & 2.5      & 9.0     \\
CH$_3$CCH$^{(\dag)}$                & 8        & 50       &      1.5e+16 & 3.0      & 9.0     \\
CH$_3^{18}$OH$^{(\dag)}$            & 40       & 100      &      2.2e+14 & 2.0      & 7.0     \\
                                    & 40       & 150      &      2.2e+14 & 4.0      & 6.0     \\
HC$_7$N$^{(\dag)}$                  & 15       & 100      &      6.0e+12 & 4.0      & 8.5     \\
HDCO$^{(\dag)}$                     & 20       & 50       &      2.0e+14 & 4.0      & 8.5     \\
$^{18}$OCS$^{(\dag)}$               & 40       & 100      &      3.0e+13 & 3.0      & 7.0     \\
$^{33}$SO$^{(\dag)}$                & 15       & 100      &      3.0e+15 & 15.0     & 6.0     \\
c-C$_3$H$_2^{(\dag)}$               & 15       & 150      &      2.0e+14 & 4.0      & 8.5     \\
\hline
\end{tabular}
\end{table}

\subsection{LTE fitting and analysis}\label{sec_rrlfit}
We followed the method of \citet{2022ApJS..263...13L} to reproduce the neutral RRLs ($Z=1$) in the Ka band.
Under the typical electron density of Orion KL \citep[$n_e\sim 10^4$ cm$^{-1}$;][]{2002ASSL..282.....G}, the
departure coefficients ($b_n$)  are close to unity, corresponding to the case
of local thermodynamic equilibrium (LTE).
The emission measure ($EM$) of X$^{(n-1)+}$ is defined as $\int n_{X^{n+}}n_e{\rm d}l$,
and $n_{X^{n+}}$  and $n_e$ are the volume densities of $X^{n+}$ and free electron, respectively.
For RRLs with small $\Delta n=n_2-n_1$, the  absorption oscillator strengths ($f_{n_1,n_2}$) 
can be well approximated by the formula of \citet{1968Natur.218..756M},
as what we have adopted for the fitting of Q-band RRLs  \citep{2022ApJS..263...13L}.
However, in the Ka-band survey, RRLs with $\Delta n$ as large as 20 were detected (Section \ref{sec_ldn}).
For such large $\Delta n$, the empirical values of $f_{n_1,n_2}$ \citep{1968Natur.218..756M} can derivate from the
precise values by more than 50 percent (Appendix \ref{sec_os_precise}).
Thus, in this work,  
we numerically  tabulated the precise values of $f_{n_1,n_2}$ (Appendix \ref{sec_os_precise}), 
instead of adopting the values derived from the empirical formula of \citet{1968Natur.218..756M}.
We note that the $EM$ will not be changed under the two different set of $f_{n_1,n_2}$, since it 
mainly depends on the electron temperature and the intensities of strong lines with small $\Delta n$.

\begin{figure*}[!t]
\centering
\includegraphics[width=0.995\linewidth]{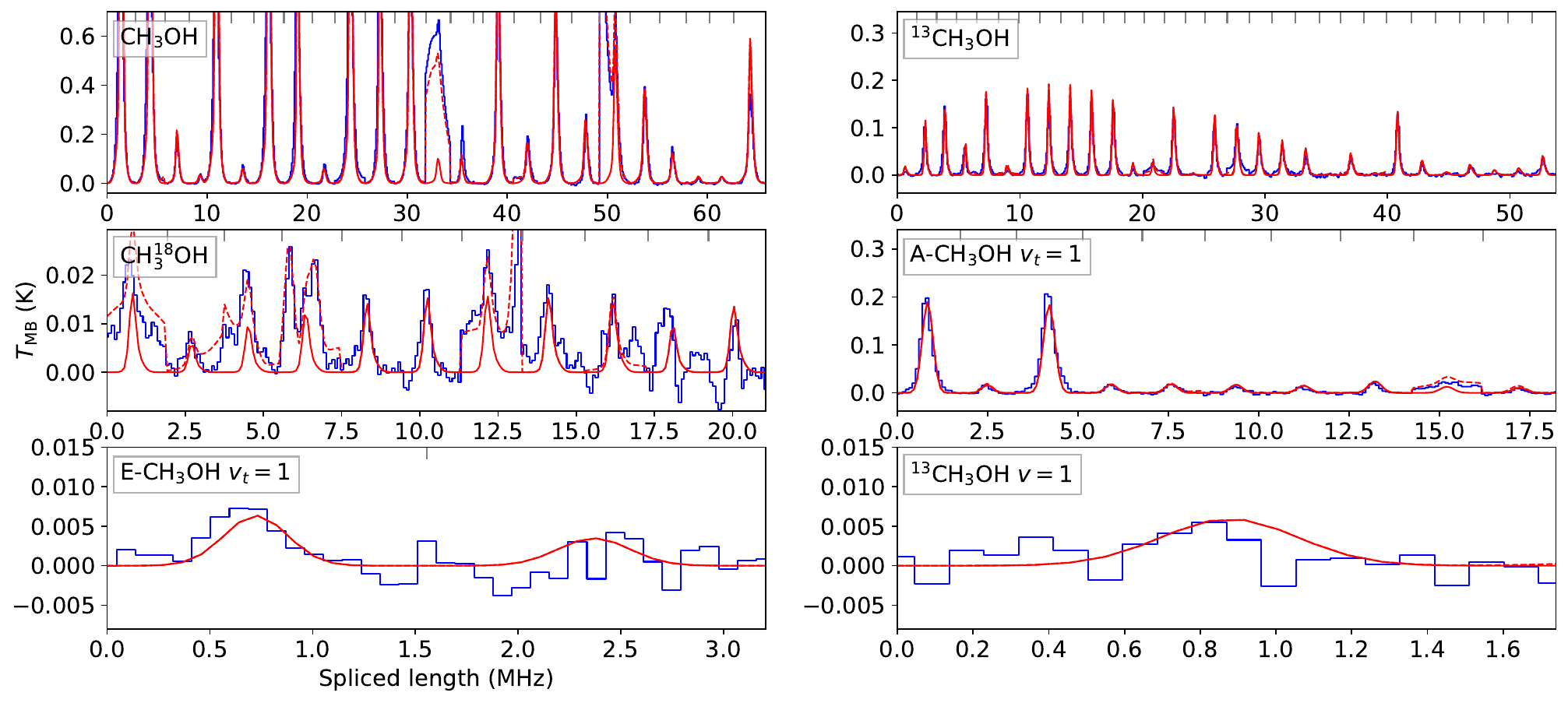}
\caption{ 
The blue lines are the spliced spectra of  CH$_3$OH and its isotopologues
as well as these species in vibrationally excited states.
The spectrum between two neighbouring upper ticks are independent frequency segment. 
The red solid lines show the result of the model fitting of the corresponding species.
The red dashed lines include the modeled emission of all species and RRLs.
See Fig. \ref{continued_fit_spec} for other molecular species detected in the Ka-band survey.
\label{example_fit_spec}}
\end{figure*} 

The electron temperature ($T_{\rm e}$) 
can be estimated from the intensity ratios between RRLs and continuum \citep{2002ASSL..282.....G},
but can not be constrained by RRLs alone.
Unfortunately, the continuum intensity was not well calibrated during our observations,
and we have to assume a  $T_{\rm e}$ to derive the emission measure.
For RRLs from \ion{H}{2} regions (neutral RRLs of H and He and ion RRLs), the $T_{\rm e}$ was fixed to be 8000 K 
\citep{1994ARA&A..32..191W,2006MNRAS.370..597L}.
The $EM$ of H was then derived to be 1.5$\times$10$^{6}$ cm$^{-6}$ pc,
and the $EM$ of He is one tenth the value of H.
The neutral RRLs of C mainly originate from the PDRs. Assuming a $T_{\rm e}$ of 300 K, 
the Ka-band spectra of the neutral RRLs of C can be reproduced by adopting an emission measure ($EM$) of
1.8$\times$10$^2$ cm$^{-2}$ pc  (Figure \ref{allspectra_first}).
Table 2 of \citet{2023AA...671L...1L} has already listed the fitted $EM$, velocity ($V_{\rm LSR}$), and line widths ($\Delta V$) 
of both neutral RRLs and ion RRLs basically based on this Ka-band survey of Orion KL,
and thus will not be repeated in this work.
In \citet{2023AA...671L...1L}, we  briefly compared those parameters to justify the detection of ion RRLs of X$^+$ with X referred to
be C and/or O. Below, we will quote those parameters for more detailed analysis.

Figure \ref{fig_Tratio_rrls} shows the intensity ratios between the observation and
the model fitting ($T_{\rm obs}/T_{\rm model}$) for different transitions of H RRLs. 
For $\Delta n<10$, $T_{\rm obs}$ and $T_{\rm model}$ are consistent with each other
within 50 percent, mainly contributed by the uncertainties of  calibration
and the different beam dilution at different frequencies.  
H96$\delta$ is an exception with a  $T_{\rm obs}/T_{\rm model}$ of $\sim 1.6$.
For $\Delta n>10$, the observed intensities tend to be lower than the model fitting.
We do not know the specific reason for this discrepancy.
One possibility is that the RRLs with large $\Delta n$ (and thus relatively large $n$ within a given frequency range) 
may deviate from LTE. Alternatively, RRLs with large $\Delta n$ tend to be weak,
and  the procedure of reducing spectral baseline through polynomial fitting  may suppress broad and weak line features
(Section \ref{sec_datareduct}).
Further observations with higher S/Ns would help to explore this issue.

\subsection{High-order neutral RRLs} \label{sec_ldn}
Among the RRLs detected towards Orion KL in this survey, H148$\upsilon$ has the maximum $\Delta n$ of 20 (Figures \ref{fig_eelexample}
and \ref{fig_Tratio_rrls}).
The line feature associated with H148$\upsilon$ has an intensity of larger than five sigma (Table \ref{linelist_rrls}),
but it is slightly blended with two transitions of the 
vibrationally excited  C$_2$H$_5$CN $\varv_{13}/\varv_{21}$ (Figure \ref{fig_eelexample}).
Through modeling the emission of RRLs (Section \ref{sec_rrlfit}) and  C$_2$H$_5$CN $\varv_{13}/\varv_{21}$ (Section \ref{secc2h5cn}),
we found that this line feature should be dominated by H148$\upsilon$
(Figure \ref{fig_eelexample}). 
For each $\Delta n<20$, there are one or more H RRLs that have been detected by this survey
with intensities consistent with the model prediction (Figure \ref{fig_eelexample}). 
Thus, the  detection of H148$\upsilon$ should be  reliable.
The maximum $\Delta n$ of this survey  is the largest compared with previous blind line surveys at frequency bands (Ku, K, and Q) 
close to this survey  (Table \ref{tabmaxn}).
\citet{2011ApSS.333..377B} reported the detection of  a 2-$\sigma$ signal H RRLs with $\Delta n=25$
and a 3-$\sigma$ signal of H RRLs with $\Delta n=21$ (H274$\phi$) from a narrow-band (5.98--6.05 GHz) survey towards Orion KL 
using the NRAO 140-foot (43 m) telescope. 
RRLs  tend to be weaker in the Ka band than in the C band.
The broad bandwidth and much higher angular resolution
of this survey enable us to detect spectrally resolved RRLs  with 
a maximum $\Delta n$ close to that of \citet{2011ApSS.333..377B}.

For He and C RRLs of this survey, the He147$o$ and C105$\epsilon$
reach the maximum $\Delta n$ of  15 and 5,
respectively (Figure \ref{fig_eelexample}).
Besides He147$o$, He114$\xi$ (with $\Delta n=14$ and $f_{\rm rest}=26866.5$ MHz) was also detected (Figure \ref{allspectra_first}
and Table \ref{linelist_rrls}),
and their intensities are consistent with those from LTE model.
Thus, the line feature of He147$o$ is considered to be a  firm detection under a 4$\sigma$
significance level (Table \ref{linelist_rrls}).
The maximum $\Delta n$ of He and C RRLs of this survey are the largest compared with previous surveys 
towards Orion KL (Table \ref{tabmaxn}).


\subsection{High-order ion RRLs}\label{sec_irrls_ldn}
Because of the blending with helium RRLs,  the wide component ($\Delta V \sim 20$ km s$^{-1}$) of carbon RRLs
originating from the ionized regions are difficult to be detected.   
The narrow component ($\Delta V \sim 5$ km s$^{-1}$; Table 2 of \citet{2023AA...671L...1L}) 
of carbon RRLs can be spectrally resolved, but it  traces the PDRs instead of the ionized regions.
The PDRs traced by narrow C RRLs have a $V_{\rm LSR}$ of $\sim 8$ km s$^{-1}$,
which is more than 10 km s$^{-1}$ redder than that of the ionized gas of M42 traced by H/He RRLs 
($V_{\rm LSR}\sim -4$ km s$^{-1}$; \citealt{Rizzo2017,2023AA...671L...1L}).
The interaction between M42 and the molecular cloud may be responsible for the discrepancy of the 
$V_{\rm LSR}$ of PDRs located at the interface between them \citep{2015A&A...575A..82C,Rizzo2017}. 
The ion RRLs provide us an opportunity to
trace the doubly ionized element  heavier than helium in the ionized regions.
In \citet{2023AA...671L...1L}, only the $\alpha$ lines of RRLs of C$^+$ and/or O$^+$ were reported.
There was no report of high-order ion RRLs (even for He$^+$) with $\Delta n>1$ in the literature. 
The ionization energy of \ion{C}{2}, 24.383 eV, is close to the value of \ion{He}{1}, 24.587 eV,
and thus the \ion{C}{3} region should be comparable with the \ion{He}{2} region in size.
Thus, these $\alpha$ lines were assigned to C$^+$ RRLs \citep{2023AA...671L...1L}.
However, the contribution of oxygen can not be excluded, because 
oxygen is as abundant as carbon, and the rest frequency shift between C$^{+}$ and O$^+$ RRLs
is small ($\sim 3.5$ km s$^{-1}$).
Interferometers may help to break the coupling between  RRLs of C$^+$ and O$^+$
since they have different ionization energies \textbf{(and therefore are presumably at different location)}, while single dish 
is difficult to distinguish them.
Below,  we tried to improve the S/Ns of ion RRLs through stacking, aiming to
search for ion RRLs with $\Delta n>1$ from this Ka-band survey of the TMRT
(Section \ref{sec_statmrt}) and to find signals of ion RRLs from independent telescopes beyond the TMRT (Section \ref{sec_effel}).

\begin{figure*}[!thb]
\centering
\includegraphics[width=0.53\linewidth]{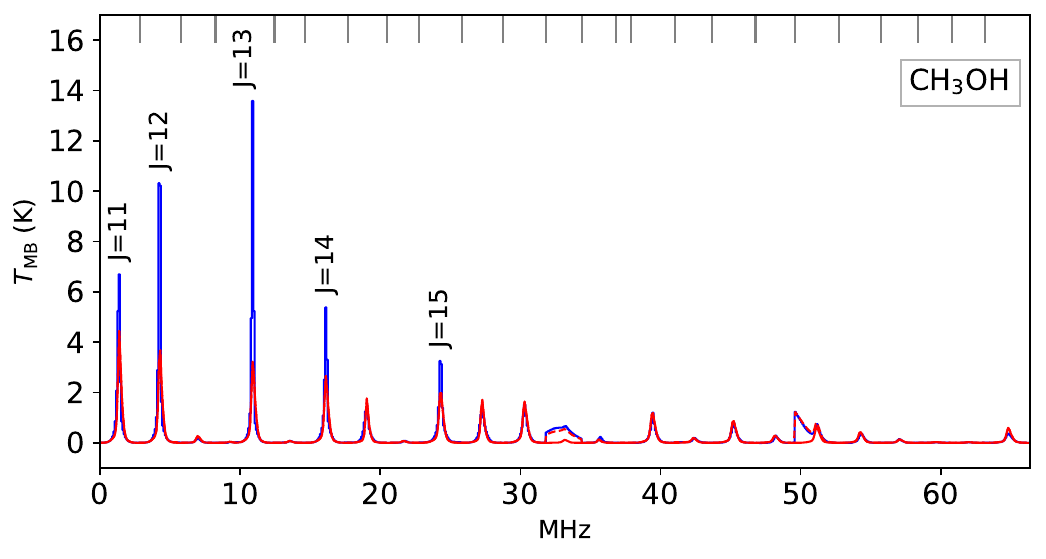}
\includegraphics[width=0.455\linewidth]{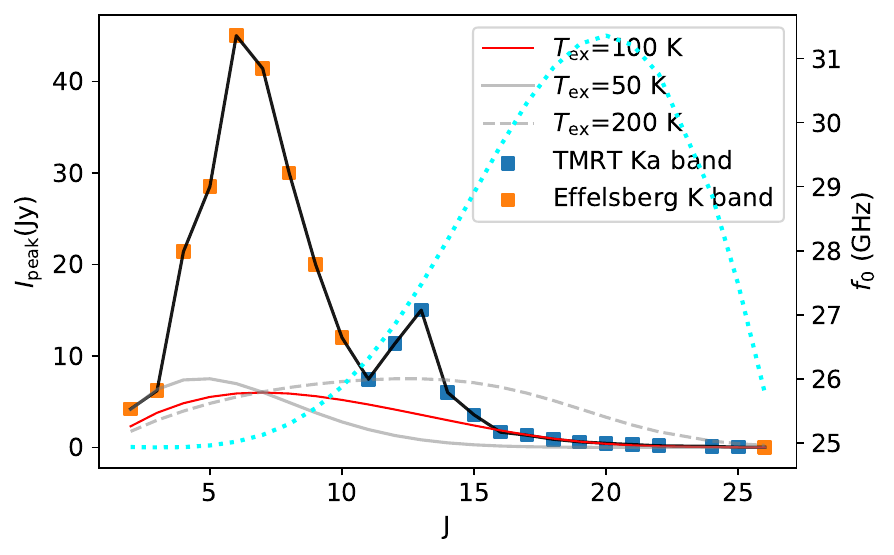}
\caption{Left: The spliced spectrum of CH$_3$OH, same as the upper left panel of Figure \ref{example_fit_spec}, but with 
a different y-axis scale.  The labeled transitions are $J_{K_a,K_c}=J_{2,J-1}-J_{1,J-1}$.
Right: The peak intensities ($I_{\rm peak}$) of CH$_3$OH $J_{2,J-1}-J_{1,J-1}$ detected in this survey (blue squares) and in the K-band survey of \citet{Gong2015}
(orange squares). The red solid, gray dashed, and gray solid lines show the predicted
peak intensity distribution along $J$ under LTE with a $T_{\rm ex}$ of 100 K, 200 K, and 50 K, respectively. 
The cyan line shows the rest frequency ($f_0$) of the transitions with different $J$.
\label{fig_ch3oh_zoomin}}
\end{figure*}

\begin{figure}[!htb]
\centering
\includegraphics[width=0.99\linewidth]{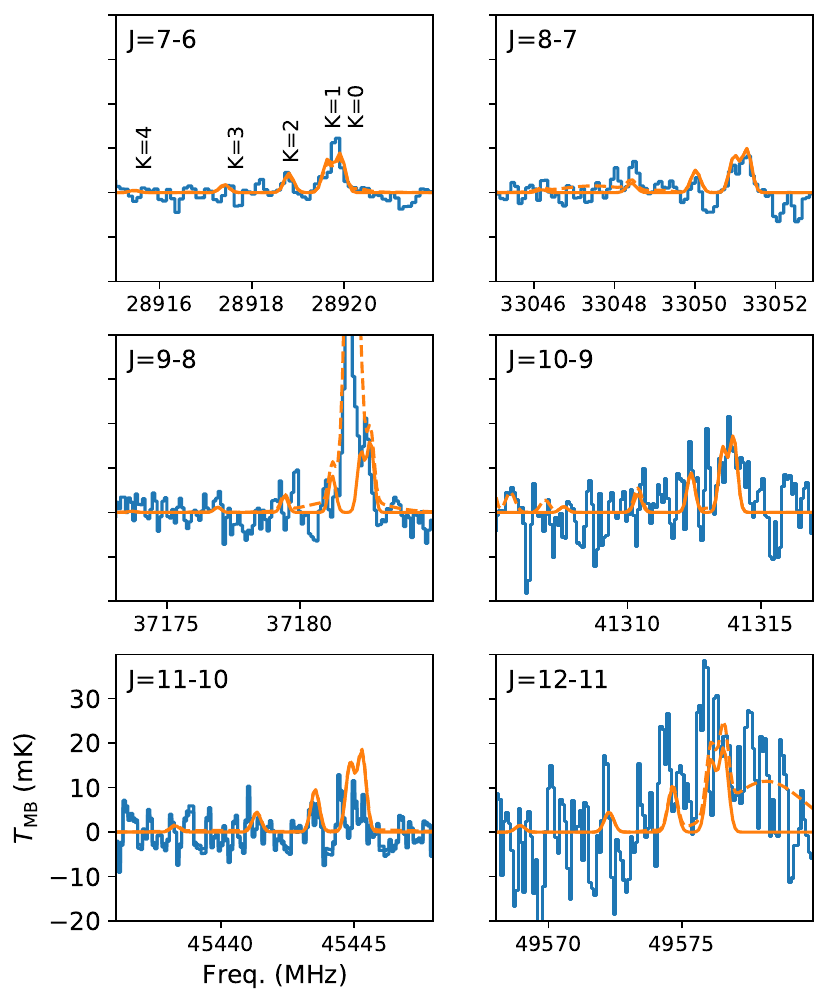}
\caption{The spectra of CH$_3$C$_3$N (blue) in the Ka and Q bands. The y-axis in all panels have the same scale.
Each rotational line is split into several transitions by the value of $K$, as marked in the upper-left panel. 
The orange solid lines are the model fitting of CH$_3$C$_3$N, and
the orange dashed lines have taken  all species into account.
All transitions of CH$_3$C$_3$N expected to be stronger than 5 mK have been shown here. 
\label{fig_ch3c3h_kaq}}
\end{figure}

\begin{figure*}[!thb]
\centering
\includegraphics[width=0.9\linewidth]{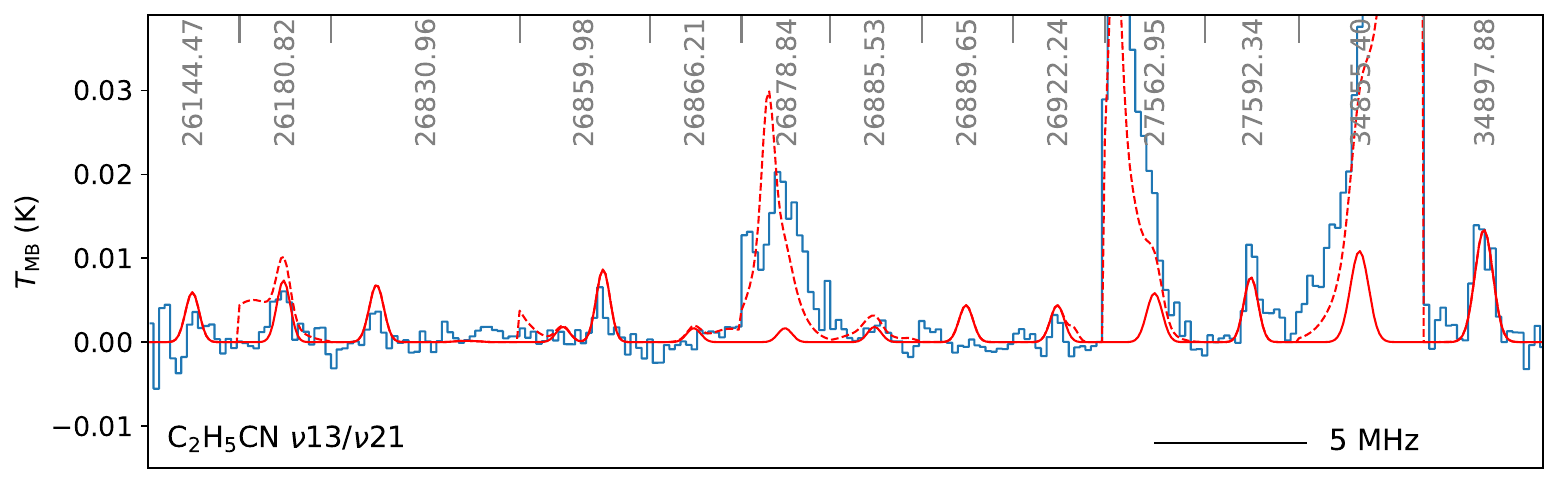}
\caption{The spliced spectrum of C$_2$H$_5$CN $v_{13}/v_{21}$. The meaning of solid and dashed red lines
are same as that of the upper left panel of Figure \ref{example_fit_spec}. 
The center frequencies of each segment are marked by the gray text labels.
\label{fig_c2h5cn}}
\end{figure*}

\subsubsection{Stacking of the Ka-band data of TMRT} \label{sec_statmrt}
For  X$^+$n$\beta$ with an even $n=2m$, Equation 1 leads to
\begin{align} \label{eq_rrlfreq_1}
\nu^{\rm RRL}_{\rm rest}(2m+2,2m) & = \frac{m_e'({\rm X})}{m_e}R_\infty  c2^2
\left( \frac{1}{(2m)^2}-\frac{1}{(2m+2)^2} \right) \notag \\
 & \sim \frac{m_e'({\rm He})}{m_e}R_\infty  c \left( \frac{1}{(m)^2}-\frac{1}{(m+1)^2} \right),
\end{align} 
and thus they should be blended with the $\alpha$ lines of neutral RRLs of helium (upper panel of Figure \ref{fig_betagammalong}).
For odd $n$, four $\beta$ lines of ion RRLs ($n=$ 125,  129, 117, and 115) show tentative line features.
The 121$\beta$ and 123$\beta$ can not be recognized because of limited S/Ns,
but do not contradict  the model (Figure \ref{fig_betagammalong}). 
Thus, it is very likely that we have detected the $\beta$ lines of ion RRLs.

To confirm the detection of $\beta$ lines and to search for higher-order ion RRLs, 
we stacked the spectra of different transitions to improve the S/N. 
We first check the Ka-band data, including both the observed  and
the modeled spectrum (Section \ref{sect_linefeature}), to 
select the spectral segments around the frequencies of ion RRLs (Equation \ref{eq_rrlfreq}) 
that are not blended by any other transitions within 
20 km s$^{-1}$ (Figure \ref{fig_betagammalong}).
For a given $\Delta n$,
we stacked all the corresponding  unblended segments
(with different $n$) aligned by velocity.
The stacking was conducted with equal weights for different segments.
The left panels of Figure \ref{fig_highorder} show the stacked Ka-band spectra of the   
$\alpha$, $\beta$ and $\gamma$ lines of ion RRLs.

It's natural to see the high-S/N signal of the stacked $\alpha$ line since each unblended $\alpha$ line
of ion RRLs can be clearly recognized \citep{2023AA...671L...1L}. 
Besides the $\alpha$ lines,
the stacked $\beta$ line shows clear detection (6$\sigma$; Figure \ref{fig_highorder}), and 
the stacked $\gamma$ line shows very marginal signal.
This is the first firm detection of ion RRLs with $\Delta n>1$ originated from the interstellar medium. 
We stacked the modeled data with the same procedure to obtain the stacked spectra (shown as the orange lines in
Figure \ref{fig_highorder}).  
In \citet{2023AA...671L...1L}, the fitting of ion RRLs was conducted based on  $\alpha$ lines
in three bands (Ku, Ka and Q).
Here, we fitted them based on the Ka-band data only, but taking into account all the $\alpha$, $\beta$, and $\gamma$ lines.
The fitted values of $EM$ (1.3$\times $10$^{3}$ cm$^{-6}$ pc),  the abundance of $X^{2+}$ (8.8$\times$10$^{-3}$) 
and the line width (17 km s$^{-1}$) are not altered compared with the values reported in  \citet{2023AA...671L...1L}.
The $V_{\rm LSR}$ fitted from the stacked spectra is -6.5 km s$^{-1}$ (with respect to the rest frequencies of C$^+$ RRLs), slightly bluer 
than the value (-5.5 km s$^{-1}$) in \citet{2023AA...671L...1L}.

Comparing the relative intensities between the observation and model can provide an 
empirical test of the departures from LTE.
The stacked $\gamma$ line seems to be slightly stronger than the modeled spectrum 
(bottom left panel of Figure \ref{fig_highorder}),
hinting at the
deviation from LTE. 
However, the S/Ns of the stacked $\beta$ and $\gamma$ lines are not high enough to dig into this issue,
and further observations with deeper sensitivities are needed to confirm it.

\subsubsection{Comparison with the archived data of Effelsberg} \label{sec_effel}
We also checked the archive data of the K-band survey of Orion KL observed by \citet{Gong2015} using the
Effelsberg 100 m telescope to search for the signals of ion RRLs.
All $\alpha$ lines of ion RRLs in the K band can be recognized marginally or with moderate S/Ns, except 
for the blended X$^+$109$\alpha$ and the X$^+$102$\alpha$ which is located at the frequency gap of the K-band survey
(see Appendix \ref{sec_effl} and Figure \ref{fig_efalphas}).  
The intensities and velocities of the $\alpha$ lines of ion RRLs in the K band are consistent with the model 
prediction based on the Ka-band data. 
Using the same procedure in Section \ref{sec_statmrt}, 
we stacked the unblended ion RRLs in the K band.
No clear signals of $\beta$ and $\gamma$ lines can be seen on the
stacked spectra of the Effelsberg K-band survey because of the limited sensitivity.
The signal of the $\alpha$ line can be clearly seen on the stacked spectrum (see the top right panel of Figure \ref{fig_highorder}).
This provides a very important cross check of ion RRLs from an independent
facility.


\section{Molecular lines} \label{secmols}
In total, 318 molecular lines were identified (Table \ref{tablineover}),
and they can be assigned to 37 species. Among them, 10 species (marked by $^{(\dag)}$ in Table \ref{model_pars_table}) have no 
transitions being detected by the Q-band survey of the TMRT, because of the limited S/Ns 
or the lacking of  transitions in the Q band.
The vibrationally excited states ($v_t>0$) of nine species were detected,
including the eight ones listed in Table \ref{model_pars_table} and C$_2$H$_5$CN $v_{13}/v_{21}$ (Section \ref{secc2h5cn}).
For each detected transition, 
some transition parameters were quoted from the CDMS  \citep{2001A&A...370L..49M}
and JPL \citep{1998JQSRT..60..883P} through the Splatalogue \footnote{\url{https://splatalogue.online/}}. 
These parameters are the rest frequency ($f_0$), upper-level energy ($E_{\rm u}$), 
Einstein coefficients ($A_{ij}$), upper-level degeneracy ($g_{u}$), and 
partitial function ($Q$). 
The $f_0$ and $A_{ij}$ for transitions of C$_2$H$_5$CN $v13/v21$ were obtained
from \citet{2021JMoSp.37511392E}.
Through Equations 9, 10 and 11 of \citet{2022ApJS..263...13L},
we modeled the emission of molecules.
For most species,  constraining $T_{\rm ex}$ is challenging due to the similar upper level energies of their different transitions. 
Therefore, $T_{\rm ex}$ is manually set to 50 K, 100 K, 150 K, 200 K, or, 300 K.
We have also referred to the literature \citep{2010A&A...517A..96T,Gong2015,Rizzo2017,2023AA...671L...1L} 
to choose proper $T_{\rm ex}$. 
To reproduce the emission of isotopologues or species with similar chemical characteristics, 
we tend to adopt the same fixed values of $T_{\rm ex}$,
and use a similar initial guesses for their fitted parameters, including 
the column density ($N$), line width ($\Delta V$) and velocity ($V_{\rm LSR}$).
We try to model the spectrum with as few velocity components as possible.
The velocity resolution of the Ka-band survey is worse than that of the Q band,
and species showing multiple velocity components may only show a single velocity component in the Ka band.  
For species detected in the Q band, if the modeled spectrum  does not deviate much (less than 20 percent) 
from the observed spectrum in the Ka band, the number of the velocity components and the physical 
parameters fitted from the Q-band data will be directly used with no modification.
The fitted parameters of molecular species are shown in Table \ref{model_pars_table}.

\subsection{Methanol emission} \label{sec_ch3oh}
In total, 67 line features are assigned to methanol and its isotopologues, CH$_3$OH,
$^{13}$CH$_3$OH, CH$_3^{18}$OH, A-CH$_3$OH $v_t=1$, E-CH$_3$OH $v_t=1$, and $^{13}$CH$_3$OH $v_t=1$ (Figure \ref{example_fit_spec}).
Among them, CH$_3^{18}$OH and $^{13}$CH$_3$OH $v_t=1$ were not detected in the Q-band survey \citep{2022ApJS..263...13L}. 
Methanol contributes  more about half the Ka-band molecular emission of Orion KL (Figure \ref{fig_flux_kaq}).
Besides thermal components (Section \ref{sec_mathanol_thermal}), the maser emssion of 
methanol is also evident, as we discuss in Section \ref{sec_mathanol_maser}.

\begin{table*}[!thb]
\centering
\caption{The species detected in the Ka-band survey \label{allspecies}}
\begin{tabular}{ccccccc}
\hline
\hline
H$^{+}{}^{(\dagger)}$  & $^{34}$SO & HCC$^{13}$CN & CH$_3$NH$_2{}^{(*)}$ & H$_2^{13}$CO & CH$_3$COCH$_3$ & CH$_3^{18}$OH$^{(\ddagger)}$\\
He$^{+}{}^{(\dagger)}$ & SO$_2$ & HC$_3$N $\varv_6=1$ & CH$_3$OH & H$_2$CCO & NH$_3$ & HC$_7$N$^{(\ddagger)}$\\
C$^{+}{}^{(\dagger)}$ & $^{34}$SO$_2$ & HC$_3$N $\varv_7=1$ & $^{13}$CH$_3$OH & CH$_3$CHO & NH$_2$D & HDCO$^{(\ddagger)}$\\
X$^{2+}{}^{(\dagger)}$ & HCN $\varv_2=1$ & HC$_3$N $\varv_7=2$ & A-CH$_3$OH $\varv_t=1$ & CH$_3$OCHO & $^{15}$NH$_3{}^{(\ddagger)}$ & $^{18}$OCS$^{(\ddagger)}$\\
H$_2$CS & HC$_3$N & HC$_5$N & E-CH$_3$OH $\varv_t=1$ & CH$_3$OCHO $\varv_t=1$ & $^{13}$CH$_3$OH $\varv=1{}^{(\ddagger)}$ & $^{33}$SO$^{(\ddagger)}$\\
CCS & H$^{13}$CCCN & C$_2$H$_3$CN & C$_2$H$_5$OH & c-CH$_2$OCH$_2$ & CH$_3$C$_3$N$^{(\ddagger)}$ & c-C$_3$H$_2{}^{(\ddagger)}$\\
SO & HC$^{13}$CCN & C$_2$H$_5$CN & H$_2$CO & CH$_3$OCH$_3$ & CH$_3$CCH$^{(\ddagger)}$ & C$_2$H$_5$CN $v_{13}/v_{21}$\\
\hline
\end{tabular}\\
{\raggedright 
Note: The species marked by $^{(\dagger)}$ ions are traced by RRLs, e.g., X$^{2+}$ was traced by X$^{+}$ RRLs.
The species marked by $^{(\ddagger)}$ are detected in this Ka-band survey but not in the Q-band survey of TMRT.
The species marked by $^{(*)}$ are only tentatively detected.
\par}
\end{table*}

\subsubsection{Methanol thermal emission}\label{sec_mathanol_thermal}
We adopted three velocity components to fit the Ka-band spectra of CH$_3$OH under the assumption of LTE as in 
the Q band \citep{2022ApJS..263...13L}.
The Ka-band emission of CH$_3$OH are dominated by strong lines ($T_{\rm peak}>1$ K),
with sharp peaks that are not fully resolved under the current spectral resolution (the left panel of 
Figure \ref{fig_ch3oh_zoomin}). Thus, the emission of CH$_3$OH were reproduced with
the narrow ($\Delta V=2$ km s$^{-1}$), moderate ($\Delta V=4$ km s$^{-1}$), and broad velocity components
($\Delta V=10$ km s$^{-1}$). Under the assumption of LTE,
$T_{\rm ex}$ of the  moderate  component (150 K) is larger than that of the narrow component (100 K; Table \ref{model_pars_table}).
The column density ratio of CH$_3$OH/$^{13}$CH$_3$OH and $^{13}$CH$_3$OH/CH$_3^{18}$OH are 20 and
3.6, respectively. The CH$_3$OH/$^{13}$CH$_3$OH is  slightly larger than the values (5--15) of \citet{Rizzo2017}, 
and smaller than the typical  abundance ratio of Galactic  $^{12}$C and $^{13}$C \citep[several tens;][]{1994ApJ...428..693W}.
This maser effects may influence this abundance ratio (Section \ref{sec_mathanol_maser}).
The $^{13}$CH$_3$OH/CH$_3^{18}$OH is slightly lower than the  abundance ratio of $^{13}$CO/C$^{18}$O (X$^{13/18}$)
in local ISM \citep{1999RPPh...62..143W,2014A&A...564A..68S}.

The spectral lines of CH$_3$OH $v_t=1$ (in both  A and E asymmetry) have line widths
of $\sim 4$ km s$^{-1}$. Adopting the excitation temperature ($T_{\rm ex}=$ 150 K) and
the column density of the moderate component fitted from the emission of $^{13}$CH$_3$OH (Table \ref{model_pars_table}),
all line features of CH$_3$OH $v_t=1$ can be reproduced with a deviation no more than 10 percent (Figure \ref{example_fit_spec}).  
It implies that the vibrationally excited CH$_3$OH is mainly associated with the moderate component (instead of the
narrow component). The narrow, moderate, and broad velocity components
may originate from the compact ridge, hot core, and plateau, respectively \citep{Rizzo2017}.

\subsubsection{Methanol masers}\label{sec_mathanol_maser}
Several strong lines of CH$_3$OH ($T_{\rm peak}>3$ K) can not be successfully reproduced properly with any $T_{\rm ex}$ (Figure \ref{fig_ch3oh_zoomin}). Those lines belong to the same group of the
transitions of E-type methanol (E-CH$_3$OH $J_{K_a,K_c} = J_{2,J-1}-J_{1,J-1}$) with 
$11\le J\le 15$, 
and their rest frequencies are 
located within 26--29 GHz.
It is consistent with the observations of  \citet{1996A&A...307..209W},
which detected narrow line components with $\Delta V\sim 0.8$ km s$^{-1}$  
for $12\le J\le 17$\footnote{E-CH$_3$OH $J_{K_a,K_c} = J_{2,J-1}-J_{1,J-1}$ 
were denoted as E-CH$_3$OH $J_{K}=J_2-J_1$ with $K_c$ being omitted in \citet{1996A&A...307..209W}.}
in the Ka band and assigned them to low gain maser regions.
In \citet{1996A&A...307..209W}, the peak intensity of the maser component
is comparable with that of the thermal component. 
The observed intensity of the transition with $J=13$ by TMRT is four times stronger than 
the value of LTE prediction, showing the most significant deviation (Figure \ref{fig_ch3oh_zoomin}).

The transitions of E-CH$_3$OH in this group with $2\le J\le 9$ are well-known 
Class I maser transitions\footnote{\url{https://maserdb.net/transitions.pl?molecule=MET1&}}
with rest frequencies located in the K  band
\citep{1986A&A...157..318M,Gong2015,2019AJ....158..233L}.
Combining the data of this work and \citet{Gong2015},
we plotted the profile of the peak intensities ($I_p$) along $J$
with $2\le J\le 16$   in the right panel of Figure \ref{fig_ch3oh_zoomin}.
Under the assumption of LTE, for compact source with an angular size smaller than the beam size, 
the profile of $I_{\rm peak}$ along $J$ can be calculated through
\begin{equation}
I_p \propto A_{ij} g_u  \exp(-E_u/T_{\rm ex}).
\end{equation}
Here, $E_u$ is in units of K.
Three different values of $T_{\rm ex}$ (50, 100, and 200 K) were adopted, and
all three profiles of $I_p$ can not fit all the Ka-band data points simultaneously
(Figure \ref{fig_ch3oh_zoomin}). 
The data points with $J>15$ could  be fitted adopting a $T_{\rm ex}$ of 100 K, and thus should be dominated by thermal emission.
Note that the rest frequencies  increase with $J$ for $J<20$, but decrease with $J$ for $J>20$ (see the cyan line
in Figure \ref{fig_ch3oh_zoomin}).
The rest frequency of the transition with a $J$ of 26 falls back to the K band, and its peak intensity 
is consistent with value extrapolated from the Ka-band data (Figure \ref{fig_ch3oh_zoomin}).
Thus, the line of  $J=26$ can be explained by the LTE fitting,
consistent with the interpretation of \citet{Gong2015}.
  
Previously, it is known that 
the maser effect would increase with $J$ from 2 to 6 and decrease with $J$ from 6 to 10
\citep{1986A&A...157..318M,Gong2015}.
The Ka-band data of this survey shows that the maser effect may increase  with $J$ again from 11 to 13,
and then decrease with $J$ from 13 to 15. 
These masers are likely class I methanol masers as 
well \citep{1992ApJ...385..232J,1996A&A...307..209W}.
The second peak can not simply be  explained by the uncertainty of intensity calibration. 
The thermal lines of CH$_3$OH and other species 
in Ka band can
all be well reproduced by the LTE model (Figures  \ref{allspectra_first} and \ref{example_fit_spec}).
Especially, the intensities of all the transitions 
of $^{13}$CH$_3$OH (upper right panel of Figure \ref{example_fit_spec})
are consistent with the LTE model 
within 20 percent (the calibration uncertainty; Section \ref{secobs}).
To model the emission of $^{13}$CH$_3$OH, we adopted the same physical parameters as those for CH$_{3}$OH, 
except for the column densities, which have been scaled down  (Table \ref{model_pars_table}).
The reason of the existence of two bumping peaks for E-CH$_3$OH $J_{K_a,K_c} = J_{2,J-1}-J_{1,J-1}$, at $J\sim 6$ and $J\sim 13$, is still unknown.
Interferometer observations are needed to confirm
this issue through spatially resolving their different emission components.  

\subsection{N-bearing species}
\subsubsection{Methylcyanoacetylene (CH$_3$C$_3$N)}
Methylcyanoacetylene (CH$_3$C$_3$N) was first detected in TMC-1 by \citet{1984ApJ...276L..25B},
with a column density of 4.5$\times$10$^{11}$ cm$^{-1}$ under a $T_{\rm ex}$ of 4 K \citep{2006ApJ...637L..37L}.
It has also been detected in high-mass protostellar molecular clumps \citep{2018ApJS..236...45G}
and the Titan's atmosphere \citep{2020DPS....5241402T}.
However,  CH$_3$C$_3$N has not been detected in  Orion KL, even by  
the ALMA interferometer \citep{2017A&A...604A..32P}. 

In this Ka-band survey, we matched two groups of the emission lines of CH$_3$C$_3$N 
($J=7-6$ and $J=8-7$; see Figure \ref{continued_fit_spec} and the upper two panels of 
Figure \ref{fig_ch3c3h_kaq}).
The transitions of $K=0$ (blended with $K=1$) were detected for both groups.
The transitions of K=2 ($E_{\rm u}\sim 35$ K) and  K=3 ($E_{\rm u}\sim 75$ K) were
tentatively detected. Thus, a $T_{\rm ex}$ of 50 K was adopted to fit the emission of CH$_3$C$_3$N,
yielding a column density of 2$\times$10$^{12}$ cm$^{-2}$.
Here, we assumed that the emission of CH$_3$C$_3$N is extended with a beam filling factor of unity.
Further, we compared the modeled and 
observed spectra in the Q band, and found that they are also consistent with each other (Figure \ref{fig_ch3c3h_kaq}).

The fitted velocity ($V_{\rm LSR}$) is $9$ km s$^{-1}$,  slightly redder than the systematic velocity (6 km s$^{-1}$) but 
consistent with that of CH$_3$CCH (Table \ref{model_pars_table}), C RRLs (Section of \ref{sec_rrlfit}),
and the ambient gas of Orion KL \citep[the extended ridge;][]{1986ApJS...60..357B,1997ApJS..108..301S}.
It implies that the CH$_3$C$_3$N emission  of Orion KL may be associated with the PDRs or ambiment gas of Orion KL. This is 
reasonable since CH$_3$C$_3$N should be chemically young, given that it is an unsaturated species with its transitions mainly detected 
in chemically young environments, e.g., the starless core TMC-1  \citep{1984ApJ...276L..25B,2006ApJ...637L..37L}.

\subsubsection{Ethyl cyanide (C$_2$H$_5$CN)} \label{secc2h5cn}
Ethyl cyanide (C$_2$H$_5$CN) is a common tracer of hot molecular gas \citep[e.g.,][]{2003ApJ...593L..51C,2022MNRAS.511.3463Q},
and its vibrational satellites can be intense in sub-millimeter bands \citep{2021JMoSp.37511392E}. 
In this survey, besides C$_2$H$_5$CN $v=0$,
rotational transitions of C$_2$H$_5$CN $v_{13}/v_{21}$ \citep[excited in bending or torsion modes;][]{2013ApJ...768...81D} 
can also be detected (Figure \ref{fig_c2h5cn}).
In total, four lines of C$_2$H$_5$CN $v_{13}/v_{21}$ were detected, less than the number of  lines of C$_2$H$_5$CN  $v_{13}/v_{21}$ detected
in the Q band \citep{2022ApJS..263...13L}.  
Due to the higher spectral sensitivity of the Ka-band survey, the transition of C$_2$H$_5$CN $v_{13}/v_{21}$ 
at 34897.89 MHz, which was not detected in the Q band, can be clearly seen on the
Ka-band spectrum (Figure \ref{fig_c2h5cn}). The other three transitions are located within 26--28 GHz.
This marks the first detection of   C$_2$H$_5$CN $v_{13}/v_{21}$ in the centimeter band.

To fit the emission of C$_2$H$_5$CN $v_{13}/v_{21}$, we adopted the transition parameters 
of C$_2$H$_5$CN $v_{13}/v_{21}$ from \citet{2021JMoSp.37511392E} and assumed a $T_{\rm ex}$ of 50 K as in \citep{2022ApJS..263...13L}. 
Since we do not know the partition function,
the column density can not be derived and the fitted parameters are thus  not listed in Table \ref{model_pars_table}. 
Here, we adopt the partition function of C$_2$H$_5$CN quoted from the CDMS.
The velocity and line width are fitted to be 5 km s$^{-1}$ and 6 km s$^{-1}$, respectively, consistent with the result fitted
from the Q-band data \citep{2022ApJS..263...13L}. 
The fitted column density of C$_2$H$_5$CN is 2$\times$10$^{15}$ cm$^{-2}$.
All transitions can be well fitted except for that in 26889 MHz. We not that this is not 
influenced by the $T_{\rm ex}$, because this transition and the strong line at 27592 MHz have similar $E_u$ ($\sim 300$ K).


\subsubsection{Cyanopolyynes (HC$_{2n+1}$N)}
Cyanopolyynes (HC$_{2n+1}$N) are abundant in kinds of environments, especially the cold cores 
\citep{1989ApJS...69..271H,2019MNRAS.488..495W} and the envelopes of late-type carbon stars \citep{2020A&A...640L..13P},
but not prominent in Orion KL. HC$_7$N was not detected by the Q-band \citep{2022ApJS..263...13L} 
and K-band \citep{Gong2015} line surveys of Orion KL.
In this survey, spectral lines of  HC$_{7}$N were detected with a peak intensity of $\sim$5 mK (Figure \ref{continued_fit_spec}).
The LTE fitting with a $T_{\rm ex}$ of 100 K yields a column density of 6$\times$10$^{12}$ cm$^{-2}$
and a $V_{\rm LSR}$ of 8.5 km s$^{-1}$. The  $V_{\rm LSR}$ of HC$_{2n+1}$N is consistent with that of HC$_5$N (8.5 km s$^{-1}$), 
and the $V_{\rm LSR}$ of the
reddest velocity component of HC$_3$N (9 km s$^{-1}$).
The column density ratio of $N$(HC$_3$N)/$N$(HC$_5$N)  is 35 (Table \ref{model_pars_table}), close to the value 
derived by \citet{Gong2015},
but much smaller than the column density ratio of $N$(HC$_5$N)/$N$(HC$_7$N), 6.6. 
However, if we only take the reddest velocity component of HC$_3$N into account, the
abundance ratio of $N$(HC$_3$N)/$N$(HC$_5$N) would be 7.5, close to the value of $N$(HC$_5$N)/$N$(HC$_7$N).
Thus, it is likely that the emission of HC$_5$N, HC$_7$N, and the reddest velocity component of HC$_3$N originates from the
same region, probably the extended ridge of Orion KL.

\subsubsection{Ammonia (NH$_3$ and NH$_2$D)}
In total, 19 lines of ammonia were detected in this survey,
including two lines of deuterated ammonia (NH$_2$D). 
Among them, NH$_3$ inversion line of $J_K=14_{13}$ (27772 MHz) has the highest $E_{\rm u}$ of 2090 K.
The emission lines of NH$_3$ should be optically thick and we did not try to precisely fit them.
To explain the high-$E_{\rm u}$ lines,
a component with a $T_{\rm ex}$ of 300 K was adopted. 
Among the two lines of NH$_2$D, one is for the $para$-NH$_2$D (30979 MHz), and the other one for the $ortho$-NH$_2$D (29188 MHz).
Under the equilibrium state, the abundance ratio of $ortho$-NH$_2$D/$para$-NH$_2$D ($o$/$p$) should be 3, 
the ratio of their statistical
weights.  When we fitted the  NH$_2$D  emission, the $T_{\rm ex}$ was fixed to be 300 K and 
the $o$/$p$ was fixed to be 3 since we did not distinguish the two types of NH$_2$D.
Hence, the free parameters are only the column density ($N$), line width ($\Delta V$) and velocity ($V_{\rm LSR}$).
The two lines can be fitted simultaneously (Figure \ref{continued_fit_spec}), and thus the assumption of LTE should be acceptable.
Only one line of $^{15}$NH$_3$ is detected, and the column density of $^{15}$NH$_3$ is calculated by adopting a $T_{\rm ex}$ of 300 K.
The abundance ratio of NH$_2$D/$^{15}$NH$_3$ is derived to be 4, but it highly depends on the $T_{\rm ex}$
and the assumption of LTE for $^{15}$NH$_3$.

\subsection{Other species}
For S-bearing species, $^{33}$SO and $^{18}$OCS, which have not been detected in the Q-band survey of TMRT, 
were detected in the Ka band (Table \ref{model_pars_table}). The abundance ratio of $^{34}$SO/$^{33}$SO
is derived to be 4.8, which does not deviate much from the value of \citet{2013A&A...556A.143E}, $\sim$6.

All O-bearing complex organic molecules detected in this survey (Table \ref{allspecies}), 
except for the isotopologues of CH$_3$OH (Section \ref{sec_ch3oh}), have been detected by the Q-band survey of the TMRT 
\citep{2022ApJS..263...13L}.
HDCO was not detected in the Q-band survey.
For HDCO and H$_2^{13}$CO, only one transition was detected for each in this survey.
Adopting the same value of $T_{\rm ex}$ of 50 K, the abundance ratio of H$_2^{13}$CO/HDCO is $\sim$1.
Forty lines of CH$_3$OCH$_3$ were detected in this survey. In the Ka band, the total emission flux of CH$_3$OCH$_3$ is
comparable with that of C$_2$H$_5$CN (Figure \ref{fig_flux_kaq}).  In contrast, in the Q band, the total emission flux of CH$_3$OCH$_3$
is only 60 percent of that of C$_2$H$_5$CN.  In the K band, only two lines of CH$_3$OCH$_3$ were detected by \citet{Gong2015}.
It implies that the Ka band has superiorities in detecting some species of complex organic molecules
compared to  the K and Q bands.

\section{Summary}\label{secsummary}
We have conducted a Ka-band line survey (26.1--35 GHz) 
towards Orion KL using the TMRT, following the Q-band survey with the same instrument \citep{2022ApJS..263...13L}. 
This survey achieved a line sensitivity of
mK level (1--3 mK) at a spectral resolution of $\sim$1 km s$^{-1}$.
This is the first blind line survey in the Ka band, and many transitions were detected for the first time.
Further, we explore the Ka-band spectrum of Orion KL and present the overall preliminary results of the survey.
The main results  include:
\begin{itemize}
\item[1] We have obtained and release the Ka-band spectrum of Orion KL by the TMRT.
In the spectrum, 592 Gaussian features are detected in total,
including 257 unblended RRLs and 318 unblended molecular lines.
The maximum $\Delta n$ of RRLs of H, He, and C reach 20, 15, and 5, respectively.
In total, 37 molecular species  were detected,
and the vibrationally excited states ($v_t>0$) of nine of these species were detected.
Ten of these species have not been detected in the Q-band survey of TMRT,
including  CH$_3$C$_3$N which was not detected in  Orion KL previsously even by ALMA.
A catalog of these lines is presented.
\item[2] Through stacking, we have detected the $\beta$  lines of ion RRLs (RRLs of C$^+$ and/or O$^+$) for the first time,
following our previous report of the $\alpha$ lines of ion RRLs. 
Tentative signal of the $\gamma$ lines of ion RRLs can  be seen on the stacked spectrum.
We also successfully matched the $\alpha$ lines of ion RRLs from the archive data
of the K-band survey of Orion KL by the Effelsberg 100 m \citep{Gong2015},
providing a very important cross check of ion RRLs from an independent
facility besides the TMRT.
\item[3] We modeled the spectral emission under the assumption of LTE.
Parameters including the column densities ($N$) of molecular species and the emission measures ($EM$) of RRLs were derived. 
Adopting the precise values of the absorption oscillator strength of RRLs ($f_{n1,n2}$), the intensities of nearly all 
RRLs can be reproduced within
a deviation of 50 percent. For $\Delta n>10$, the observed intensities tend to be slightly lower than model fitting.   
\item[4] Most species can be well fitted under the assumption of LTE except for CH$_3$OH.
The intensities of some transitions of  E-CH$_3$OH $J_2-J_1$
can be at most four times the values of LTE prediction. 
The maser effect may explain this deviation \citep{1996A&A...307..209W}. 
Combining the data in Ka (this survey) and K band \citep{Gong2015},
we found that besides the bumping peak at $J\sim 6$,  there is another one at  $J\sim 13$.
\end{itemize}

Overall, this work emphasizes that the TMRT is able to conduct deep line surveys in the Ka band,
and this band is very useful for surveying RRLs and molecular lines simultaneously.


\begin{acknowledgements}\small 
We wish to thank the staff of the TMRT 65 m for their
help during the observations. This work has been supported by the National Key R\&D Program of China (No. 2022YFA1603100).
X.L. acknowledges the supports by NSFC No. 12203086 and No. 12033005 and CPSF No. 2022M723278. T.L. acknowledges the supports by National Natural Science Foundation of China (NSFC) through grants No.12073061 and No.12122307, the international partnership program of Chinese Academy of Sciences through grant No.114231KYSB20200009, Shanghai Pujiang Program 20PJ1415500 and the science research grants from the China Manned Space Project with no. CMS-CSST-2021-B06.
D.S. acknowledges the support from Ramanujan Fellowship (SERB) and PRL, India.
We show warm thanks to the anonymous referee for providing many deep-insight comments for improving the paper.
\end{acknowledgements}

\facilities{Tianma 65 m (TMRT)}
\software{GILDAS/CLASS \citep{2000ASPC..217..299G}, astropy \citep{2013A&A...558A..33A}}

\bibliography{ms}
\bibliographystyle{aasjournal}

\clearpage   
\appendix 
\section{The absorption oscillator strength of RRLs} \label{sec_os_precise}
\begin{figure*}[htb]
\centering
\includegraphics[width=0.48\linewidth]{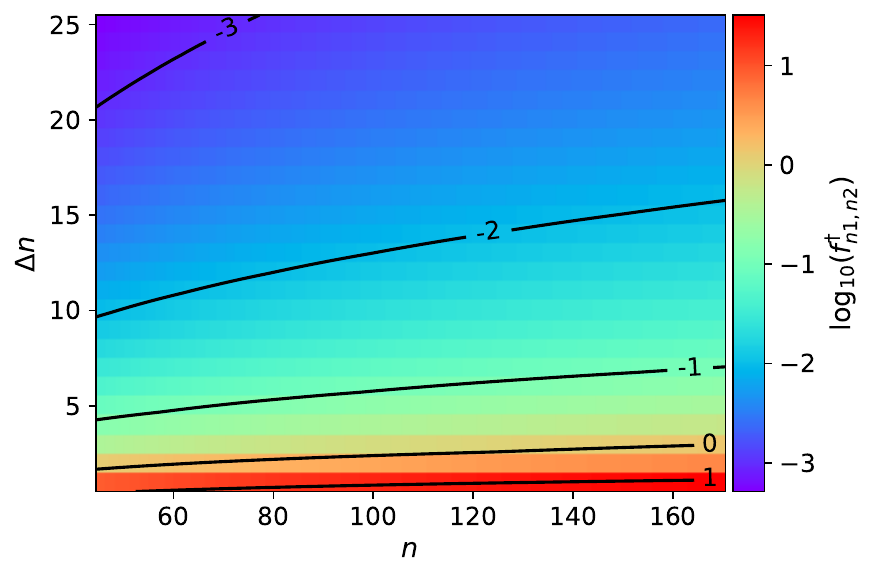}
\includegraphics[width=0.48\linewidth]{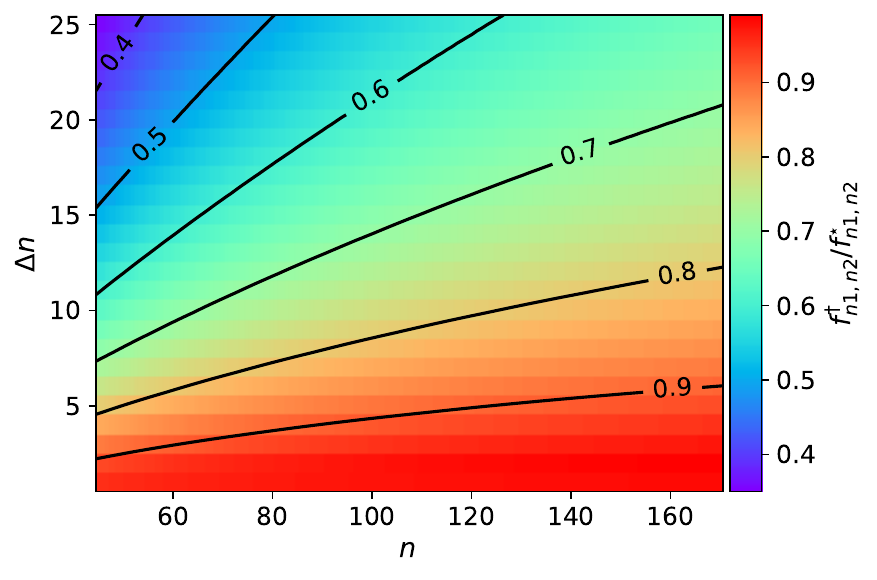}
\caption{The $\log_{10}(f^{\dag}_{n_1,n_2})$ (left) and $f^{\dag}_{n_1,n_2}/f^{\star}_{n_1,n_2}$ (right)
for RRLs with different $n$ and $\Delta n$. \label{fn1n2fig}}
\end{figure*}

For hydrogenic RRL emitters, the absorption oscillator strength, $f_{n1,n2}$, is independent of $Z$ ignoring the effect of the 
slightly different reduced mass of electron \citep{1959SvA.....3..813K,1967ApJ...149...15G}.
\citet{1968Natur.218..756M}  gave a useful
approximation for the oscillator strength $f^{\star}_{n_1,n_2}$ for RRLs: 
\begin{equation}
f^{\star}_{n_1,n_2} \sim n_1M_{\Delta n} \left( 1+1.5\frac{\Delta n}{n_1} \right) \label{eq_tau_fnn}
\end{equation}
In \citet{2022ApJS..263...13L}, we empirically fitted the values of $M_{\Delta n}$ tabulated in \citet{1968Natur.218..756M} as
\begin{equation}
M_{\Delta n} \sim 0.1905 \left(1/\Delta n\right)^{2.887} \label{eq_Mdn}
\end{equation}
Actually, the oscillator strength can be accurately calculated through the Gordon formula \citep{1929AnP...394.1031G}.
Following the Eqs. (3.1)--(3.7) of \citet{1971MNRAS.153..471B},
we numerically tabulated  the precise oscillator strength (denoted as 
$f^{\dag}_{n1,n2}$) \footnote{\url{https://gitee.com/liuxunchuan/tmrtsurvey/blob/master/fn1n2.py}}.
The $\log_{10}(f^{\dag}_{n_1,n_2})$ and $f^{\dag}_{n_1,n_2}/f^{\star}_{n_1,n_2}$ are shown in
Figure \ref{fn1n2fig}. 
The $f^{\star}_{n_1,n_2}$ deviates from $f^{\dag}_{n_1,n_2}$ by less than 20 percent 
when $\Delta n<5$.

\clearpage

\section{The $\alpha$ lines of X$^+$ RRLs in K band by Effelsberg}\label{sec_effl} 
\begin{figure*}[!htb]
\centering
\includegraphics[width=0.9\linewidth]{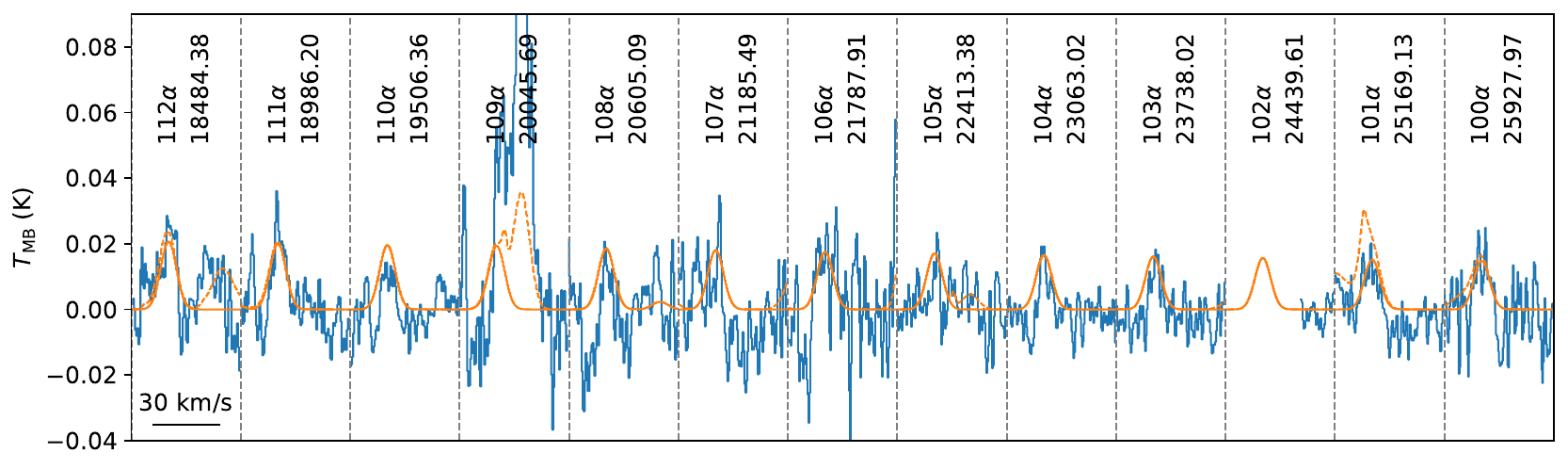}
\caption{The K-band $\alpha$ lines of ion RRLs (blue) from the archive data of Effelsberg \citep{Gong2015}.
The orange lines are the modeled spectrum. The  orange  dashed line takes all species into account,
and the orange solid line only considers the ion RRLs.   \label{fig_efalphas}}
\end{figure*}

Figure \ref{fig_efalphas} shows the $\alpha$ lines of ion RRLs from the archive data of the 
K-band survey of Orion KL by the Effelsberg 100 telescope \citep{Gong2015}.
To exclude blending, we first modeled the K-band spectrum of Orion-KL.
The RRLs are modeled using the parameters ($EM$, $\Delta V$, and $V_{\rm LSR}$) of this work.
All molecular species detected in K band \citep{Gong2015} can be detected in Q band \citep{2022ApJS..263...13L} and Ka band (this work).
Thus,  to model the emission of molecular species,  
the parameters ($N$, $\Delta V$,  $V_{\rm LSR}$ and $T_{\rm ex}$) obtained based on the Q-band and Ka-band data  are used with no modification.
Fortunately, most of the K-band $\alpha$ lines of ion RRLs are not blended (Figure \ref{fig_efalphas}).

\clearpage

\begin{figure*}[!htb] \caption{The Orion KL spectrum \label{allspectra}}
\includegraphics[width=0.99\linewidth]{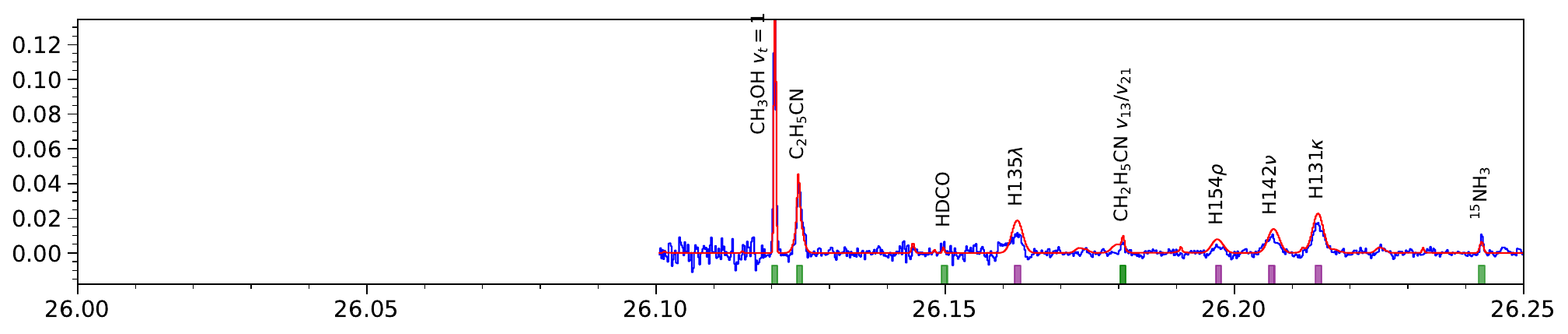}\\
\includegraphics[width=0.99\linewidth]{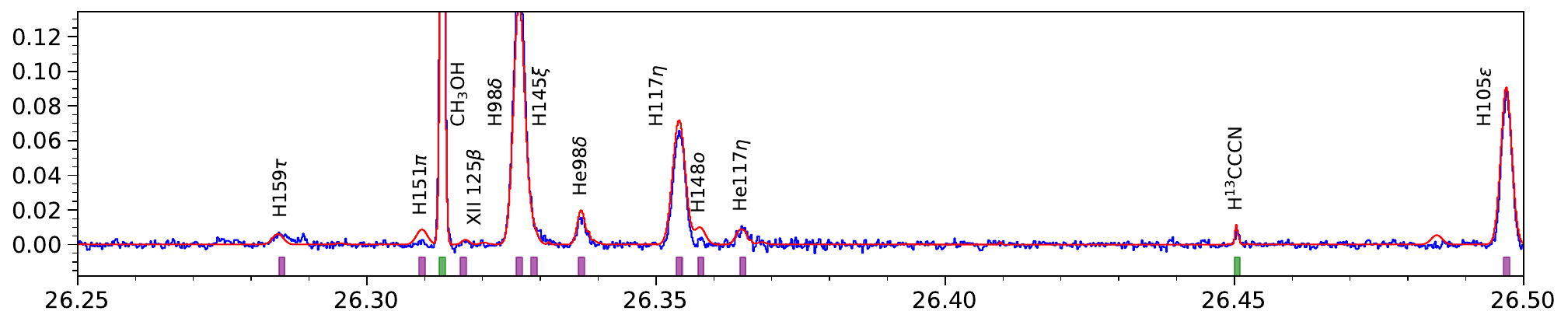}\\
\includegraphics[width=0.99\linewidth]{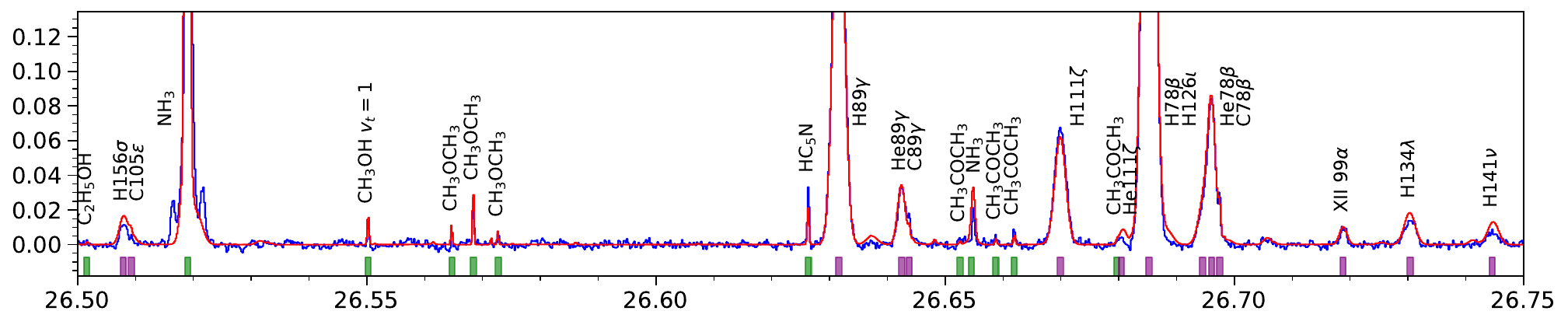}\\
\includegraphics[width=0.99\linewidth]{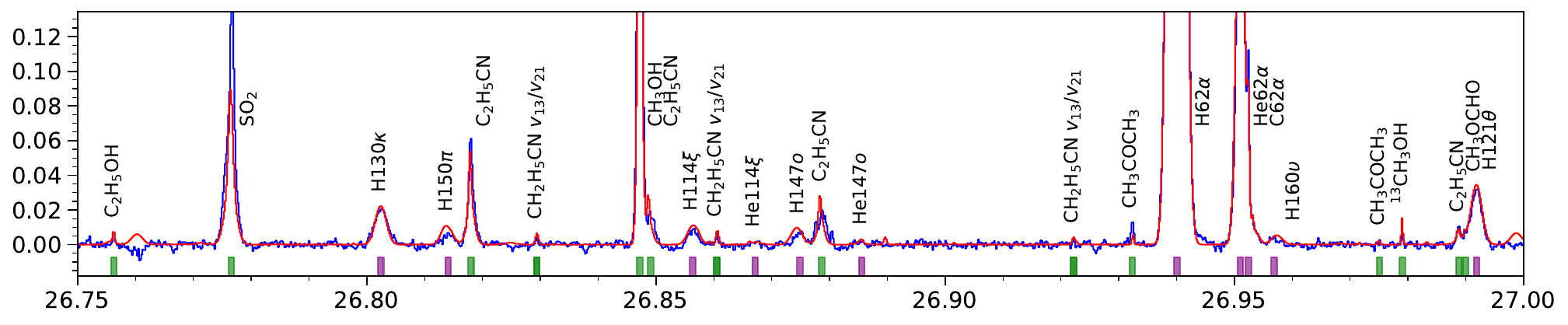}\\
\includegraphics[width=0.99\linewidth]{{27000_27250}.pdf}\\
{\raggedright \footnotesize \textbf{Note}: The blue line is the Orion KL spectrum 
observed by the TMRT 65 m,
which has been smoothed to have a frequency resolution of 183 kHz ($\sim$1.8 km s$^{-1}$ at 30 GHz). 
The red line represents the modeling fitting. The purple strips denote the detected RRLs.  
The green strips denote the 
molecular lines.
The gray strips mark the U lines. 
\par}
\end{figure*} 
        
\begin{figure*}[!htb]
\centering
{\raggedright \centering  \textbf{Figure \thefigure} {\it (continued)} \par}
\includegraphics[width=0.99\linewidth]{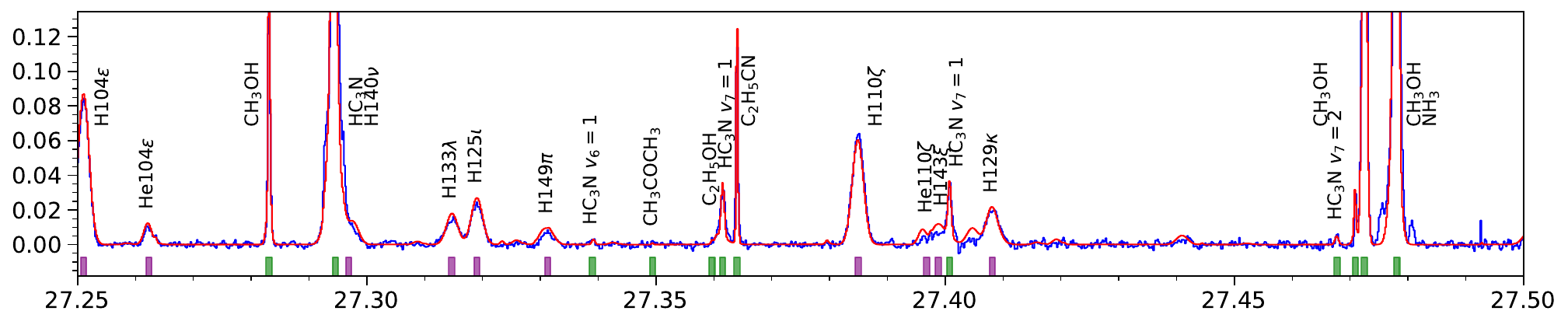}\\
\includegraphics[width=0.99\linewidth]{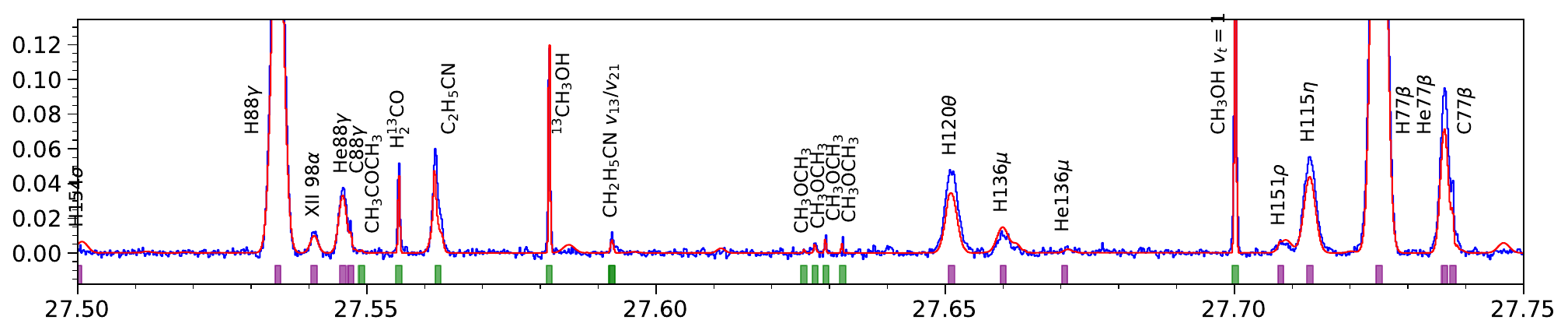}\\
\includegraphics[width=0.99\linewidth]{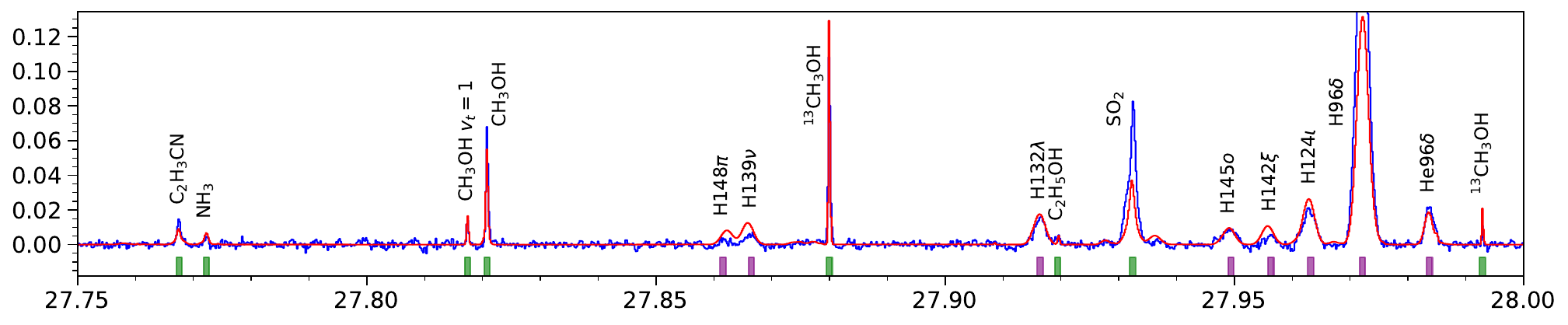}\\
\includegraphics[width=0.99\linewidth]{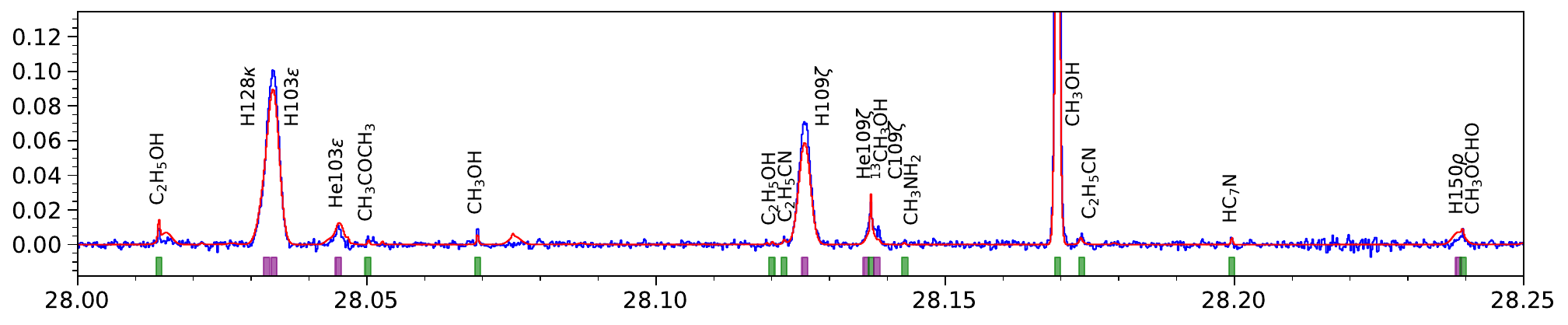}\\
\includegraphics[width=0.99\linewidth]{{28250_28500}.pdf}\\
\includegraphics[width=0.99\linewidth]{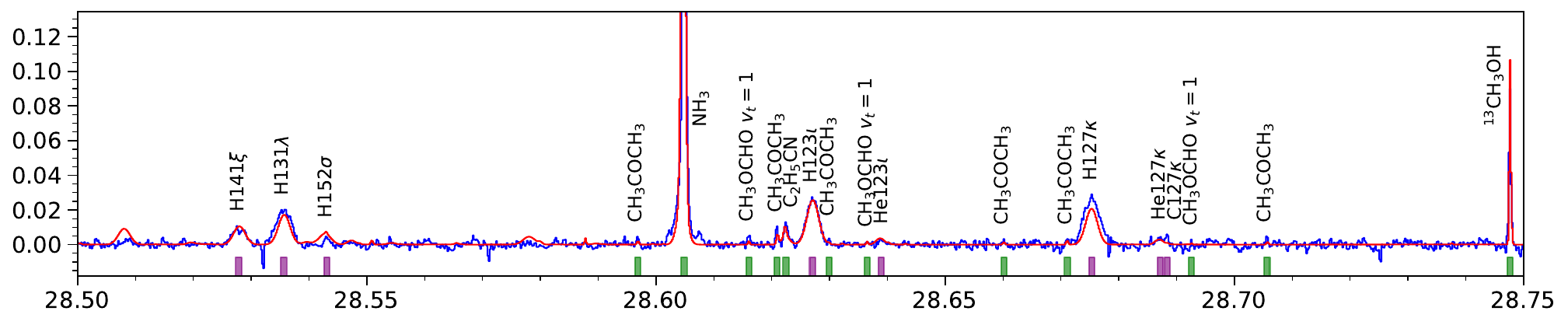}\\
{\raggedright \footnotesize \textbf{Note}: The blue line is the Orion KL spectrum 
observed by the TMRT 65 m,
which has been smoothed to have a frequency resolution of 183 kHz ($\sim$1.8 km s$^{-1}$ at 30 GHz). 
The red line represents the modeling fitting. The purple strips denote the detected RRLs.  
The green strips denote the 
molecular lines.
The gray strips mark the U lines. 
\par}
\end{figure*} 
        
\begin{figure*}[!htb]
\centering
{\raggedright \centering  \textbf{Figure \thefigure} {\it (continued)} \par}
\includegraphics[width=0.99\linewidth]{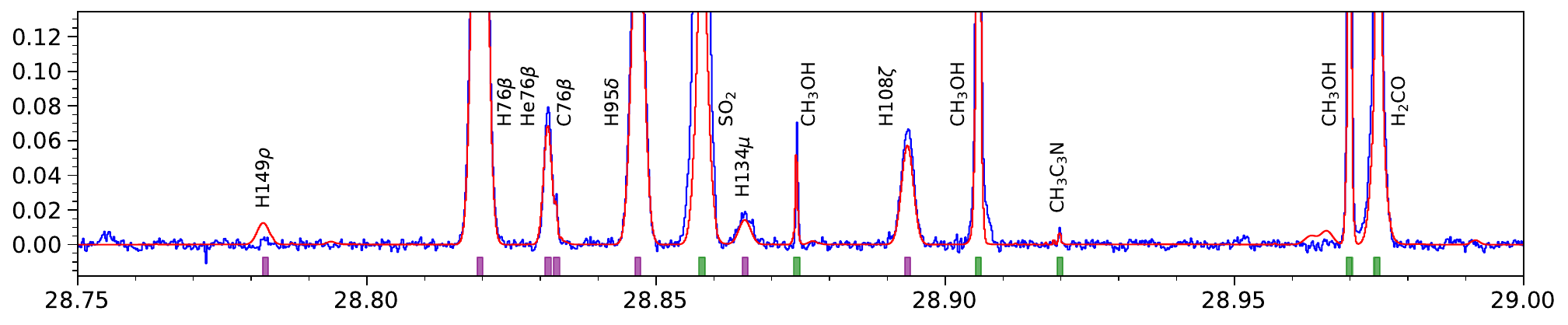}\\
\includegraphics[width=0.99\linewidth]{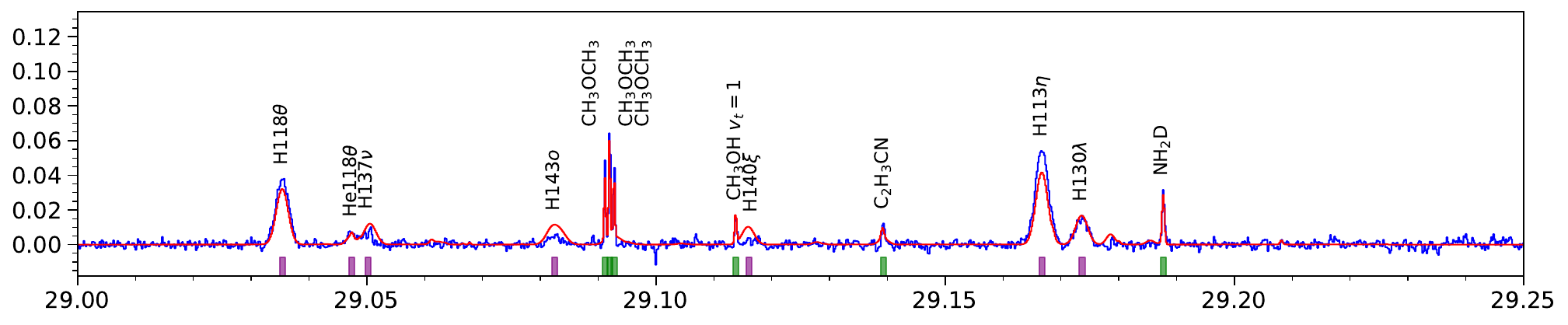}\\
\includegraphics[width=0.99\linewidth]{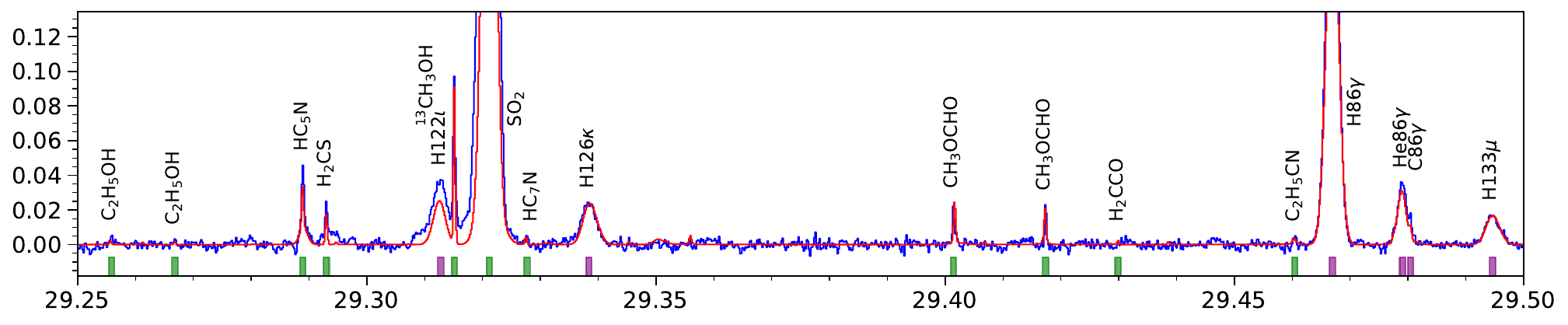}\\
\includegraphics[width=0.99\linewidth]{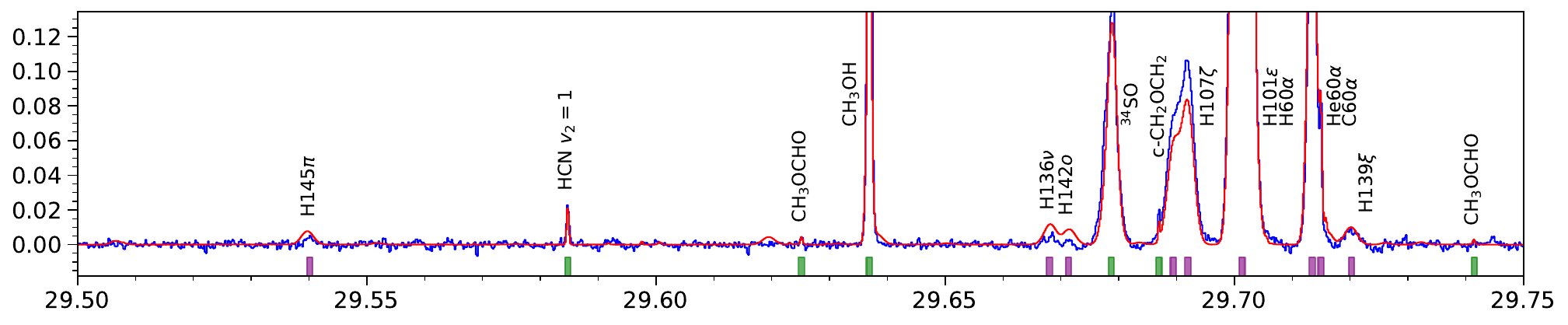}\\
\includegraphics[width=0.99\linewidth]{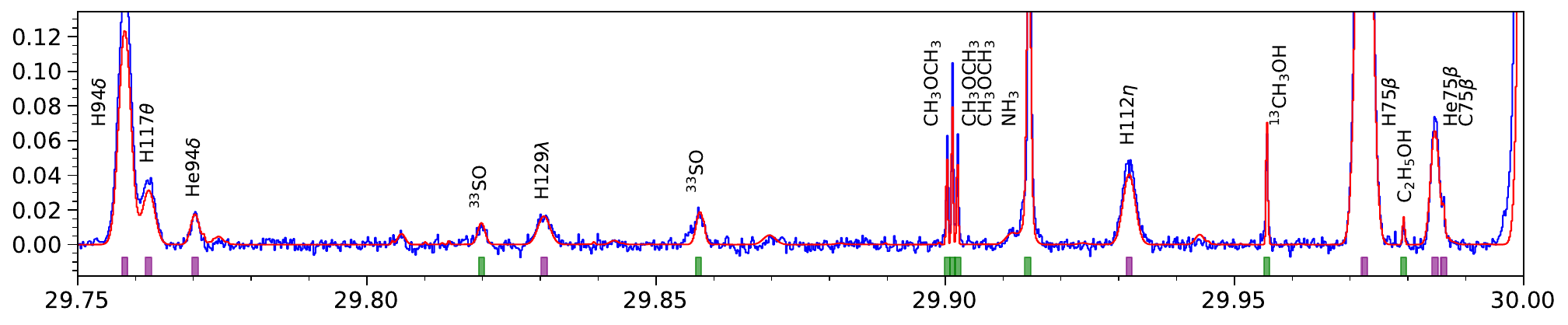}\\
\includegraphics[width=0.99\linewidth]{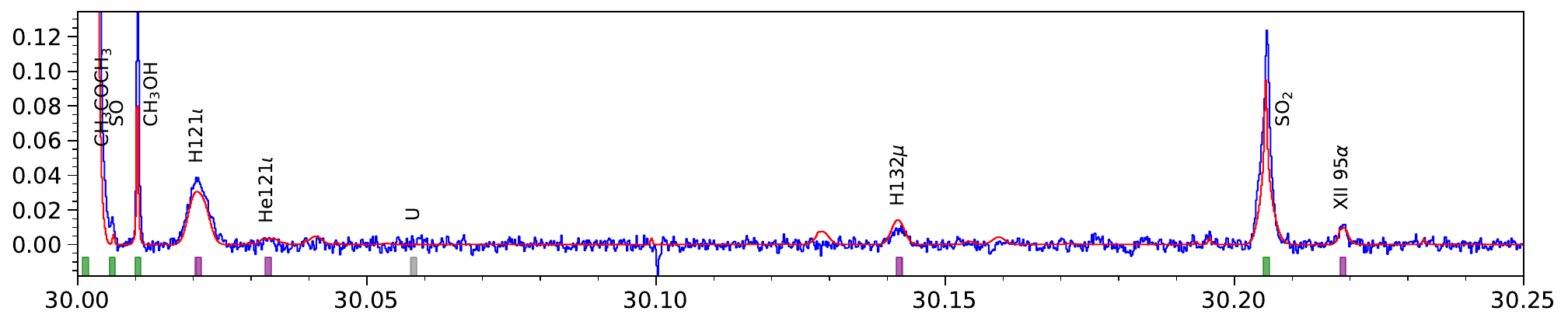}\\
{\raggedright \footnotesize \textbf{Note}: The blue line is the Orion KL spectrum 
observed by the TMRT 65 m,
which has been smoothed to have a frequency resolution of 183 kHz ($\sim$1.8 km s$^{-1}$ at 30 GHz). 
The red line represents the modeling fitting. The purple strips denote the detected RRLs.  
The green strips denote the 
molecular lines.
The gray strips mark the U lines. 
\par}
\end{figure*} 
        
\begin{figure*}[!htb]
\centering
{\raggedright \centering  \textbf{Figure \thefigure} {\it (continued)} \par}
\includegraphics[width=0.99\linewidth]{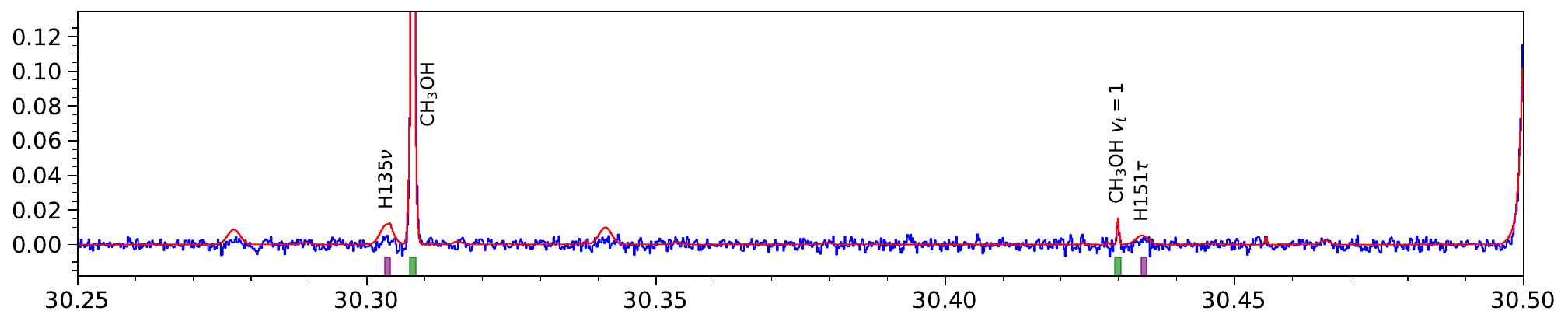}\\
\includegraphics[width=0.99\linewidth]{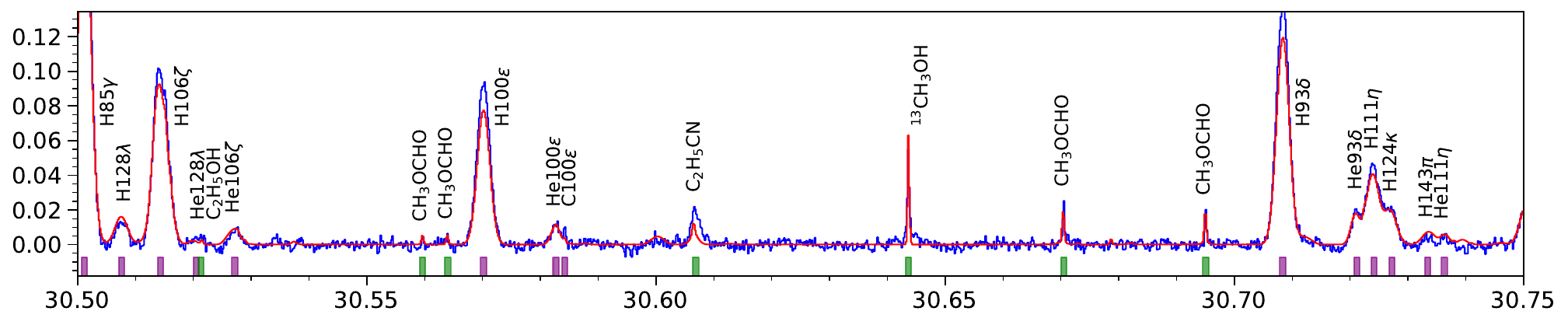}\\
\includegraphics[width=0.99\linewidth]{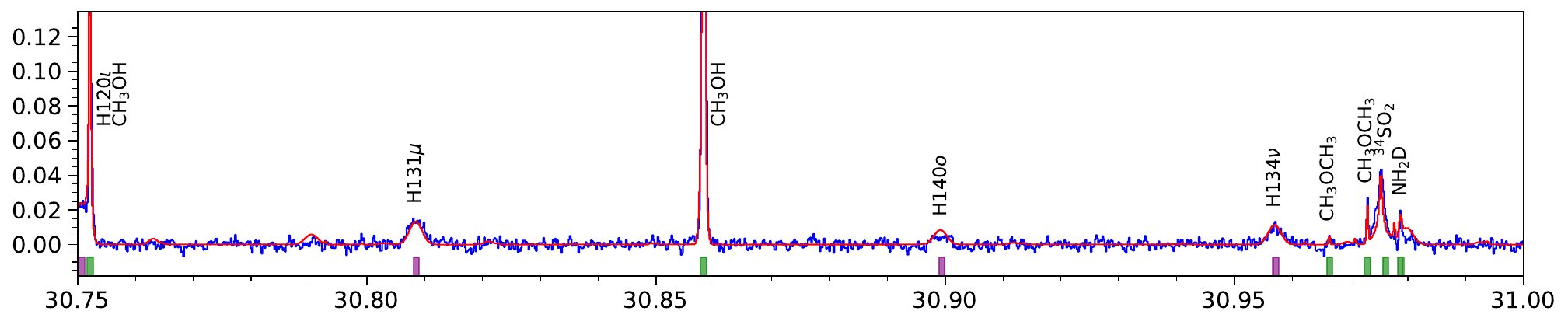}\\
\includegraphics[width=0.99\linewidth]{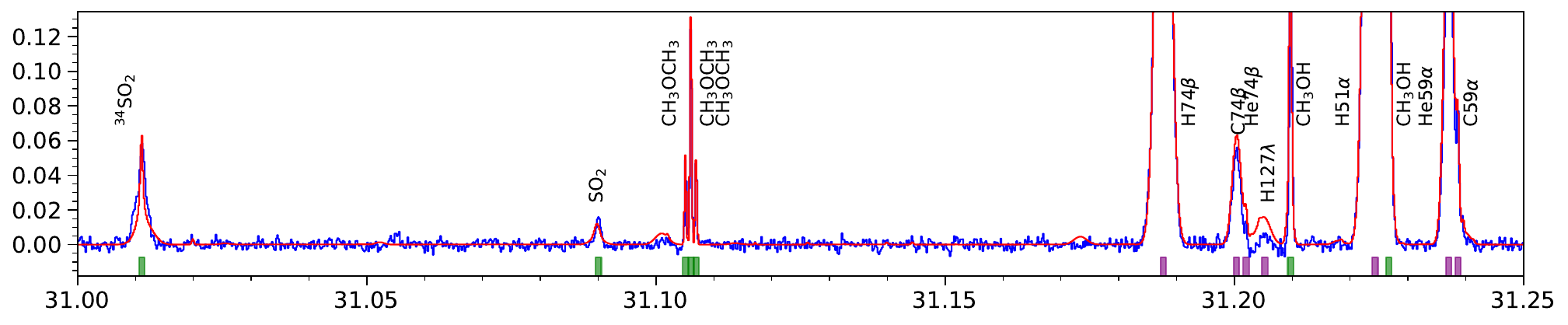}\\
\includegraphics[width=0.99\linewidth]{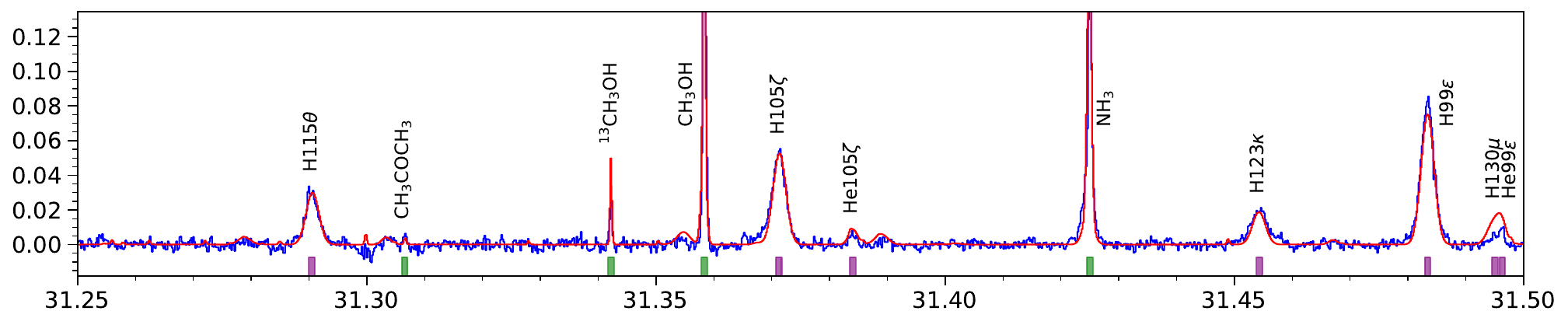}\\
\includegraphics[width=0.99\linewidth]{{31500_31750}.pdf}\\
{\raggedright \footnotesize \textbf{Note}: The blue line is the Orion KL spectrum 
observed by the TMRT 65 m,
which has been smoothed to have a frequency resolution of 183 kHz ($\sim$1.8 km s$^{-1}$ at 30 GHz). 
The red line represents the modeling fitting. The purple strips denote the detected RRLs.  
The green strips denote the 
molecular lines.
The gray strips mark the U lines. 
\par}
\end{figure*} 
        
\begin{figure*}[!htb]
\centering
{\raggedright \centering  \textbf{Figure \thefigure} {\it (continued)} \par}
\includegraphics[width=0.99\linewidth]{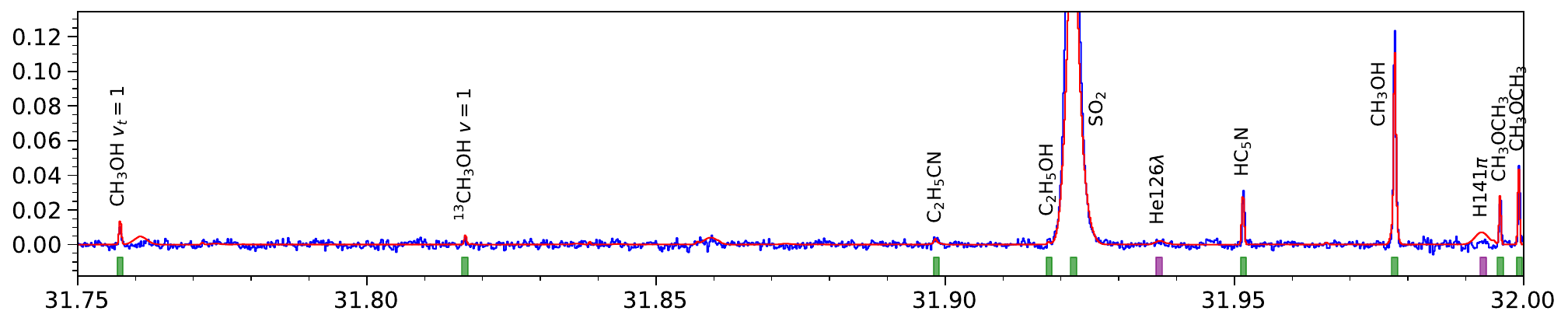}\\
\includegraphics[width=0.99\linewidth]{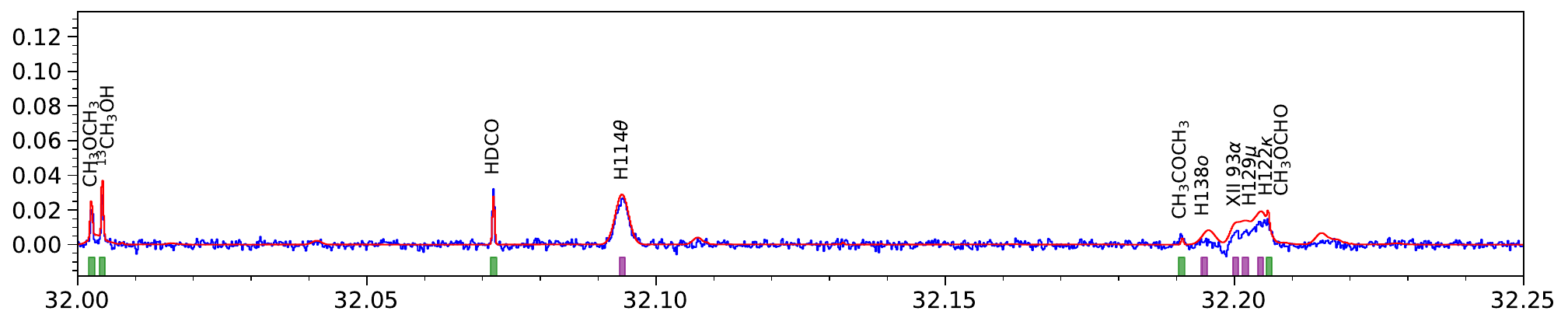}\\
\includegraphics[width=0.99\linewidth]{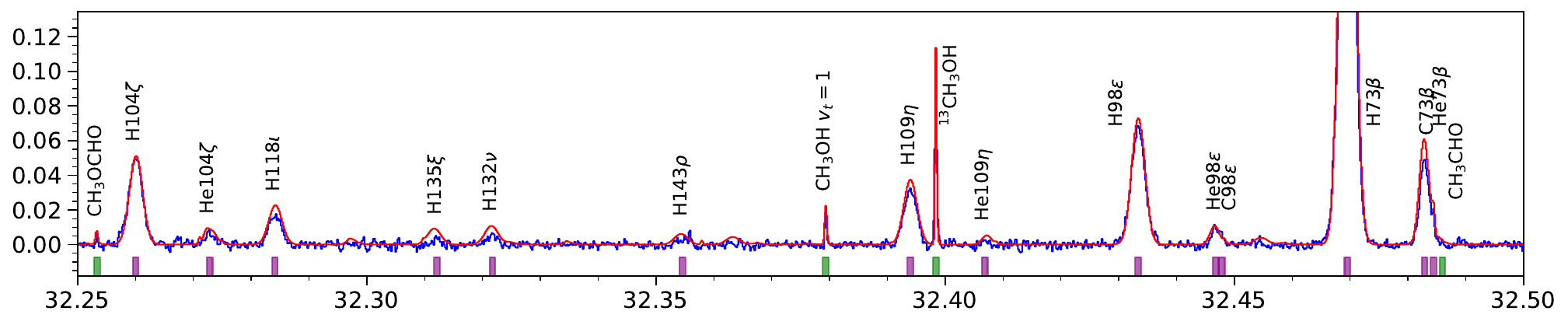}\\
\includegraphics[width=0.99\linewidth]{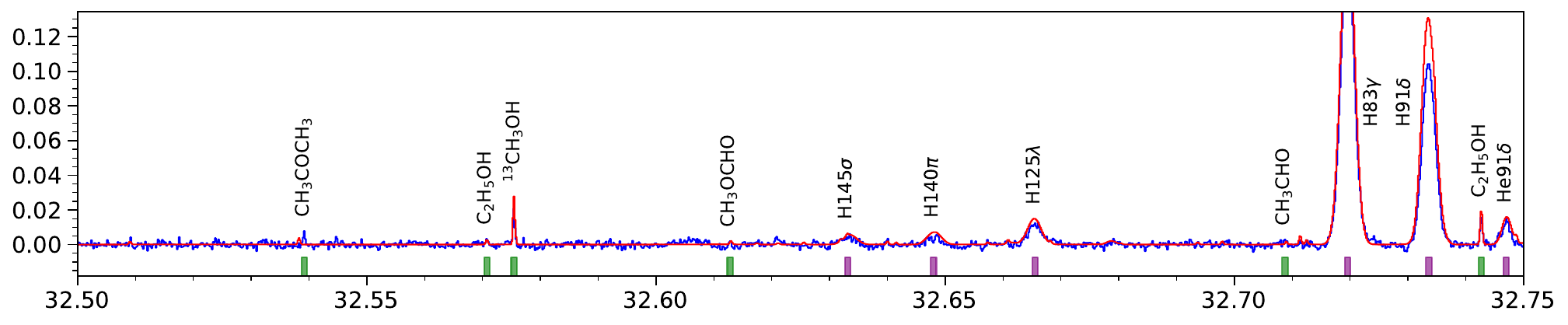}\\
\includegraphics[width=0.99\linewidth]{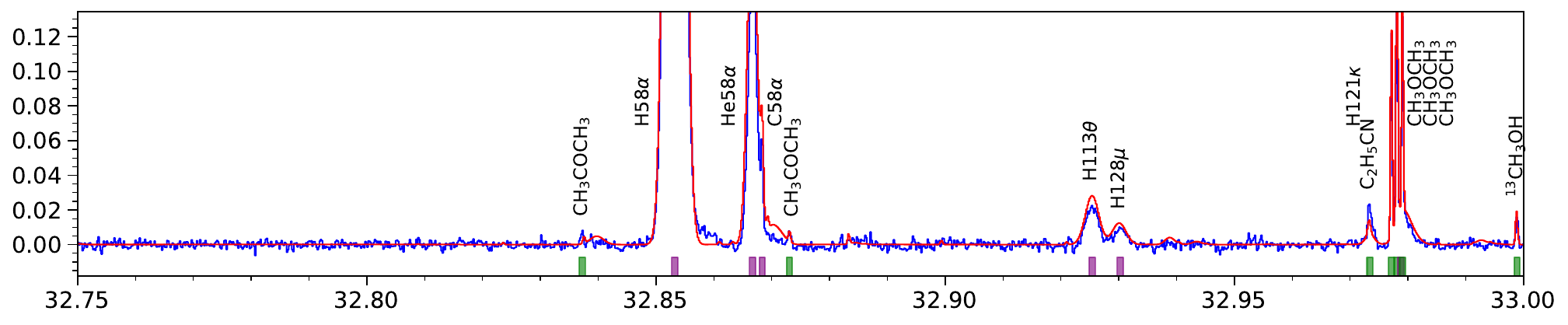}\\
\includegraphics[width=0.99\linewidth]{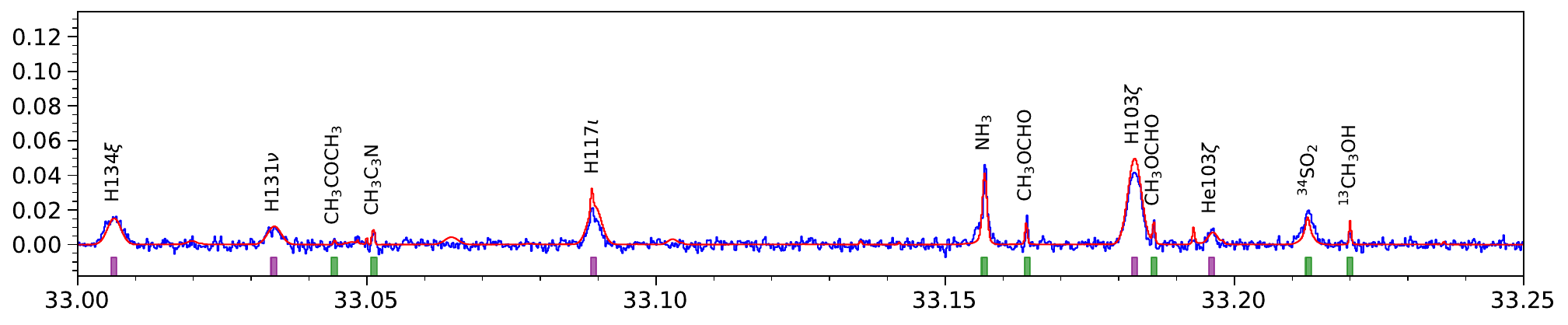}\\
{\raggedright \footnotesize \textbf{Note}: The blue line is the Orion KL spectrum 
observed by the TMRT 65 m,
which has been smoothed to have a frequency resolution of 183 kHz ($\sim$1.8 km s$^{-1}$ at 30 GHz). 
The red line represents the modeling fitting. The purple strips denote the detected RRLs.  
The green strips denote the 
molecular lines.
The gray strips mark the U lines. 
\par}
\end{figure*} 
        
\begin{figure*}[!htb]
\centering
{\raggedright \centering  \textbf{Figure \thefigure} {\it (continued)} \par}
\includegraphics[width=0.99\linewidth]{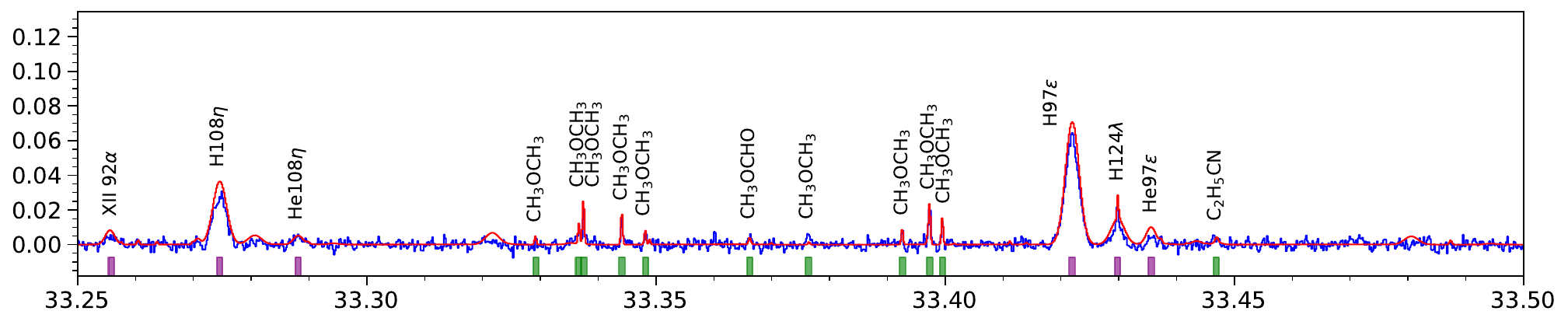}\\
\includegraphics[width=0.99\linewidth]{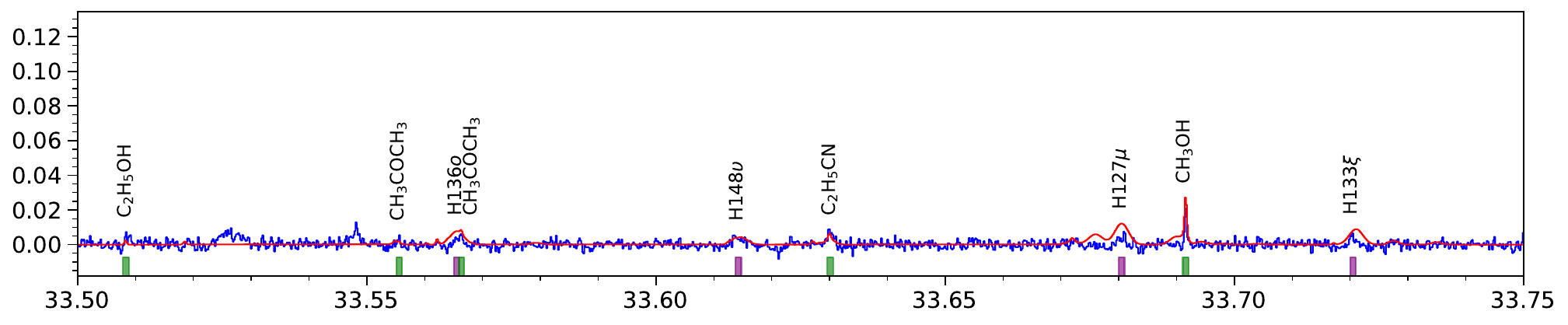}\\
\includegraphics[width=0.99\linewidth]{{33750_34000}.pdf}\\
\includegraphics[width=0.99\linewidth]{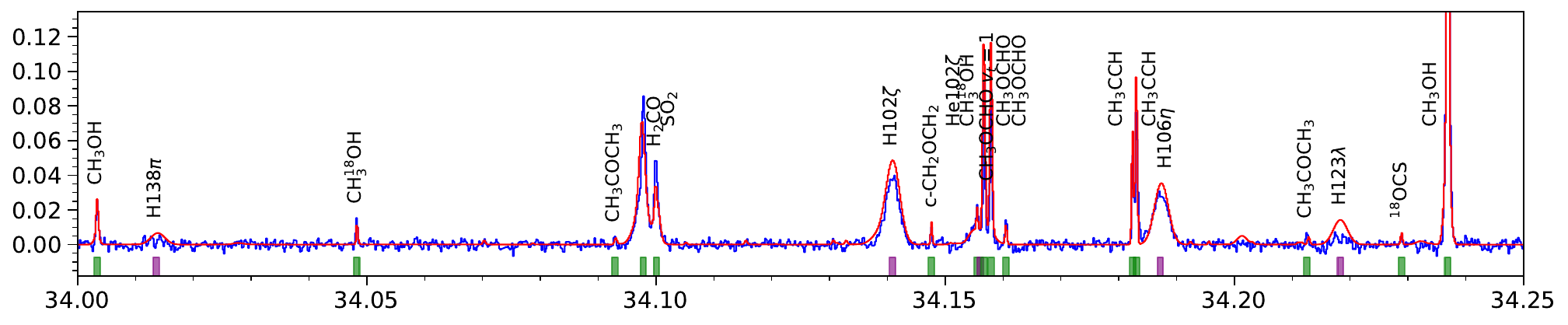}\\
\includegraphics[width=0.99\linewidth]{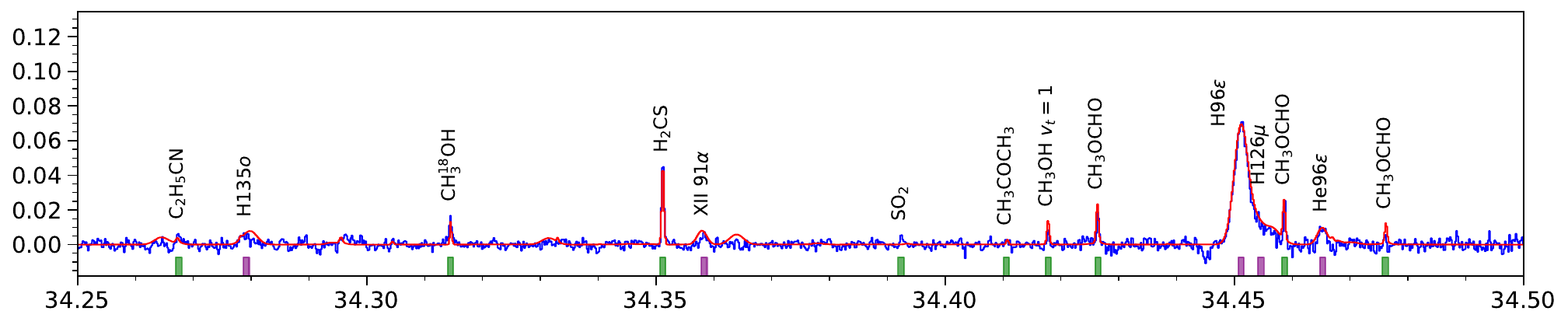}\\
\includegraphics[width=0.99\linewidth]{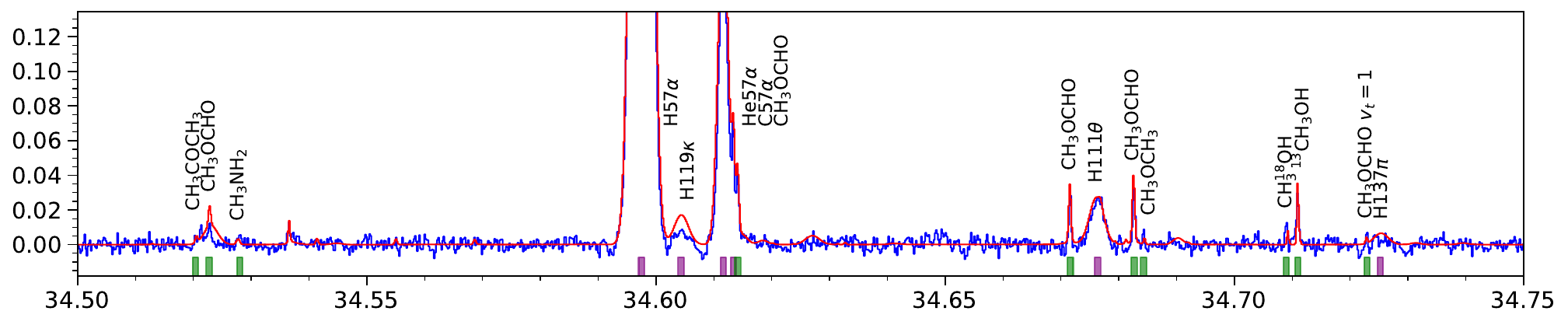}\\
{\raggedright \footnotesize \textbf{Note}: The blue line is the Orion KL spectrum 
observed by the TMRT 65 m,
which has been smoothed to have a frequency resolution of 183 kHz ($\sim$1.8 km s$^{-1}$ at 30 GHz). 
The red line represents the modeling fitting. The purple strips denote the detected RRLs.  
The green strips denote the 
molecular lines.
The gray strips mark the U lines. 
\par}
\end{figure*} 
        
\begin{figure*}[!htb]
\centering
{\raggedright \centering  \textbf{Figure \thefigure} {\it (continued)} \par}
\includegraphics[width=0.99\linewidth]{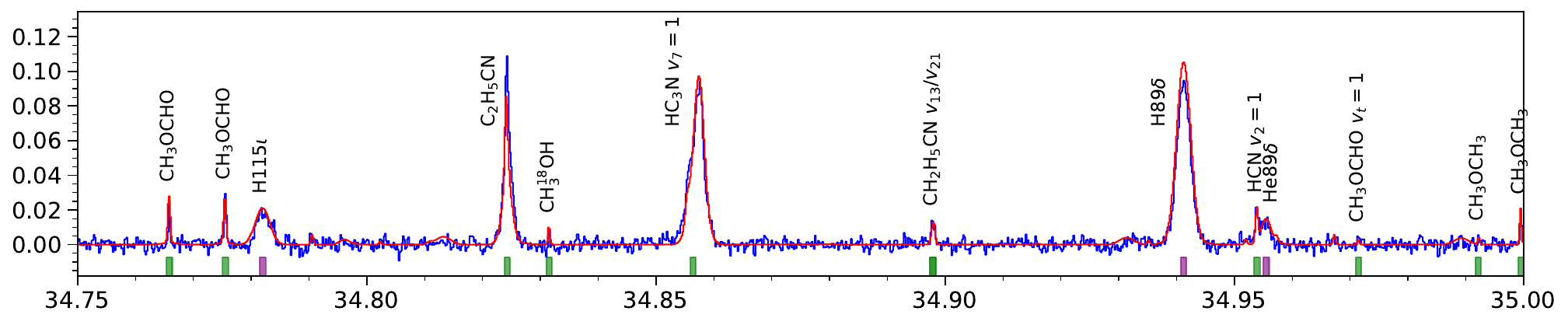}\\
{\raggedright \centering  \textbf{Figure \thefigure} {\it continued} \par}
\end{figure*}

\clearpage

\begin{figure*}[!htb] 
\centering
\caption{Model fitting for different species. 
    \label{continued_fit_spec} 
}
\includegraphics[width=0.485\linewidth]{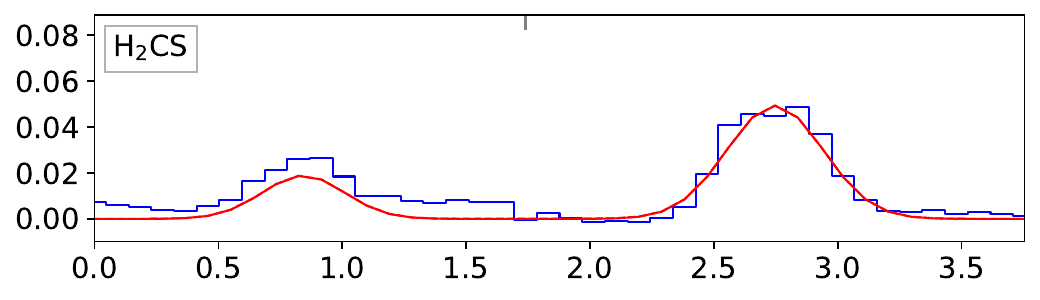}
\includegraphics[width=0.485\linewidth]{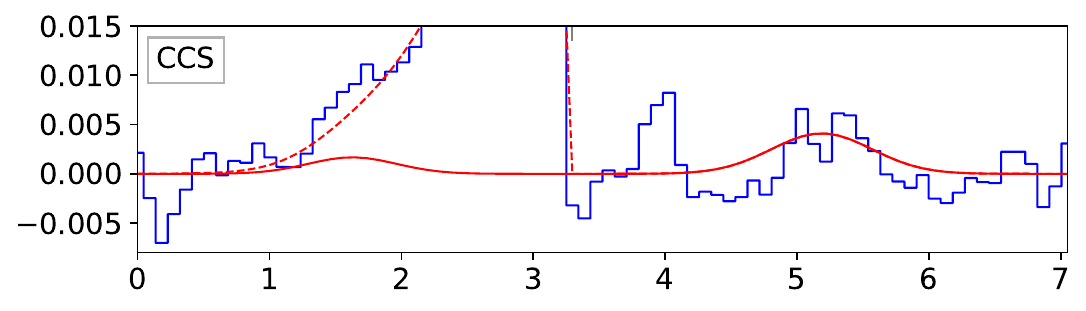}
\\
\includegraphics[width=0.485\linewidth]{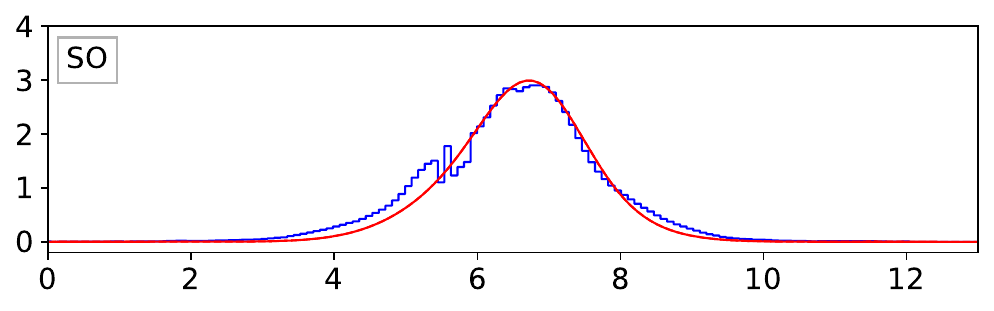}
\includegraphics[width=0.485\linewidth]{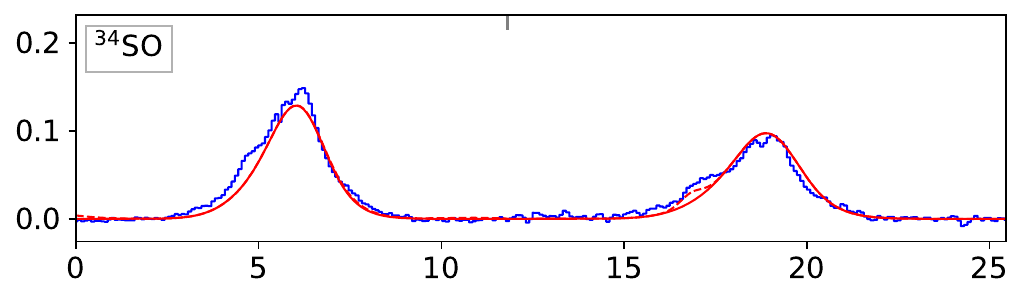}
\\
\includegraphics[width=0.485\linewidth]{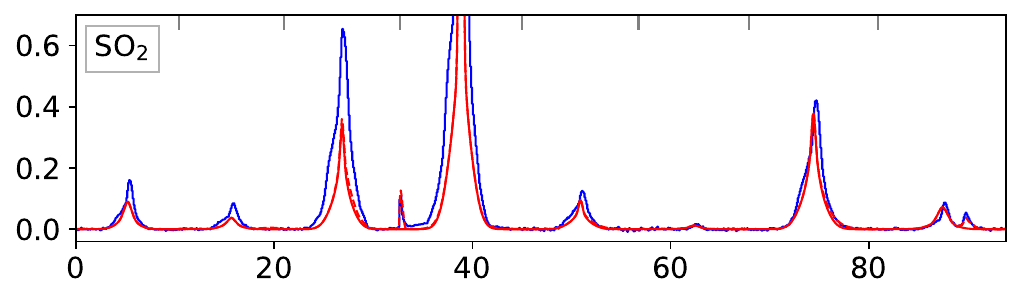}
\includegraphics[width=0.485\linewidth]{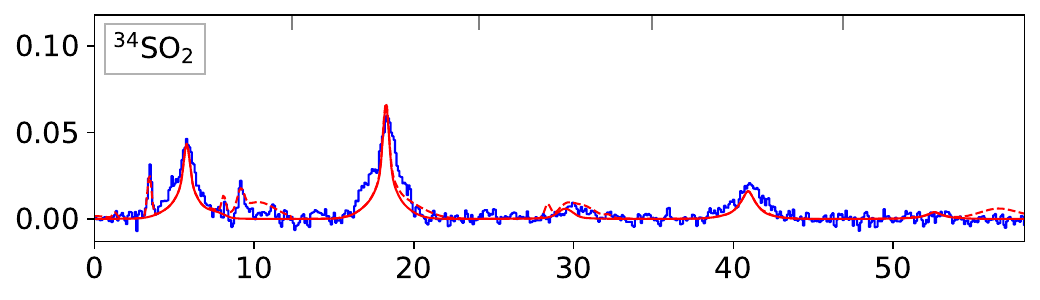}
\\
\includegraphics[width=0.485\linewidth]{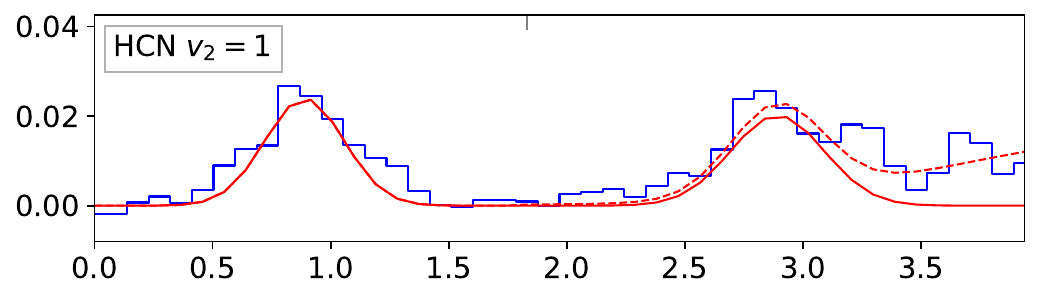}
\includegraphics[width=0.485\linewidth]{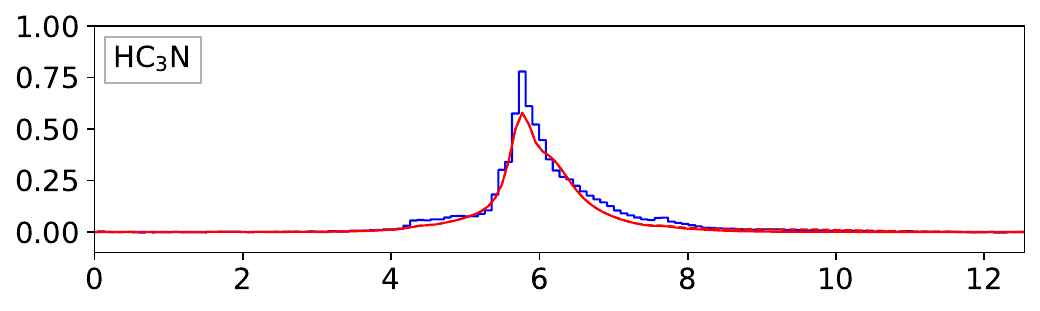}
\\
\includegraphics[width=0.485\linewidth]{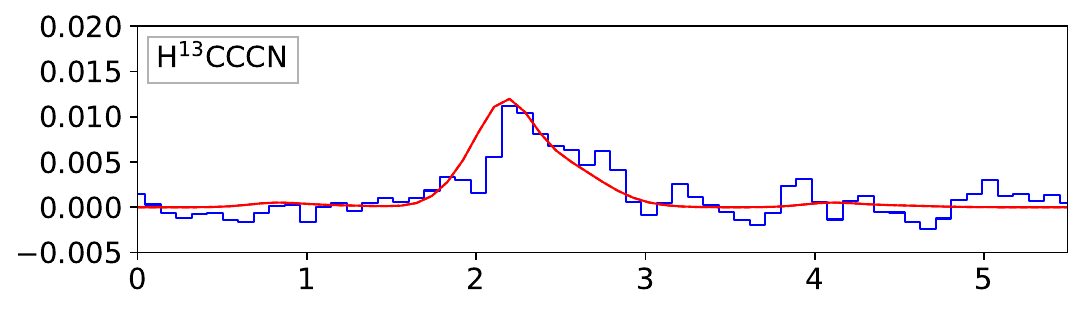}
\includegraphics[width=0.485\linewidth]{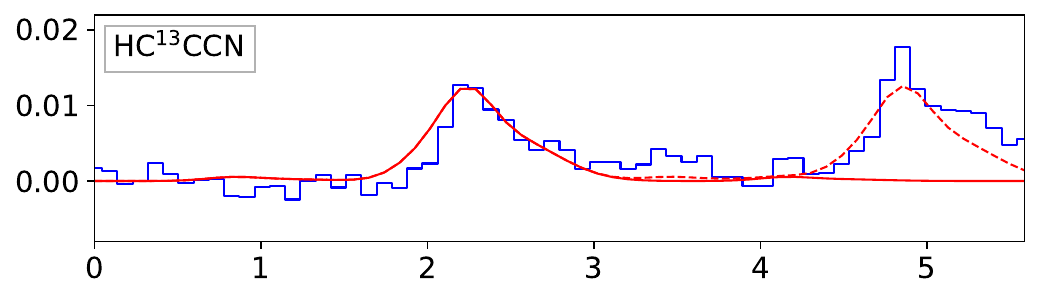}
\\
\includegraphics[width=0.485\linewidth]{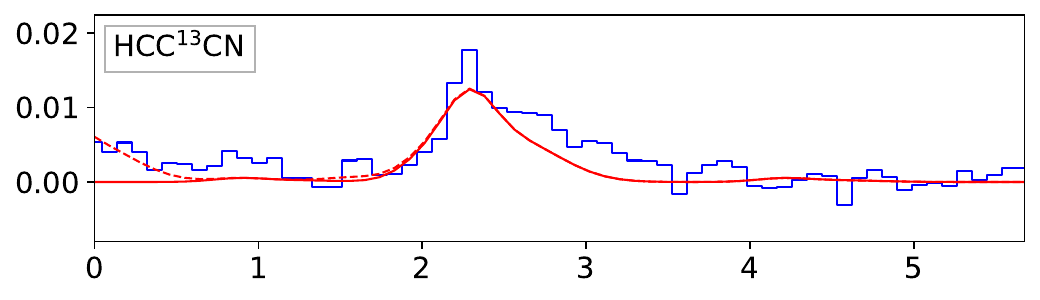}
\includegraphics[width=0.485\linewidth]{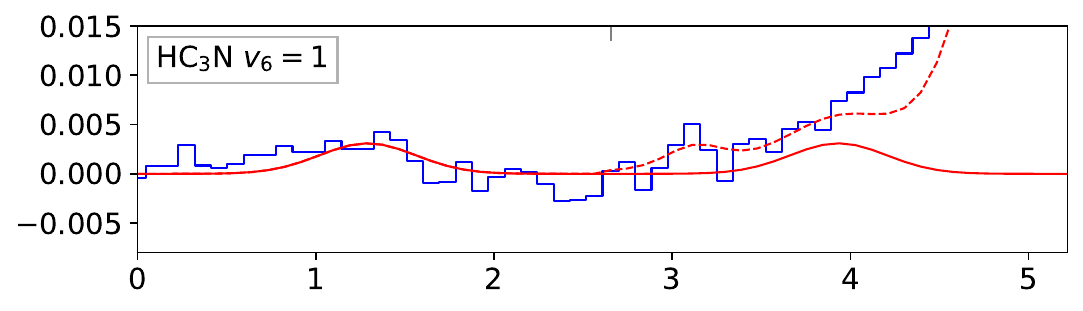}
\\
\includegraphics[width=0.485\linewidth]{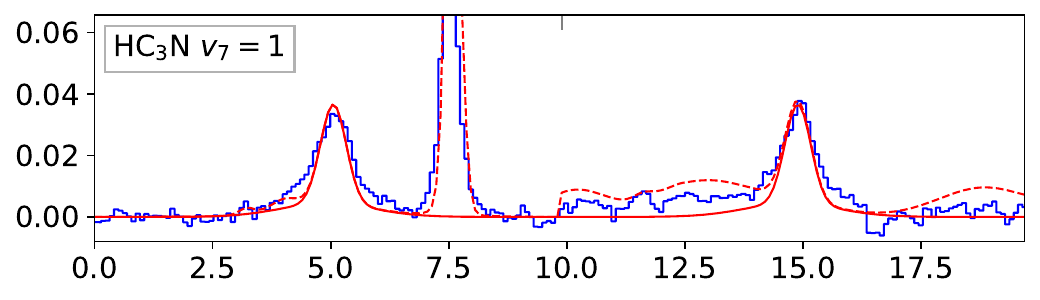}
\includegraphics[width=0.485\linewidth]{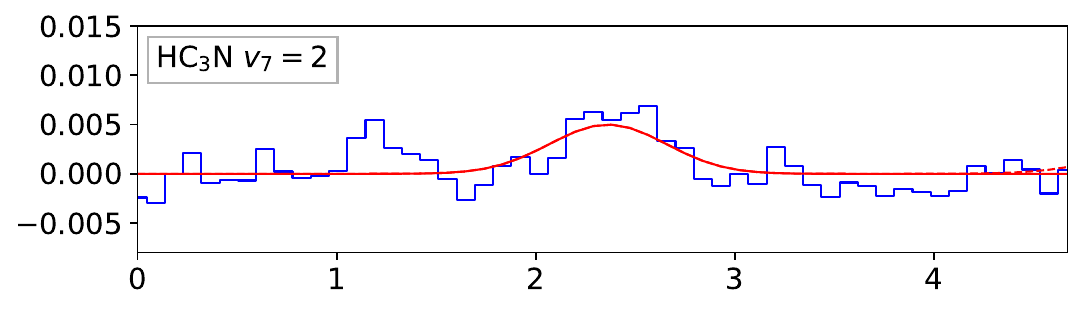}
\\
\vspace{1ex}
{\raggedright \footnotesize Note: In each panel, the spectra between two neighbouring 
upper ticks are independent frequency segments
containing transitions of the corresponding species. The $x$ axis is in unit of MHz.
The $y$ axis is $T_{\rm MB}$ in unit of K. 
See Figure  \ref{example_fit_spec} for the spliced spectra of  methanol.
\par}
{\raggedright \centering  \textbf{Figure \thefigure} {\it continued} \par}
\end{figure*}
        
\begin{figure*}
\centering
{\raggedright \centering  \textbf{Figure \thefigure} {\it (continued)} \par}
\includegraphics[width=0.485\linewidth]{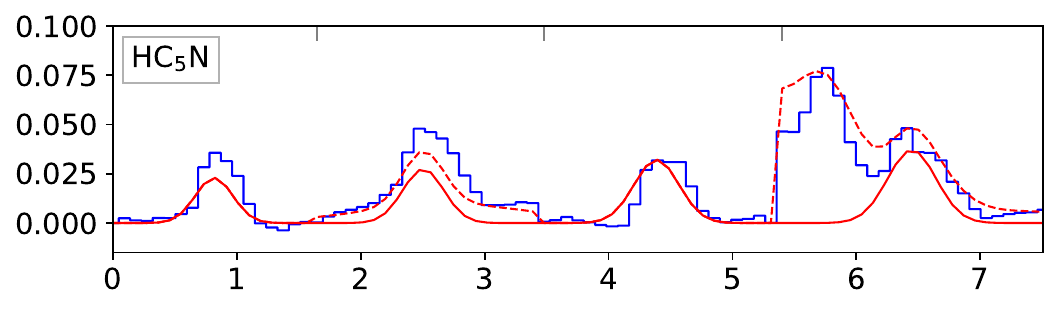}
\includegraphics[width=0.485\linewidth]{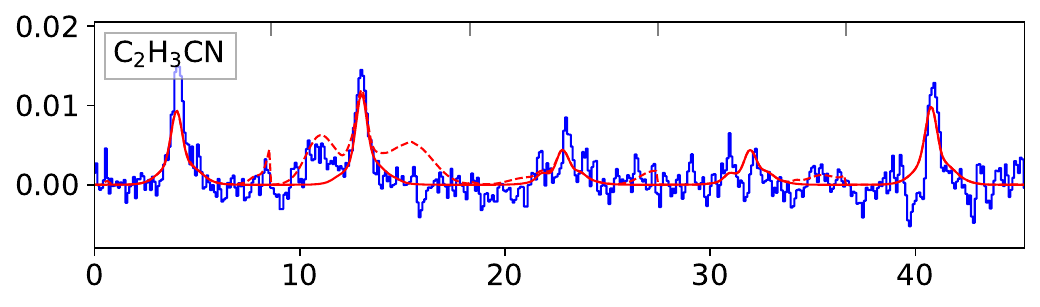}
\\
\includegraphics[width=0.485\linewidth]{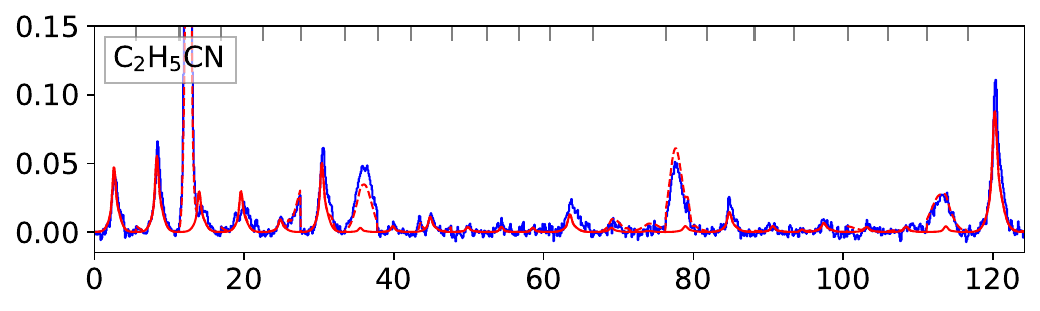}
\includegraphics[width=0.485\linewidth]{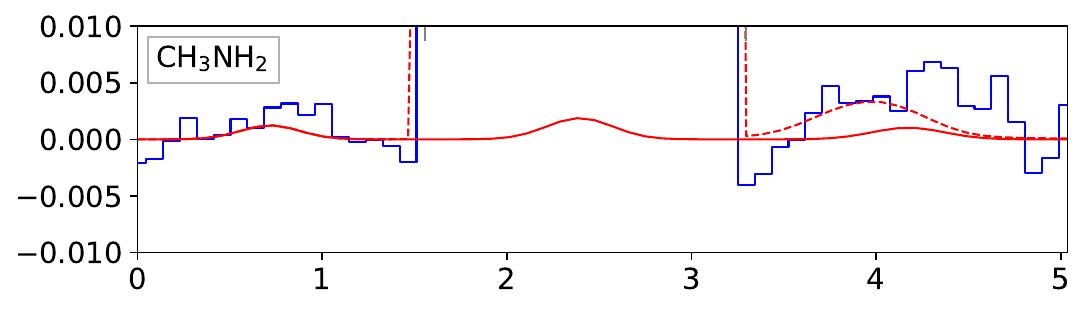}
\\
\includegraphics[width=0.485\linewidth]{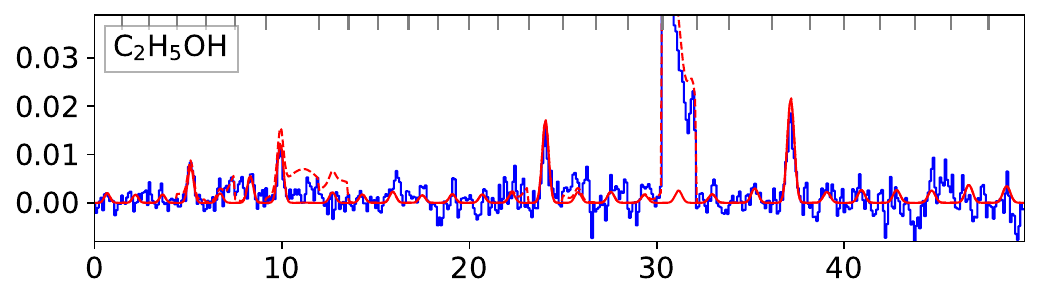}
\includegraphics[width=0.485\linewidth]{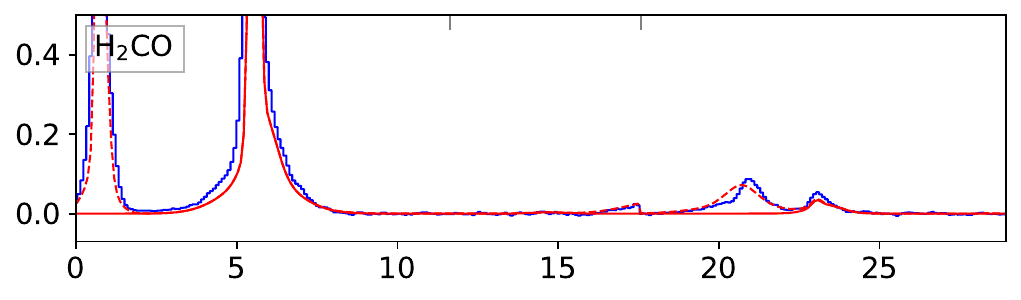}
\\
\includegraphics[width=0.485\linewidth]{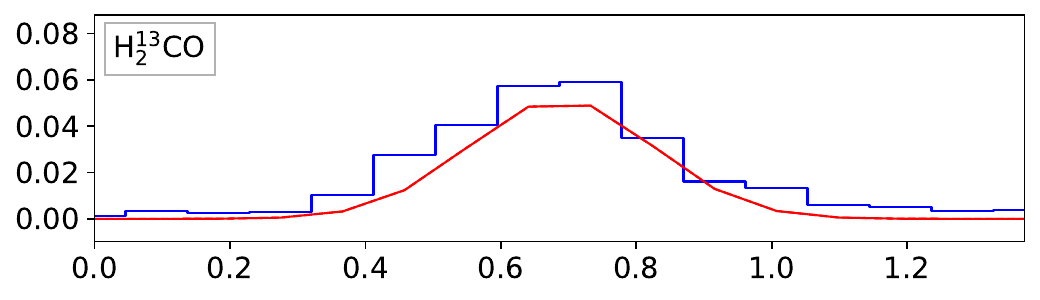}
\includegraphics[width=0.485\linewidth]{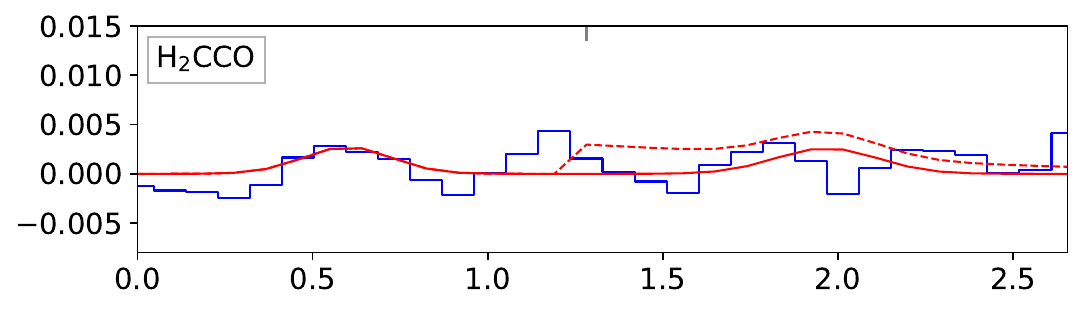}
\\
\includegraphics[width=0.485\linewidth]{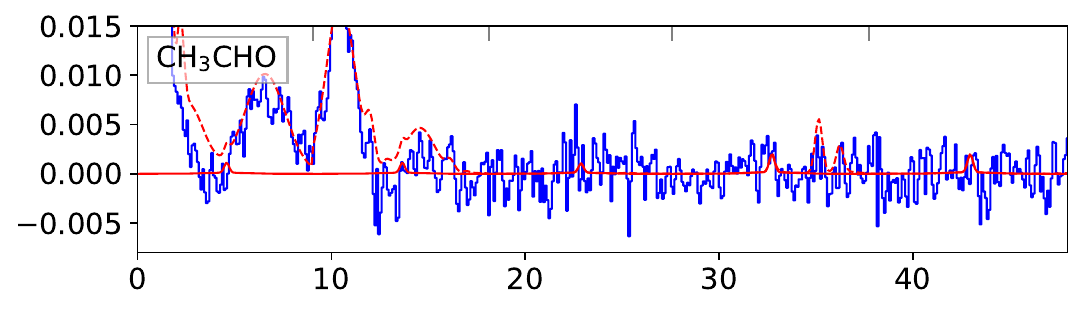}
\includegraphics[width=0.485\linewidth]{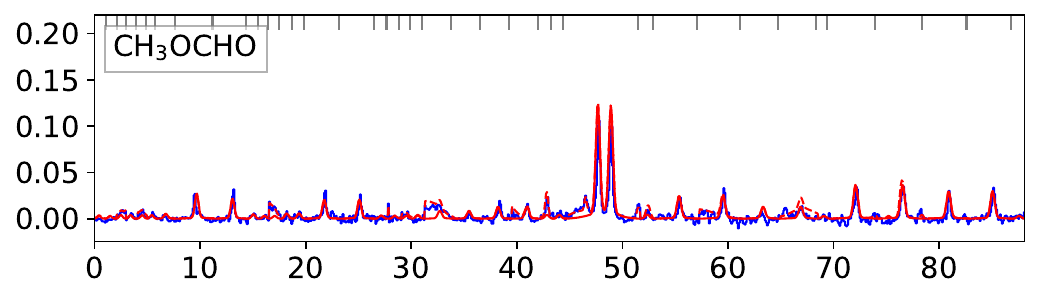}
\\
\includegraphics[width=0.485\linewidth]{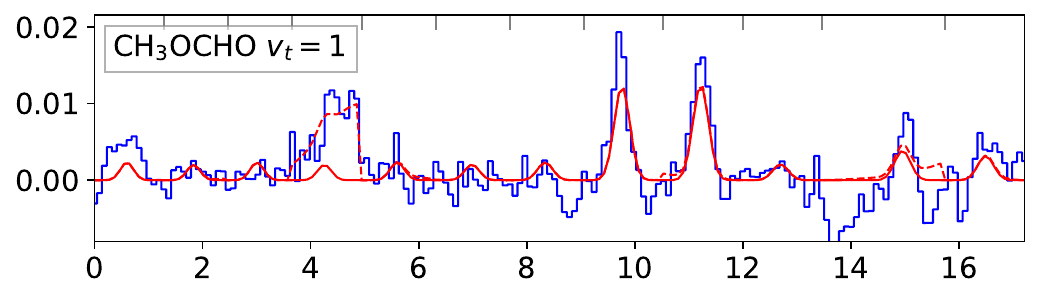}
\includegraphics[width=0.485\linewidth]{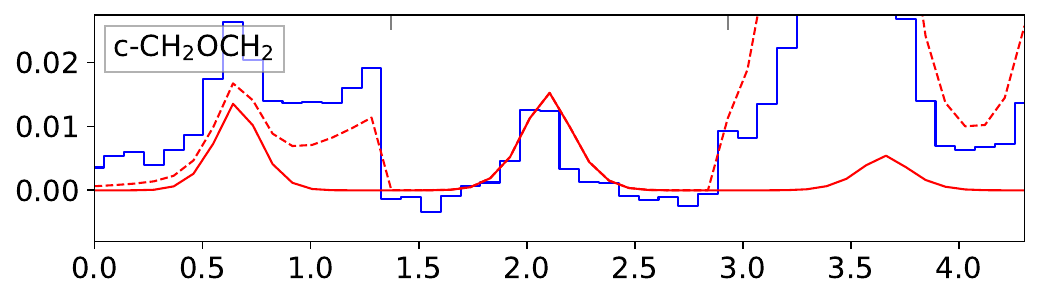}
\\
\includegraphics[width=0.485\linewidth]{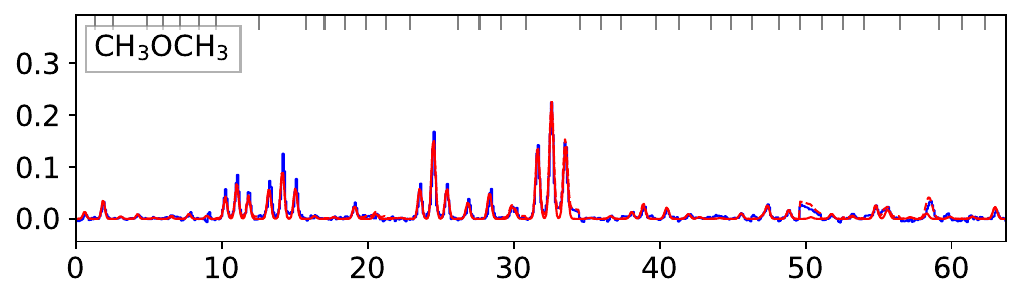}
\includegraphics[width=0.485\linewidth]{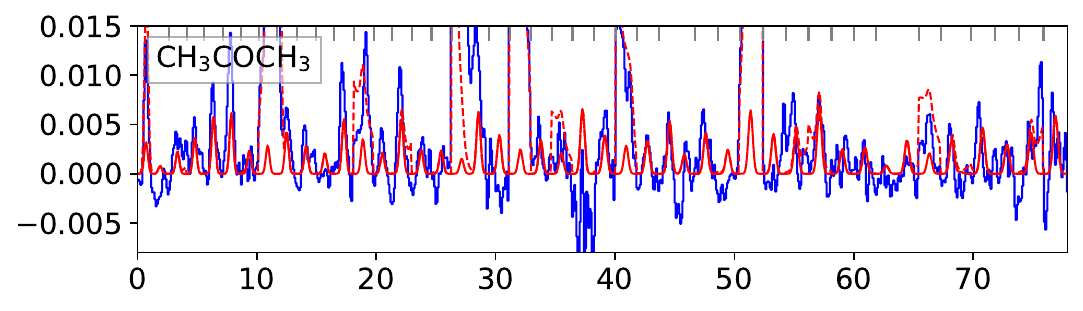}
\\
\includegraphics[width=0.485\linewidth]{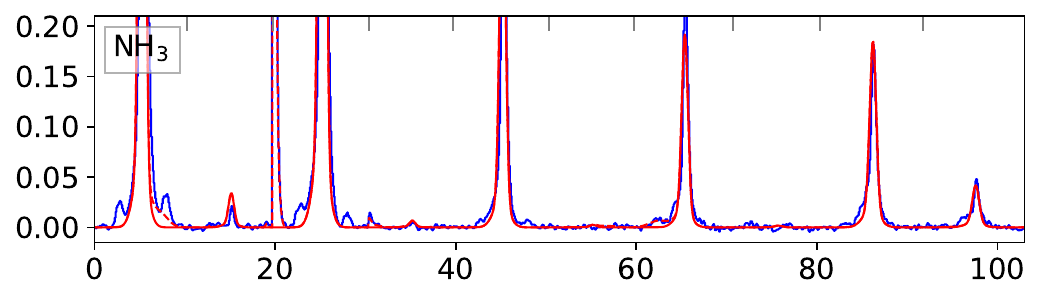}
\includegraphics[width=0.485\linewidth]{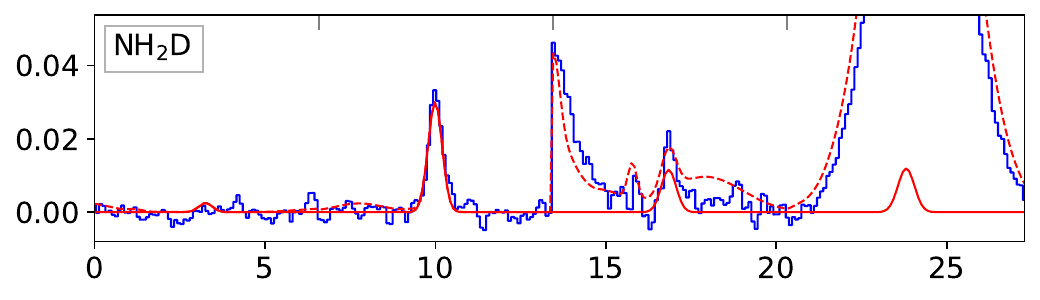}
\\
{\raggedright \centering  \textbf{Figure \thefigure} {\it continued} \par}
\end{figure*}
        
\begin{figure*}
\centering
{\raggedright \centering  \textbf{Figure \thefigure} {\it (continued)} \par}
\includegraphics[width=0.485\linewidth]{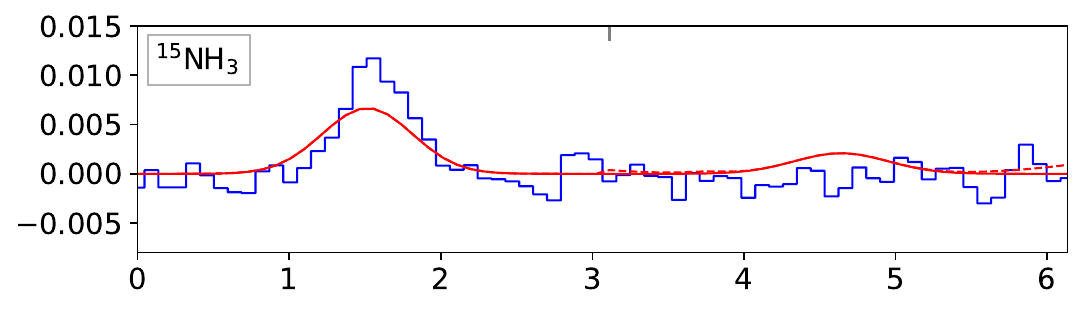}
\includegraphics[width=0.485\linewidth]{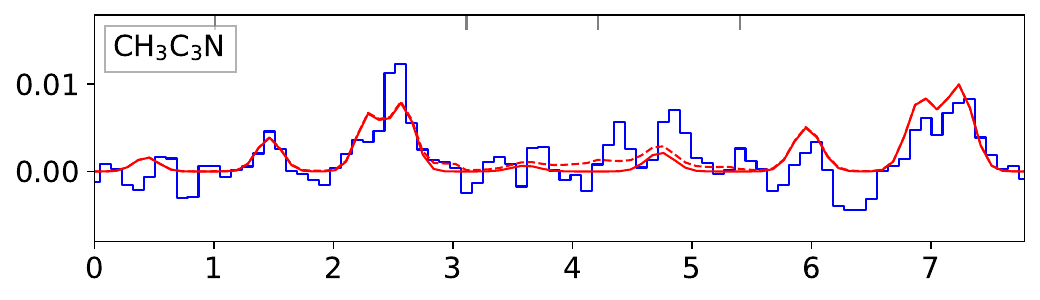}
\\
\includegraphics[width=0.485\linewidth]{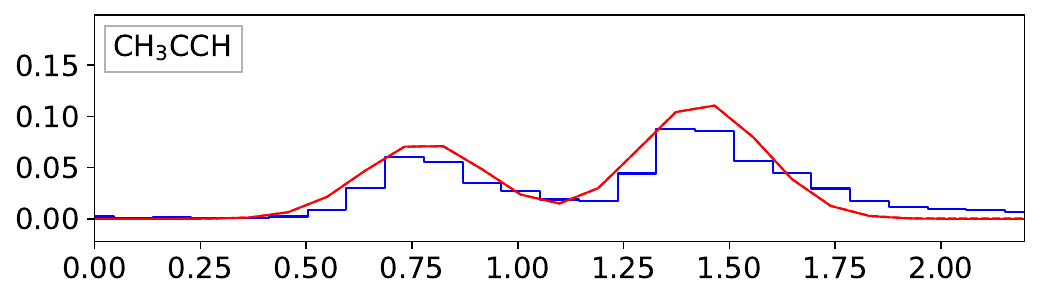}
\includegraphics[width=0.485\linewidth]{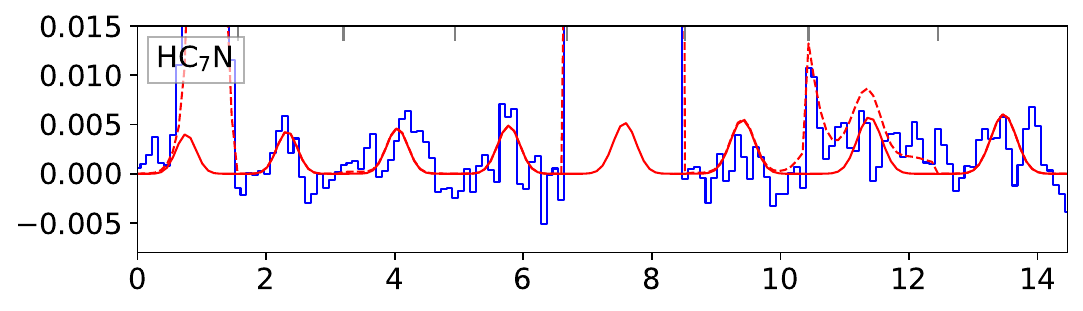}
\\
\includegraphics[width=0.485\linewidth]{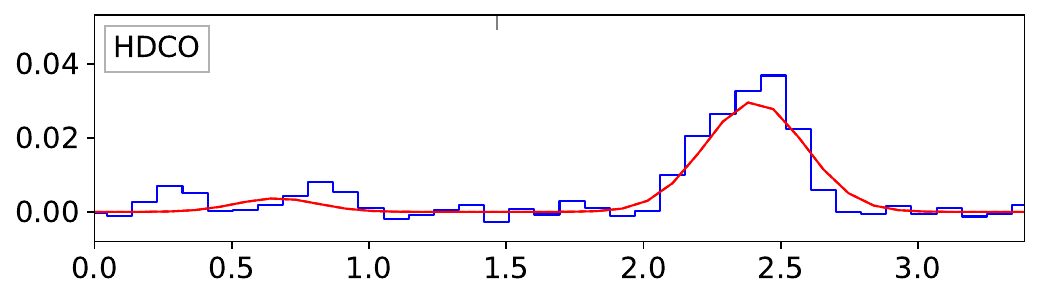}
\includegraphics[width=0.485\linewidth]{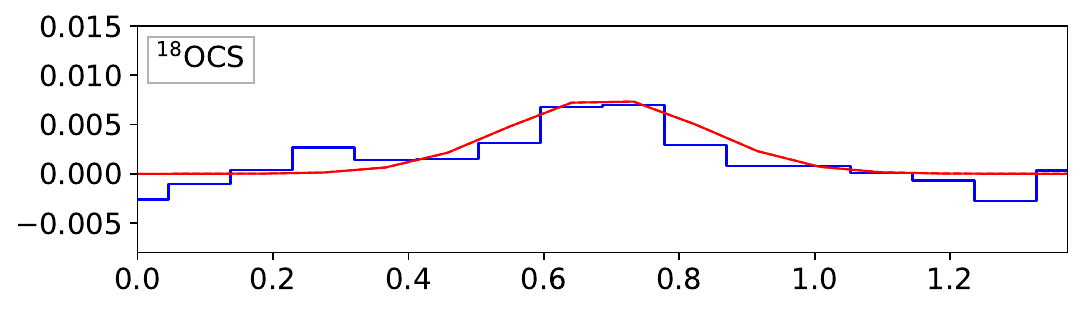}
\\
\includegraphics[width=0.485\linewidth]{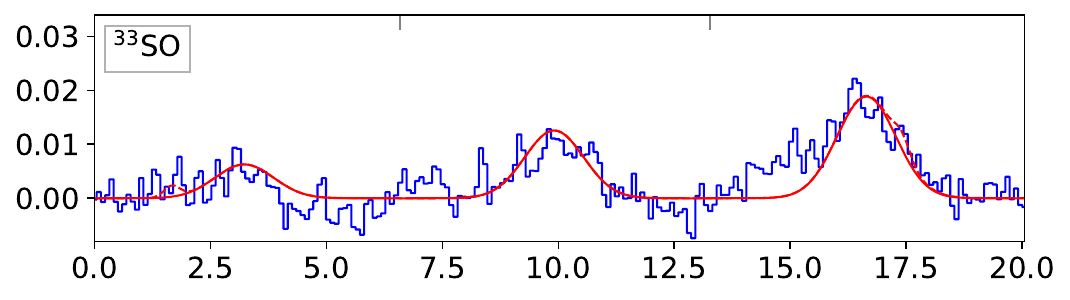}
\includegraphics[width=0.485\linewidth]{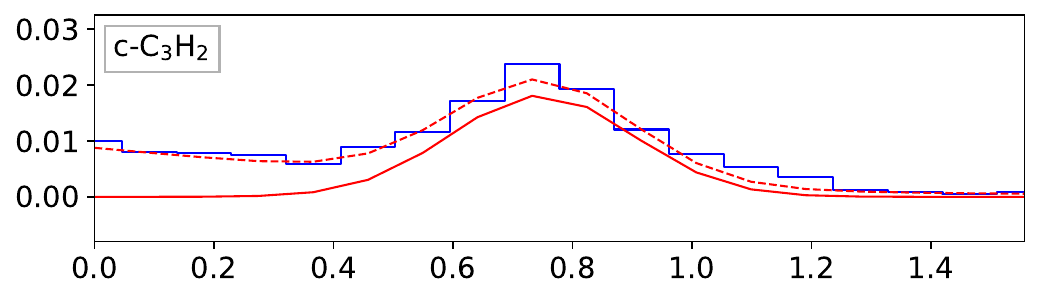}
\\
{\raggedright \centering  \textbf{Figure \thefigure} {\it continued} \par}
\end{figure*}

\clearpage

\section{Line list of molecular lines}
Table \ref{linelist} shows the results of identification and Gaussian fitting  of 
molecular lines, including RRLs which are blended with molecular lines.
The unblended RRLs are listed in Table \ref{linelist_rrls}. \\
{
\footnotesize
\textbf{Table Notes}:
\begin{itemize}\setlength{\itemsep}{0pt} \setlength{\parskip}{0pt}
\item[(1)] Doppler correction has been applied to $f_{\rm obs}$ assuming a 
source velocity  of 6 km s$^{-1}$ in LSR (Sect. \ref{sec_datareduct}). \\
Rows with empty $f_{\rm obs}$ correspond to blended lines.
\item[(2)] The emitter of ion RRL is denoted as \ion{X}{2}, and the $f_{rest}$ is adopted as 
the rest frequency of the corresponding \ion{C}{2} RRL. \\
The unidentified species is denoted as U.
\item[(3)] Rows with a same species name and transition label correspond
to different emission components of a same transition.\\
The transition labels for HCN $\varv_t=1$ are $J^p$ with $p$ the parity.\\
The transition labels for CH$_3$OCH$_3$ are  $J_K^s$ with $s$ the symmetry substate.\\
The transition labels are $(J,K)$ for NH$_3$ convention lines.
\item[(4)] The numbers in brackets in the 7th and 8th columns 
represent the uncertainties of the \textbf{last digital} of corresponding parameters.\\
\end{itemize}
}

\startlongtable
\begin{deluxetable}{llllllllll}\label{linelist}
\tablecaption{Linelist}
\tablehead{
\colhead{$f_{\rm obs}$} & \colhead{Species} & \colhead{$f_{\rm rest}$} & \colhead{Transition} & \colhead{$E_u$} & \colhead{$A_{ij}$} & 
\colhead{$\int {T_{\rm MB}}{\rm d}V$} & \colhead{$\Delta$V} & \colhead{$T_{\rm MB}^{\rm peak}$}& 
\colhead{$V_{\rm LSR}$} \\
\colhead{(MHz)} & \colhead{}  & \colhead{(MHz)} & \colhead{$J_{K_a,K_c}$} & \colhead{(K)} 
& \colhead{(10$^{-6}$ s$^{-1}$)} & \colhead{(K km s$^{-1}$)} & \colhead{(km s$^{-1}$)} & \colhead{K}
& \colhead{km s$^{-1}$}
}
\startdata
26120.55 & CH$_3$OH $\varv_t=1$            & 26120.584   & $10_{1,9}-11_{2,9}$             & 452.7   & 0.06        & 0.879(9)            & 4(1)                & 1.93E$-$01& 6.4(1)      \\
26124.84 & C$_2$H$_5$CN                    & 26124.592   & $3_{1,3}-2_{1,2}$               & 3.6     & 1.17        & 0.507(9)            & 13(1)               & 3.72E$-$02& 3(1)        \\
26149.88 & HDCO                            & 26150.160   & $8_{2,6}-8_{2,7}$               & 144.3   & 0.06        & 0.021(2)            & 2(1)                & 8.16E$-$03& 9(1)        \\
26180.75                      & C$_2$H$_5$CN $\varv 13$/$\varv 21$& 26180.635   &                                         & 299.8   & 0.06        & 0.049(9)            & 8(1)                & 5.95E$-$03& 5(1)        \\
26242.78 & $^{15}$NH$_3$                   & 26242.760   & $J_K=9_{9}-9_{9}$               & 852.5   & 0.4         & 0.078(9)            & 6(1)                & 1.18E$-$02& 6(1)        \\
26289.13                      & U                               &             &                                         &         &             & 0.050(2)            & 8(1)                & 5.60E$-$03&             \\
26313.06 & E-CH$_3$OH                      & 26313.192   & $11_{2,10}-11_{1,10}$           & 175.5   & 0.1         & 9.09(2)             & 6.4(1)              & 1.33E$+$00& 7.5(1)      \\
26450.51 & H$^{13}$CCCN                    & 26450.596   & $J=3-2$                         & 2.5     & 1.28        & 0.088(7)            & 8.8(9)              & 9.47E$-$03& 7.0(3)      \\
\enddata
\end{deluxetable}

\clearpage
\section{Line list of RRLs unblended with molecular lines}
Table \ref{linelist_rrls} shows the results of identification and Gaussian fitting of
RRLs. The RRLs blended with molecular lines are not included (see Table \ref{linelist}).\\
{
\footnotesize
\textbf{Table Notes}:
\begin{itemize}\setlength{\itemsep}{0pt} \setlength{\parskip}{0pt}
\item[(1)] Doppler correction has been applied to $f_{\rm obs}$ assuming a 
source velocity  of 6 km s$^{-1}$ in LSR (Sect. \ref{sec_datareduct}). \\
$f_{\rm obs}$ with a superscript of `*' means it is only marginally detected.\\
Rows with empty $f_{\rm obs}$ correspond to blended lines.
\item[(2)] The emitter of ion RRL is denoted as \ion{X}{2}, and the $f_{rest}$ is adopted as 
the rest frequency of the corresponding \ion{C}{2} RRL. \\
The unidentified species is denoted as U.
\item[(3)] The numbers in brackets in the 4th and 5th columns 
represent the uncertainties of the \textbf{last digital} of corresponding parameters.\\
\end{itemize}
}

\startlongtable
\begin{deluxetable}{lllllll}\label{linelist_rrls}
\tablecaption{Linelist}
\tablehead{
\colhead{$f_{\rm obs}$} & \colhead{Species} & \colhead{$f_{\rm rest}$}  & 
\colhead{$\int {T_{\rm MB}}{\rm d}V$} & \colhead{$\Delta$V} & \colhead{$T_{\rm MB}^{\rm peak}$} 
&\colhead{$V_{\rm LSR}$}\\
\colhead{(MHz)} & \colhead{}  & \colhead{(MHz)} & \colhead{(K km s$^{-1}$)} & \colhead{(km s$^{-1}$)} & 
\colhead{K} & \colhead{km s$^{-1}$}
}
\startdata
26162.51                      & H135$\lambda$                   & 26161.602    & 0.21(1)             & 17(1)               & 1.16E$-$02& -4.4(6)     \\
26197.28                      & H154$\rho$                      & 26196.134    & 0.073(9)            & 20(1)               & 3.48E$-$03& -7(1)       \\
26206.55                      & H142$\nu$                       & 26205.853    & 0.313(9)            & 30(1)               & 9.93E$-$03& -2(1)       \\
26214.56                      & H131$\kappa$                    & 26213.600    & 0.439(9)            & 25(1)               & 1.62E$-$02& -5(1)       \\
26284.90                      & H159$\tau$                      & 26283.781    & 0.12(1)             & 21(3)               & 5.08E$-$03& -7(2)       \\
26309.59                      & H151$\pi$                       & 26308.667    & 0.028(8)            & 9(4)                & 2.81E$-$03& -5(2)       \\
26316.70                      & \ion{X}{2}125$\beta$            & 26315.897    & 0.04(1)             & 12(3)               & 3.13E$-$03& -3(2)       \\
26326.36                      & H98$\delta$                     & 26325.466    & 4.27(2)             & 25(1)               & 1.60E$-$01& -4(1)       \\
\enddata
\end{deluxetable}

\end{CJK*}
\end{document}